\documentclass[aoas,preprint]{imsart}
\usepackage[utf8]{inputenc}
\usepackage[margin=1.0in]{geometry}
\usepackage[parfill]{parskip}
\usepackage{amsmath}
\usepackage{wrapfig}
\usepackage{bm}
\usepackage{hyperref}
\usepackage{graphicx} 
\usepackage[dvipsnames]{xcolor} 
\usepackage{natbib}
\usepackage{verbatim}
\usepackage{booktabs,floatrow}

\makeatletter
\g@addto@macro{\UrlBreaks}{\UrlOrds}
\makeatother

\usepackage{caption}
\newcommand*{\vertbar}{\rule[-1ex]{0.5pt}{2.5ex}}

\newcommand{\undset}[2]{\underset{\scriptscriptstyle #1}{#2\strut}}

\DeclareUnicodeCharacter{2212}{-}

\begin{document}

\begin{frontmatter}
\title{Bayesian joint modeling of chemical structure and dose response curves}
\runtitle{Bayesian joint modeling of chemical structure and dose response curves}

\begin{aug}
\author{\fnms{Kelly R.} \snm{Moran}\thanksref{m1}\ead[label=e1]{kelly.r.moran@duke.edu}},
\author{\fnms{David} \snm{Dunson}\thanksref{m1}\ead[label=e2]{dunson@duke.edu}},
\author{\fnms{Matthew W.} \snm{Wheeler}\thanksref{m2}\ead[label=e3]{matt.wheeler@nih.gov}},
\and
\author{\fnms{Amy H.} \snm{Herring}\thanksref{m1}
\ead[label=e4]{amy.herring@duke.edu}}

\runauthor{K. R. Moran et al.}

\affiliation{Duke University\thanksmark{m1}}
\affiliation{
National Institutes of Health\thanksmark{m2}}

\end{aug}

\begin{abstract}
Today there are approximately 85,000 chemicals regulated under the Toxic Substances Control Act, with around 2,000 new chemicals introduced each year. It is impossible to screen all of these chemicals for potential toxic effects either via full organism \textit{in vivo} studies or \textit{in vitro} high-throughput screening (HTS) programs. Toxicologists face the challenge of choosing which chemicals to screen, and predicting the toxicity of as-yet-unscreened chemicals. Our goal is to describe how variation in chemical structure relates to variation in toxicological response to enable \textit{in silico} toxicity characterization designed to meet both of these challenges. With our Bayesian partially Supervised Sparse and Smooth Factor Analysis ($\text{BS}^3\text{FA}$) model, we learn a distance between chemicals targeted to toxicity, rather than one based on molecular structure alone. Our model also enables the prediction of chemical dose-response profiles based on chemical structure (that is, without \textit{in vivo} or \textit{in vitro} testing) by taking advantage of a large database of chemicals that have already been tested for toxicity in HTS programs. We show superior simulation performance in distance learning and modest to large gains in predictive ability compared to existing methods. Results from the high-throughput screening data application elucidate the relationship between chemical structure and a toxicity-relevant high-throughput assay. An \textbf{\textsf{R}} package for $\text{BS}^3\text{FA}$ is available online at \url{https://github.com/kelrenmor/bs3fa}.
\end{abstract}



\end{frontmatter}

\section{Introduction}

Daily life involves being exposed to a variety of chemical substances from diverse sources and at varying concentrations. A myriad of legislation and regulatory bodies work to assess consumer and industrial products for toxicity and reduce exposure risk. The Toxic Substances Control Act (TSCA), passed by Congress in 1976 and administered by the US Environmental Protection Agency (EPA), regulates the bulk of\footnote[1]{Exceptions regulated under different legislation include foods and food additives, drugs, cosmetics, pesticides, tobacco products, research substances used in small quantities, and radioactive materials and waste.} new and existing chemicals in the United States (US). When the TSCA was enacted, around 60,000 chemicals were grandfathered into the program and effectively considered safe for use. The EPA has struggled to catch up on this backlog while also keeping up with the rate of new introductions (roughly 2,000 chemicals per year) as they assess chemicals for potential toxicity. High-throughput screening methods have proved vital to this effort as they allow researchers to quickly conduct millions of tests. 

The EPA's Toxicity Forecaster (ToxCast) research program, in which thousands of chemicals are tested in more than 700 high-throughput assay endpoints, is used to prioritize, screen and evaluate chemicals for potential toxic effects \citep{dix2006toxcast, judson2009vitro, kavlock2012update}. However, even high-throughput toxicity screening (HTS) programs, which allow for the relatively cheap and fast collection of dose-response information via \textit{in vitro} studies rather than full organism \textit{in vivo} studies, are still too slow and expensive to be able to study all chemicals. \textit{In silico} studies, i.e. those performed via computer modeling rather than in the lab, can be used to guide the design of and supplement the results from lab-based studies. Specifically, the characterization of an activity relevant chemical distance \textit{in silico} enables more targeted design of further \textit{in vitro} studies, increasing the efficiency of resource allocation. In addition, predicting toxicity via such studies helps bridge the gap between the number of chemicals of interest and the number with known toxicological profiles.

\begin{figure}[!htpb]
\centering
\includegraphics[width=0.6\textwidth]{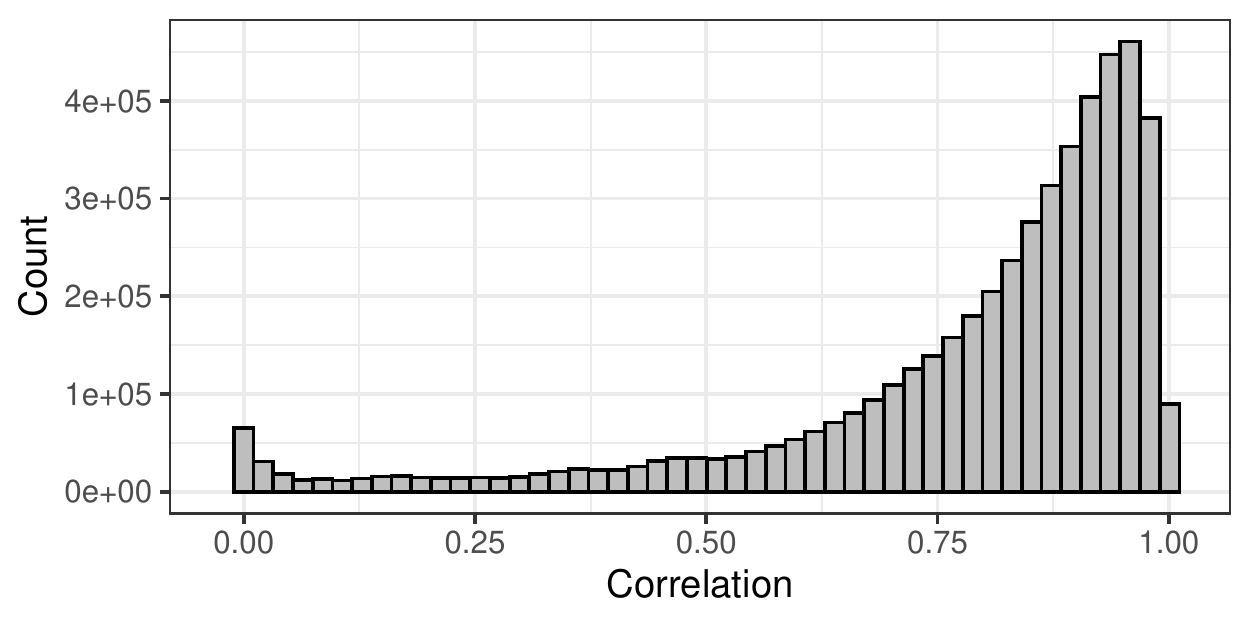}
\caption{Pairwise correlation between each of the 777 molecular descriptors in ToxCast for the chemicals profiled.}
\label{fig:mold2Correlation}
\end{figure}

The goal of this work is to make inferences about how variation in chemical structure relates to variation in toxicological response. Sparse function-on-scalars regression models \citep{chen2016variable, barber2017function, fan2017high, kowal2018bayesian} do this in a limited way by selecting the important chemical structure features and giving them appropriate coefficients or weights. Because there are many redundant and highly correlated structure features (see Figure \ref{fig:mold2Correlation}), a PCA-esque approach that introduces latent factors related to the major directions of variation in the molecular structure is more informative than such penalized regression approaches. However, simply performing PCA or other unsupervised dimension reduction approaches on the chemical structure ignores the distinction between \textit{overall} variation and \textit{toxicity-relevant} variation in the molecular structure. A supervised dimension reduction approach, on the other hand, provides a coherent and flexible framework within which to describe the relationship between molecular variation and activity variation via a shared latent subspace. 

\begin{figure}[!phtb]
\centering
\includegraphics[width=0.95\textwidth]{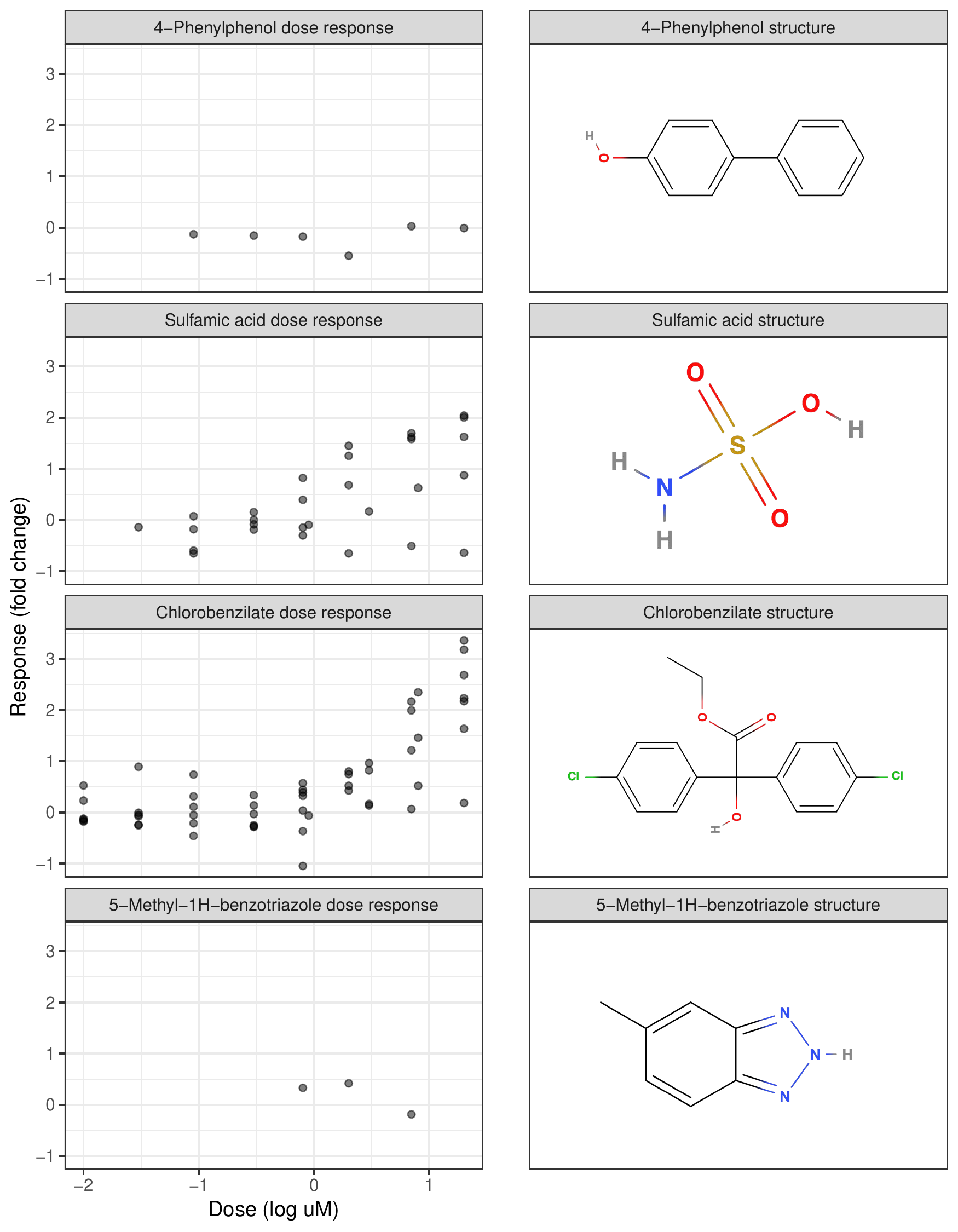}
\caption{Left: Dose response data for example chemicals from the ToxCast ATG PXR assay (i.e., $\bm{y}_i$). Right: 2D chemical structure diagrams for example chemicals (converted from SMILES into $\bm{x}_i$ using the Mold2 software).}
\label{fig:ExDRChems}
\end{figure}

Dose response data and molecular structure are the two sources of information in ToxCast relevant to addressing our goal. Explicitly, chemical $i$ in ToxCast has two relevant pieces of information: the vector of response observations at $D$ doses $\bm{y}_i = [y_i(d_1),\ldots,y_i(d_D)]'$, and the vector of $S$ molecular features $\bm{x}_i = [x_{i1},\ldots,x_{iS}]'$. Observations $\bm{y}_i$ are sparse, noisy, and not on a regular grid. For an example see Figure \ref{fig:ExDRChems}; Chlorobenzilate has 54 observations at 11 unique doses, yet 5-Methyl-1H-benzotriazole only has 3 doses with one observation each. Not all aspects of the feature space (i.e. not all entries in $\bm{x}_i$) are likely to be relevant to the toxicological response. 

Response variables in ToxCast are assay specific. Assays in ToxCast each measure a single endpoint, e.g. the binding to a certain receptor protein, or the transcription of a target gene. The specific assay endpoint considered in our real data example, the AttaGene pregnane X receptor (PXR) assay, records the fold-change values in the activity of the nuclear pregnane X receptor for drug-treated vs. control-treated human hepatic cells (specifically a HG19 subclone of HepG2). Dimethyl sulfoxide (DMSO) is used as the negative control. The response is measured via reporter RNAs that are produced proportionately to the activity of corresponding transcription factor (here, PXR). This assay has been shown to be related to the body's response to toxic substances \citep{kliewer2002nuclear}, so a higher response value for this assay endpoint can be interpreted as higher level of toxicity.

This work makes no monotonicity assumptions on the shape of the response. There is a nonzero baseline activity level of the nuclear pregnane X receptor in unstimulated hepatic cells, so both induction and suppression of the activity of this transcription factor are possible outcomes in response to stimulation, rendering a positive monotonicity constraint unsuitable. Furthermore, the lack of any monotonicity constraint (not specifically a positive one) leaves open the possibility for a biphasic response such as hormesis to be fit.

\begin{figure}[!htpb]
\centering
\includegraphics[width=0.45\textwidth]{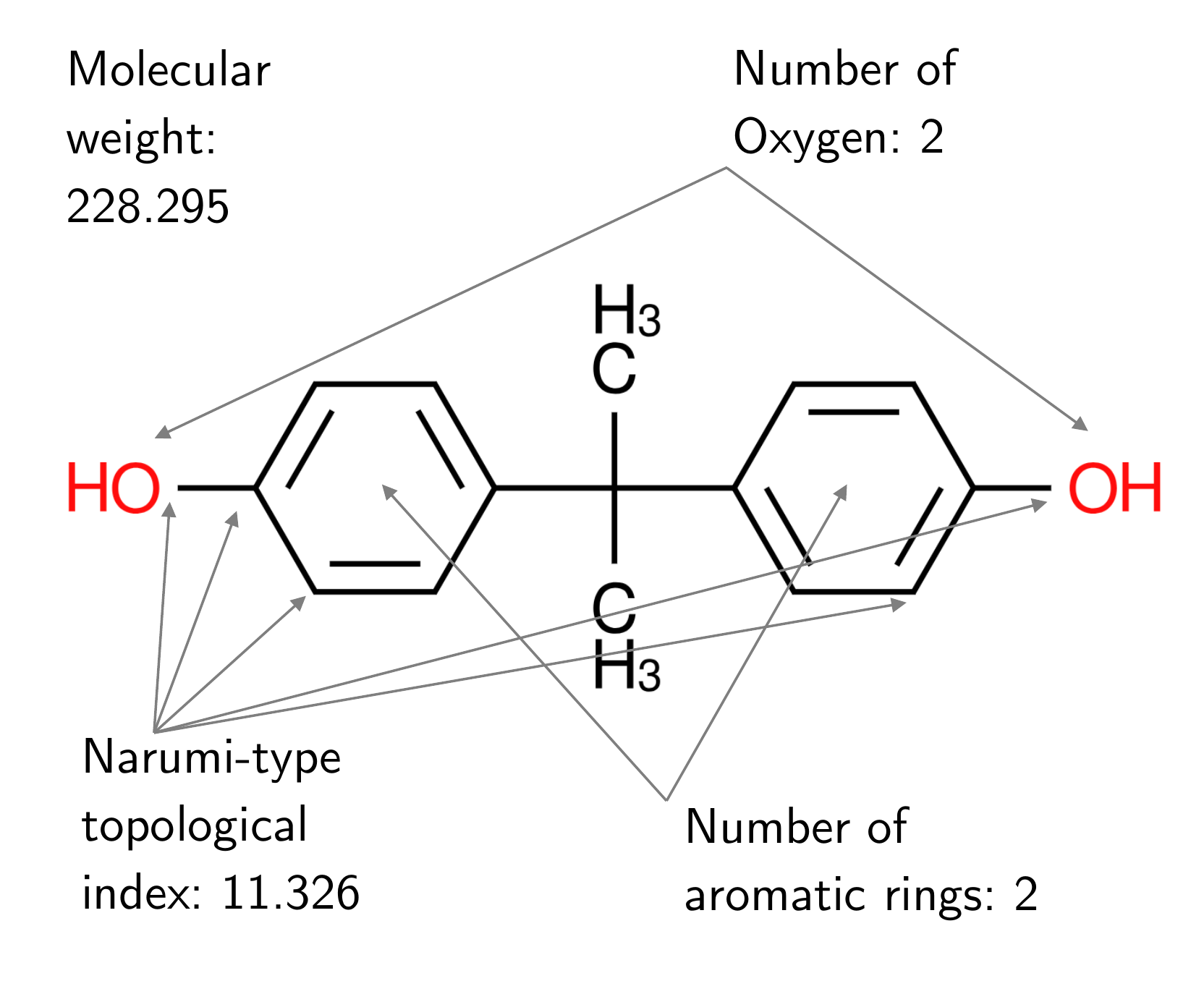}
\caption{Numeric values for select Mold2 traits of Bisphenol A (BPA). The chemical formula for BPA is $\text{C}_{15}\text{H}_{16}\text{O}_2$, and its SMILES descriptor is CC(C)(C1=CC=C(C=C1)O)C2=CC=C(C=C2)O.}
\label{bpa_example}
\end{figure}

The tool used to quantitatively summarize a chemical's molecular structure is Mold2  \citep{hong2008mold2}, which generates a set of 777 numeric descriptors using the simplified molecular-input line-entry system (SMILES) specification \citep{weininger1988smiles}. See Figure \ref{bpa_example} for select Mold2 output for an example chemical, and \citep{hong2012mold2} for a discussion of the use of Mold2 in Quantitative Structure–Activity Relationship (QSAR) models. 

In order to coherently model both structural and toxicological response variation, we propose a Bayesian partially Supervised Sparse and Smooth Factor Analysis ($\text{BS}^3\text{FA}$) model. The model assumes structured variation in the molecular features $\bm{x}_i$ is driven by two sets of latent factors: call these $F_{x\text{-specific}}$ and $F_{\text{shared}}$. $F_{x\text{-specific}}$ is unrelated to the toxicological response and is responsible for structured molecular variability that does not impact toxicity. $F_{\text{shared}}$ is assumed to drive variation in the toxicological response $\bm{y}_i$, and thus is responsible for structured molecular variability that \textit{does} impact toxicity. The directions spanned by these two sets of latent factors can be thought of as the ``toxicity-irrelevant'' and ``toxicity-relevant'' spaces, respectively.

Chemical similarity can be characterized by proximity in this latent toxicity-relevant space, enabling a measure of distance with uncertainty quantification that is adapted to the particular response space of interest. Such a metric is powerful because (1) it is based on a subspace driving variation in activity, whereas proximity with respect to the full set of molecular descriptors does not necessarily mean proximity with respect to activity \citep{martin2002structurally, nikolova2003approaches}, and (2) it is purely statistically derived, requiring no knowledge of the fundamental chemical and biological processes responsible for the activity, as such information is not always available. Such an activity-relevant distance metric could be used by toxicologists in the design of diverse chemical libraries or to select new compounds to augment a screening collection such as ToxCast. 

As with function-on-scalars regression approaches, the $\text{BS}^3\text{FA}$ model allows for the prediction of activity profiles for chemicals that have not yet been screened in ToxCast. It does so by embedding the full set of molecular features for a new chemical into the latent toxicity-relevant feature space $F_{\text{shared}}$, and then projecting this embedding out to the activity space. The predicted dose-response profiles can be used to generate point and interval estimates for common univariate toxicological outcomes of interest, such as 50\% activity concentration (AC50), maximum activity, or the area under curve (AUC), which can be used in place of the as of yet unobserved \textit{in vitro} results for that chemical.

The rest of the paper is organized as follows. First, we describe existing and potential approaches to modeling chemical structure and activity. Then, the $\text{BS}^3\text{FA}$ model is described and its performance is compared to that of existing algorithms on simulated data sets. Next, a detailed analysis of the motivating application data set is considered, where the $\text{BS}^3\text{FA}$ model is run with Mold2 chemical features and the Attagene PXR assay from the ToxCast data set as input data. Finally, the results are discussed and future areas of research are highlighted. 

\section{Background}

\begin{figure}[htp]
\centering
\includegraphics[height=0.1\textheight]{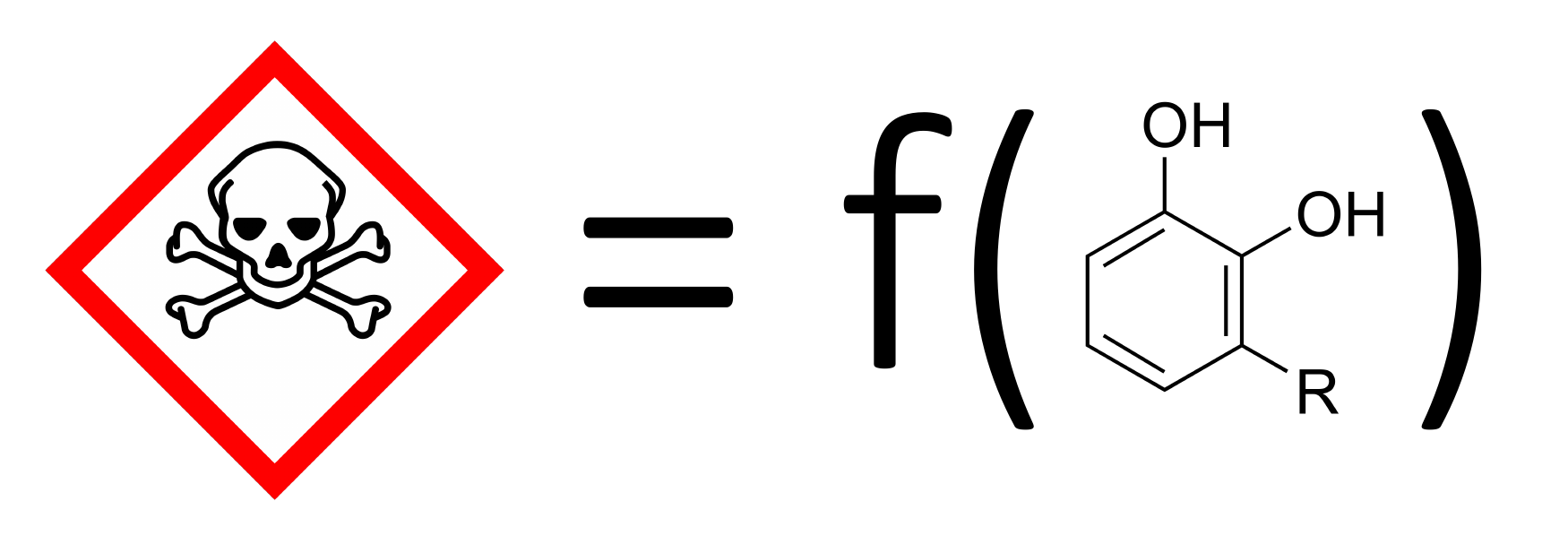}
\caption{Quantitative Structure-Activity Relationship (QSAR) models predict toxicity as a function of chemical structure.}
\label{fig:QSAR}
\end{figure}

QSAR models (see Figure \ref{fig:QSAR}) are based on the assumption that chemicals with similar features are likely to have similar effects. The ToxCast data poses two main challenges for QSAR modeling. First, it is often not trivial to characterize similarity in activity-relevant chemical feature space well \citep{martin2002structurally, nikolova2003approaches}. Second, the majority of QSAR models aim to relate structure to a \textit{summary} of the data across times/doses (e.g., \citet{liu2011classification, patel2014relating, o2019linked}) rather than to the \textit{full} dose response curves. 

We discuss these considerations in the context of existing QSAR models and other related approaches not yet applied to QSAR models. To the authors' knowledge, no existing approaches are able to address the challenge of learning a low-dimensional representation for multivariate feature data $\bm{x}_i$ partially supervised by sparse functional data $\bm{y}_i$.  

\subsection{QSAR approaches for dose response profiles}

Two existing QSAR approaches have attempted to relate molecular descriptors to full dose response curves \citep{low2015bayesian, wheeler2019bayesian}. In \citet{low2015bayesian}, a Bayesian regression tree is defined over functions where each leaf represents a different dose-response surface. This method was used to learn about the relationship between chemical properties and observed dose-response. However, the model lacks the ability to scale to the numbers of chemicals and molecular descriptors considered here. Furthermore, predictive performance was found to be lacking in leave-one-out analysis. Finally, the code was designed under the assumption that each chemical would be tested at the same doses with the same number of replicates at each dose.

The Bayesian Additive Adaptive Basis Tensor Product (BAABTP) model \citep{wheeler2019bayesian} is designed purely for prediction. It learns basis functions via independent Gaussian process (GP) priors over the molecular structure space and the dose space. In the model, step one is to perform PCA on the set of Mold2 chemical descriptors. Step two is to use the principal feature space explaining 95\% of the variation in this Mold2 descriptor set as the input to the distance kernel for the molecular structure GPs.

\begin{figure}[htp]
\centering
\includegraphics[height=0.3\textheight]{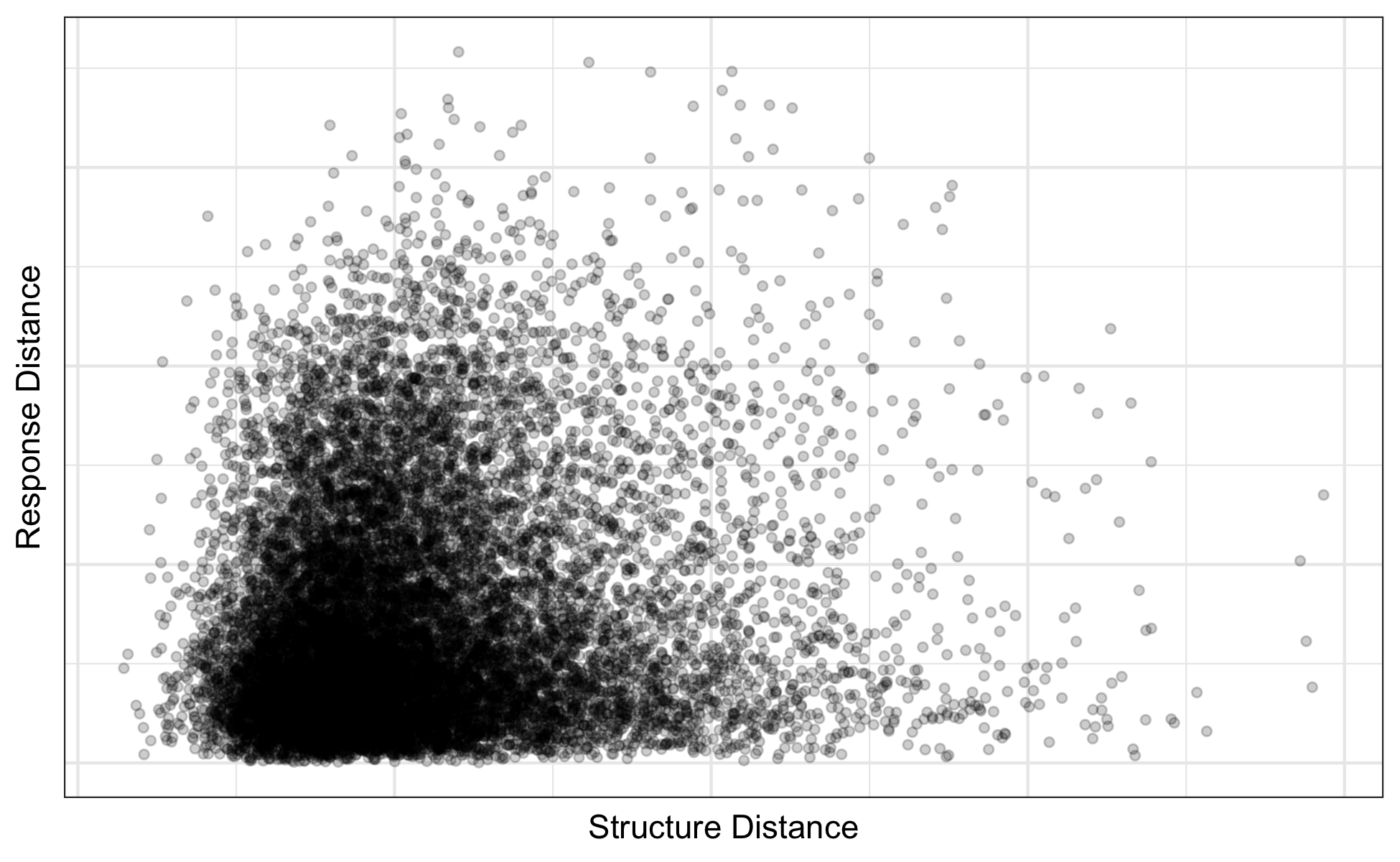}
\caption{Relationship between chemical ``structure distance'' and chemical ``activity distance'' when PCA and functional PCA (FPCA) are performed \textit{independently} on the Mold2 chemical structures and the ToxCast dose response curves. Each point on the graph shows the Euclidean distance between the (functional) principal component scores accounting for 95\% of the variability in the data for one pair of chemicals on the ($y$-) $x$-axis.}\label{fig:dist_fpca_pca}
\end{figure}

The BAABTP model has two major problems, both stemming from the Gaussian process prior over chemical structure. First, the model becomes computationally intractable when the number of chemicals increases past a few thousand-- the number of chemicals tested in the ATG PXR assay has increased from under 1,000 up to nearly 4,000 in the time since the data were analyzed in \citet{wheeler2019bayesian}, and this number will only continue to grow. Second, the GP priors over chemical structure rely on a concept of molecular distance based on total, rather than toxicity-relevant, variability. Mold2 descriptors are a numeric representation of the 2D structure of a chemical. Thus, while the leading principal components (PCs) account for the majority of structural variability across chemicals, these leading PCs may not be those most relevant to toxicity. Figure \ref{fig:dist_fpca_pca} illustrates this phenomenon, showing that chemicals may be close in PCA-based structure space but distant in response space. In other words, similarity in directions of highest structural variability does not necessarily correspond to similarity in directions of highest activity variability. 

\subsection{Towards the proposed approach}


Supervised and sparse functional PCA (supSFPCA) \citep{li2016supervised} defines a hierarchical model to provide supervision for dimension reduction of functional data by another multi-output data source. A small modification to the penalty term used in this algorithm would allow for the opposite relation (i.e., to supervise the dimension reduction of the feature data $\bm{x}_i$ by functional dose-response data $\bm{y}_i$), but the larger issue is that the supSFPCA algorithm is designed such that the algorithm finds directions that maximize \textit{unexplained} variability in $\bm{x}_i$ (that is, what isn't accounted for by $\bm{y}_i$) rather than finding directions that maximize variability \textit{explained} by $\bm{y}_i$. As such, this algorithm is of little use for prediction or for learning about directions of variability of most relevance for toxicity. 

The Bayesian latent factor regression (LFRM) model of \citet{montagna2012bayesian} characterizes $\bm{y}_i$ as a linear combination of a high-dimensional set of basis functions. The basis functions themselves are fixed rather than learned, which means that a large number of basis functions must be included in order to flexibly model many possible functional shapes. This high number of bases drives the choice to use a latent factor regression model on the basis coefficients rather than defining a regression model on the basis coefficients themselves. Thus the latent factor scores enter the model in a ``lower'' layer in \citet{montagna2012bayesian} than they do in $\text{BS}^3\text{FA}$, and are a linear \textit{function of} covariates $\bm{x}_i$ rather than being assumed to \textit{drive variability in} $\bm{x}_i$. A shrinkage prior is placed on the regression terms associated with the latent factor scores in the model, allowing covariates to impact certain facets of the functional response while others are shrunk to zero. 

Our approach was inspired by the conceptual goal of separating variability shared by $\bm{x}_i$ and $\bm{y}_i$ into toxicity-relevant, toxicity-irrelevant, and noise components, similar to the idea behind the Joint and Individual Variation Explained (JIVE) method \citep{lock2013joint}, and an analogous joint Bayesian factor model (JBFM) \citep{ray2014bayesian}. We utilize similar shrinkage tools as LFRM and the Bayesian function-on-scalars regression (B-FOSR) of \citep{kowal2018bayesian}, implementing sparsity-inducing coefficient-level priors to account for the high-dimensional predictors, and imposing ordered sparsity so as to reduce the impact of the choice of latent subspace dimension. Unlike LFRM and like B-FOSR, the $\text{BS}^3\text{FA}$ model learns a flexible basis for the functional response rather than prescribing a set of basis functions. Like LFRM, JBFM, and B-FOSR, modeling takes place within a Bayesian framework for unified parameter estimation, prediction, and uncertainty quantification about posterior summaries of interest, and model components are identifiable.

Like many of the approaches described in the previous sections, our model is able to predict the dose-response profiles for new chemicals. Unlike previous models, our model is able to learn a toxicity-relevant subspace underlying variation in both the chemical structure and toxicological response. This subspace can be used to describe a statistically-driven activity-relevant distance between chemicals. It can also provide insight into how toxicity-relevant variability manifests across both molecular structure and dose-response profiles.


\section{$\text{BS}^3\text{FA}$ model}\label{sec:methods}

\begin{figure}[!htb]
\centering
\includegraphics[width=0.7\textwidth]{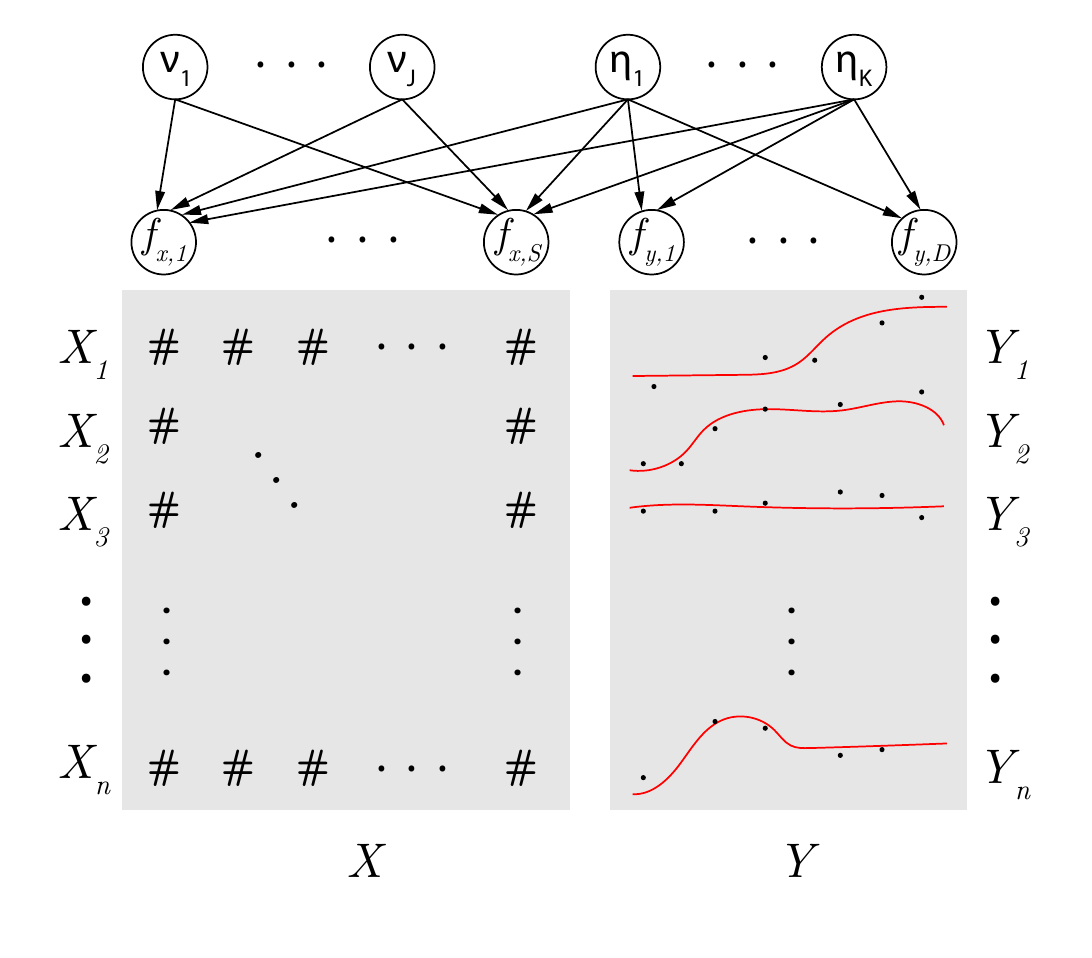}
\caption{Visual representation of the $\text{BS}^3\text{FA}$ model. Entries in $X$ are the chemical feature descriptors from Mold2, while  entries in $Y$ are noisy realizations of an underlying smooth dose response curve. Observations $f_{x,1},\ldots,f_{x,S}$ are the true mean of the chemical features and $f_{y,1},\ldots,f_{y,D}$ are the true mean of the dose response values. Latent variables $\eta_1,\ldots,\eta_K,$ the underlying toxicity-relevant factors, are shared, and  $\nu_1,\ldots,\nu_K,$ the underlying toxicity-irrelevant factors, are specific to the chemical features. Arrows denote probabilistic dependency.}
\label{mod_image}
\end{figure}

Figure \ref{mod_image} gives a visual representation of the $\text{BS}^3\text{FA}$ model. $\text{BS}^3\text{FA}$ is able to: (1) learn a linear low dimensional latent space underlying both molecular structure and activity, (2) develop a distance metric for chemicals relying only on molecular structures relevant to toxicity, (3) handle responses observed at a sparse, irregularly spaced set of doses, and (4) enable activity predictions for chemicals having no observed dose response information.

\subsection{Model specification}

\textbf{The characteristics of chemical structure and toxicological response can likely be summarized using fewer descriptors.} As shown in Figure \ref{fig:mold2Correlation}, many molecular descriptors tend to exhibit high correlation. Furthermore, their realized value can often be attributed to some underlying trait of the molecule (e.g., many descriptors are largely driven by molecule size; the descriptors number of carbon, number of oxygen, molecular weight, and number of atoms in the molecule all have pairwise correlation above 0.9). Similarly, dose response profiles exhibit a somewhat limited range of shapes, suggesting possibly a few underlying functional ``building blocks'' comprising variation in activity.

Factor modeling is a means by which to model variability in high dimensional data via an underlying lower dimensional subspace. For some set of observations $\{\bm{z}_i\}_{i=1}^N,$ where $\bm{z}_i$ is the $P-$dimensional vector of measurements for observation $i,$ the traditional (non-joint) factor model is:
\begin{align}
\begin{split}
\bm{z}_i = \Lambda \bm{\eta}_i + \bm{\epsilon}_i, \quad & \bm{\eta}_i \sim \text{N}_K(0,I), \quad \bm{\epsilon}_i \sim \text{N}_P(0, \Sigma_0), \\
& \Sigma_0 = \text{diag}(\sigma_1^2, \ldots, \sigma_P^2) \\
& i=1, \ldots, N.
\end{split}
\end{align}

The prior induced on the latent $\bm{z}_i$ by integrating out the unknown $\bm{\eta}_i$ is then:
\begin{equation}
\bm{z}_i \sim \text{N}(0, \Lambda \Lambda' + \Sigma_0),
\end{equation}
yielding a lower dimensional representation of the covariance between measurements.

\textbf{Chemical features are often non-normal (e.g. count, skewed continuous, or binary).} Many chemical descriptors from Mold2 are counts of particular elements (number of carbon, number of oxygen, etc.). 

In order to allow this framework to encompass data of mixed type, define:
\begin{align}
x_{is} &= f_s(z_{is}), \quad  i=1,\ldots, N, \quad s=1,\ldots,S.  
\end{align}
The particular link function $f_s$ depends on the feature specification, allowing for mixed scale data via selection of an appropriate link by scale and type. Let $f_s(z_{is}) = z_{is}$ or $f_s(z_{is}) = \text{log}(z_{is})$ for continuous $x_{is}$, with the latter chosen for strictly positive and positively skewed cases. Let $f_s(z_{is}) = 1(z_{is}>0)$ for binary $x_{is},$ where $1(\cdot)$ is an indicator function taking the value of 1 when the argument is true and 0 when the argument is false. Categorical variables may be incorporated under this framework by transforming the $C$ categories into $C-1$ binary variables indicating whether or not the categorical value for that individual took a given non-baseline category value; the result is that either none or one of the $C-1$ variables will take on a value of 1. Finally for count $x_{is}$, which may or may not be zero-inflated, let $f_s(z_{is})$ be a rounding operator such that $f_s(z_{is}) = 0$ if $z_{is}<0$ and $f_s(z_{is}) = t$ if $t-1 \leq z_{is} < t,$ as specified in \citet{canale2013nonparametric}.

\textbf{There is likely a shared low dimensional space underlying chemical features and activity.} $\text{BS}^3\text{FA}$ assumes that some underlying factors explain all of the variation in the dose response curves and, jointly, part of the variation in the associated chemical features. Recall $\bm{z}_i$ and $\bm{y}_i$ denote $S-$ and $D-$dimensional (latent) continuous features and observed dose response curves, respectively, for observation $i$. Assume that the indexing of $\bm{y}_i$ is such that the function is `in order' (for the ToxCast data, in order means that the $D$ unique doses are sorted such that the doses increase with index). Also, for notational convenience assume that functional data $\bm{y}_i$ are only observed once per index (the case of notationally awkward multiply observed doses is explicitly addressed in the Gibbs sampler described in the Appendix). $\text{BS}^3\text{FA}$ models
\begin{align}
\begin{split}\label{main_eq}
\undset{S \times 1}{\bm{z}_i} & = \undset{S \times 1}{\bm{\mu}^z} + \undset{S \times K}{\Theta}\text{ } \undset{K \times 1}{\bm{\eta}_i} + \undset{S \times J}{\Xi}\text{ } \undset{J \times 1}{\bm{\nu}_i} + \undset{S \times 1}{\bm{e}_i}, \\
\undset{D \times 1}{\bm{y}_i} & = \undset{D \times 1}{\bm{\mu}^y} + \undset{D \times K}{\Lambda}\text{ } \undset{K \times 1}{\bm{\eta}_i} + \undset{D \times 1}{\bm{\epsilon}_i}, \\
& i=1, \ldots, N.
\end{split}
\end{align}

The form of the above model is that of a set of linked factor models. Note that although there are three non-error component pieces aside from the global mean terms (namely $\Theta \bm{\eta}_i$, $\Xi \bm{\nu}_i$, and $\Lambda \bm{\eta}_i$), there are only two unique factor vectors: $\bm{\eta}_i$ and $\bm{\nu}_i$. These factors are highly interpretable. The term $\bm{\eta}_i$ represents the shared latent space underlying structured variability in both $\bm{z}_i$ and $\bm{y}_i$ (note that it appears in both factor models). In the expression for $\bm{y}_i$ it is the sole factor vector and in that for $\bm{z}_i$ it is one of two factor vectors. Thus, it is responsible for all structured variation in $\bm{y}_i$ but only part of the structured variation in $\bm{z}_i$. The term $\bm{\nu}_i$ represents structured variation in $\bm{z}_i$ that is \textit{unrelated} to $\bm{y}_i.$ 

By not including unique response-only factors (i.e., factors not assumed to underlay chemical structure) in the mean formulation for $\bm{y}_i$, the model assumes that the mean dose-response curves can be constructed from factors that are all also present in the chemical feature data. Making this assumption allows for easier identification and accurate estimation of parameters associated with chemical features predictive of activity, which is the main goal of this work. The trade off of this assumption is that if the feature set is inappropriate (e.g., not containing enough information by which to estimate activity) and/or the dose-response profiles are truly driven by some individual factors not related to chemical features, then the estimates for new chemicals may be overconfident and the model fit may be poor. The shrinkage prior on $\Theta$, discussed in a later section, offers some mitigation of this risk.

If $\bm{z}_i$ and $\bm{y}_i$ are mean-centered prior to analysis, fix $\bm{\mu}^z$ and $\bm{\mu}^y$ to be zero-vectors for computational savings. In practice mean centering is not sensible to perform for count or binary features (i.e., for $s$ s.t. $x_{is}$ is binary or count). For non-centered $x_{is}$, $\mu^z_{s}$ is given a Cauchy prior expressed as $\mu^z_{s} \sim \text{N}(0,\zeta_s^{-1})$, $\zeta_s\sim \text{Ga}(0.5,0.5)$. Since the Cauchy distribution has high density around zero and heavy tails, it is able to capture meaningful signals while still encouraging shrinkage. If the $\bm{y}_i$ are not mean-centered, $\bm{\mu}^y$ is given a GP prior analogous to the prior placed on the columns of $\Lambda$, discussed in the next section.


In the above model, the mean of $\bm{z}_i$ is $\bm{\mu}^z + \Theta \bm{\eta}_i + \Xi \bm{\nu}_i$, and $\bm{e}_i$ is a term for unstructured noise in $\bm{z}_i$. Similarly, the mean of $\bm{y}_i$ is $\bm{\mu}^y + \Lambda \bm{\eta}_i$, and $\bm{\epsilon}_i$ is the unstructured noise term for $\bm{y}_i$. The priors on the shared factors $\{\bm{\eta}_i\},$ the $X-$specific factors $\{\bm{\nu}_i\},$ and error terms are set to be those typically used in factor analysis:
\begin{align}
\begin{split}
\bm{\eta}_i & \sim \text{N}_K(0,I), \quad \bm{\nu}_i \sim \text{N}_J(0,I), \\
\bm{e}_i & \sim \text{N}_S(0, \Sigma_{X}), \quad \Sigma_{X} = \text{diag}(\sigma_{X,1}^2, \ldots, \sigma_{X,S}^2), \\
\bm{\epsilon}_i & \sim \text{N}_D(0, \Sigma_{Y}), \quad \Sigma_{Y} = \text{diag}(\sigma_{Y}^2, \ldots, \sigma_{Y}^2),  
\end{split}
\end{align}
Homoscedastic variance is assumed for dose-response curves $Y.$ Fix $\sigma_{X,s}^2$ to 1 if feature $s$ is binary, for reasons of identifiability.

Note that as of yet we have not fully addressed how to structure the factor model in light of the functional nature of the $\bm{y}_i,$ nor the issue of many entries in $\bm{z}_i$ likely being unrelated to $\bm{y}_i.$ The following sections will describe how structure can be imposed on $\Lambda$ and $\Theta,$ respectively, in light of these considerations. 

\textbf{The dose-response curve data are functional in nature.} Figure \ref{fig:ExDRChems} shows the noisy observations from a set of example chemicals, but the underlying signal represents a smooth curve relating chemical dose to response.

For functional data it is preferable to have each loading vector (i.e., each column of $\Lambda$) itself be functional. The desired smoothness of the mean curves underlying noisy observations $\bm{y}_i$ can thus be imposed via the choice of smooth priors on the loading matrix $\Lambda$. Let $\bm{\lambda}_k$ denote the $k$th column of $\Lambda,$ so
\begin{gather*}
\Lambda = 
\left[
  \begin{array}{cccc}
    \vertbar & \vertbar &        & \vertbar \\
    \bm{\lambda}_{1}    & \bm{\lambda}_{2}    & \ldots & \bm{\lambda}_{K}    \\
    \vertbar & \vertbar &        & \vertbar 
  \end{array}
\right].
\end{gather*}



In order to learn rather than prescribe smooth bases, columns of $\Lambda$ are modeled as $D-$dimensional Gaussian processes:
\begin{equation}
\bm{\lambda}_k \sim \mathcal{GP}(0,c_k(\cdot)), \quad
c_k(d,d') = \alpha_k^2 e^{-\frac{(d-d')^2}{2\ell^2}}, \quad k=1,\ldots K.
\label{eq:GP_kernel}
\end{equation}

The Gaussian process function variance $\alpha_k^2$ is comprised of two components: a global inverse variance term $\phi$ and a column-specific inverse variance term $\tau_k.$ The column-specific inverse variance term $\tau_k$ utilizes the multiplicative gamma process prior of \citep{bhattacharya2011sparse}, leading to stochastic shrinkage of columns of $\Lambda$ toward 0 by index.
\begin{align}
\begin{split}
\label{eq:shrink_lam}
\alpha_k^2 = \big(\phi \tau_k\big)^{-1}, \quad 
\tau_k = \prod_{h=1}^k \delta_h, \quad k=1,\ldots K, \\ 
\phi \sim \text{Ga}(g_{\phi}/2, g_{\phi}/2), \quad \delta_1 \sim \text{Ga}(a_1,1),\quad
\delta_h \sim \text{Ga}(a_2,1), \quad h \geq 2. 
\end{split}
\end{align}
Following the note by \citep{durante2017note} on hyperparameter selection, set $a_1=2.1$ and $a_2=3.1$ in equation (\ref{eq:shrink_lam}). The value of $g_{\phi},$ the hyperparameter for the global function precision of the GP, should be chosen to reflect the scale of the data.

\begin{figure}[hpt]
\centering
\includegraphics[width=0.8\textwidth]{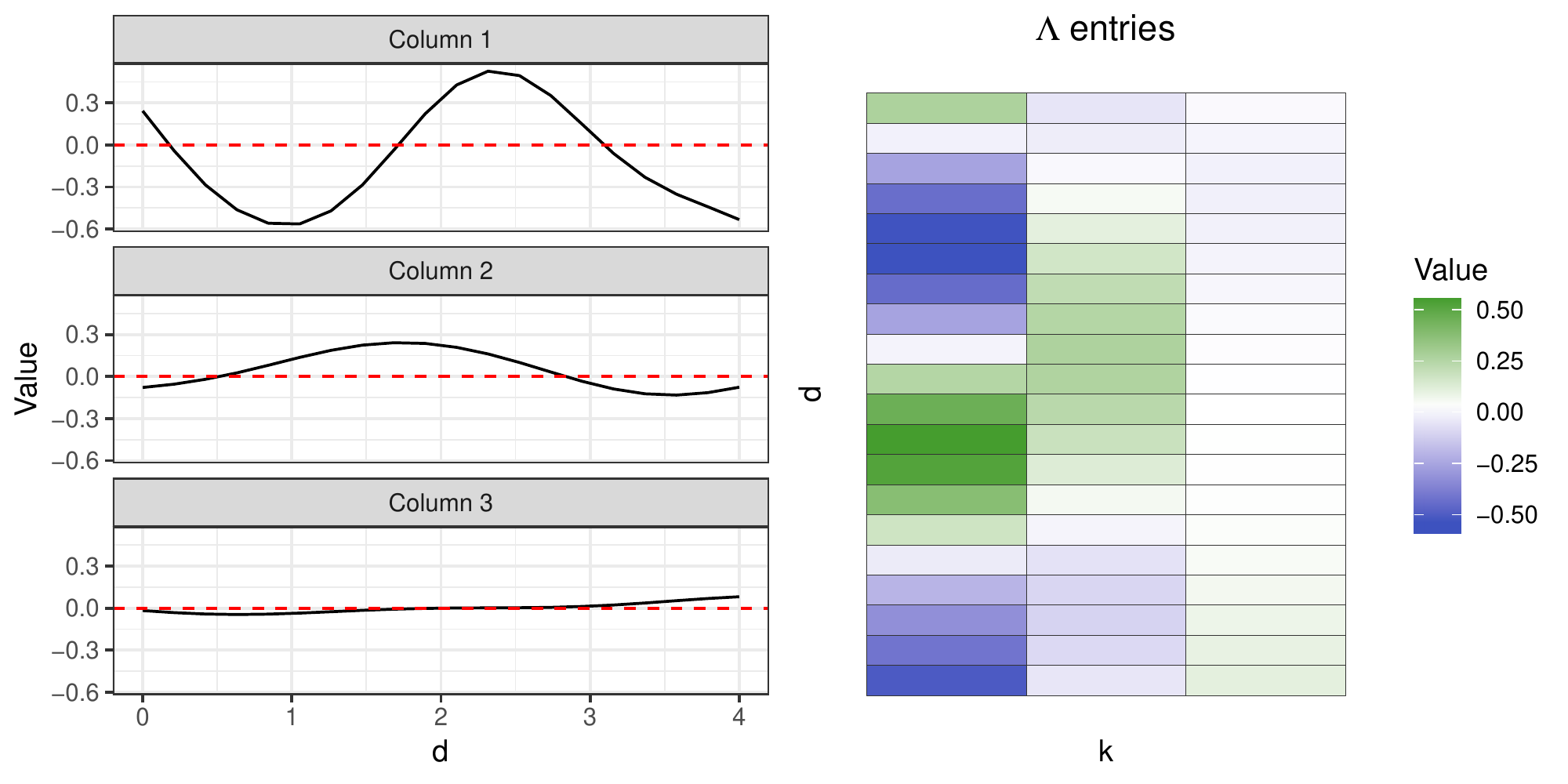}
\caption{Visualization of the smoothness and the shrinkage of columns of an example $D=20$ by $K=3$ loadings matrix $\Lambda$. The effect is an automatic truncation of the number of factors in the model and a learned latent dimension. In this figure columns of $\Lambda$ were sampled using the kernel in equation (\ref{eq:GP_kernel}) with $\ell^2=1$ and $\alpha_k^2$ being $1, \frac{1}{25},\frac{1}{400}$ for $k=1,2,3,$ and the indices $d$ corresponding to the length-20 vector from 0 to 4, inclusive. Note that as $\alpha_k^2$ decreases, the functions tend to flatten.}
\label{fig:BandDLambda}
\end{figure}

This stochastic shrinkage leads to an effective truncation of the factors and an automatically learned dimension of the latent space so long as $K$ is chosen large enough (whether $K$ is adequately large can be assessed by monitoring the convergence of $\alpha_k^2$ to 0 as $k$ approaches $K$). To see why, consider the model for $\bm{y}_i$ written in expanded form: $\bm{y}_i = \bm{\lambda}_1 \eta_{i,1} + \bm{\lambda}_2 \eta_{i,2} + \ldots + \bm{\lambda}_K \eta_{i,K} + \bm{\epsilon}_i.$ Assuming $K$ is large enough, as $k$ approaches $K$, the vector $\bm{\lambda}_k$ should be approximately the $0$-vector, meaning that many of the later terms will contribute negligibly to the mean of $\bm{y}_i.$ See Figure \ref{fig:BandDLambda} for a visualization of $\Lambda$ and column shrinkage.

The linked nature of the factor models leads to an induced prior on the covariance between elements of $\bm{z}_i$ and elements of $\bm{y}_i$; specifically, $\text{Cov}(z_{is},y_{i}(d)) = \sum_k \theta_{sk} \lambda_k(d).$ The GP prior on columns of $\Lambda$ implies that the covariance between a given feature and a given point on the dose-response profile is a smooth function of dose, a desirable model characteristic.

\textbf{Many features are likely unimportant for certain aspects of chemical activity.} That is, if a set of chemical descriptors has not been carefully selected to be toxicity-relevant, it is unlikely that all are related to the shape of the dose-response curves. Even if all features are toxicity-relevant, it is plausible that features will impact different pieces of the toxicity profile (e.g., some features may impact the steepness of the dose-response, while others may impact the height of the final plateau). The way to encourage such a relationship is via element-wise shrinkage, i.e. zeros in entries, of the factor loadings matrix $\Theta.$

Shrinkage on elements $\theta_{sk}$ of $\Theta$ is desirable because it is likely that for a given factor many features have negligible impacts on the associated component of the functional $\bm{y}_i$. Explicitly,
\begin{gather*}
\Theta = 
\begin{bmatrix}
    \theta_{11} & \theta_{12} & \theta_{13} & \dots  & \theta_{1K} \\
    \theta_{21} & \theta_{22} & \theta_{23} & \dots  & \theta_{2K} \\
    \vdots & \vdots & \vdots & \ddots & \vdots \\
    \theta_{S1} & \theta_{S2} & \theta_{S3} & \dots  & \theta_{SK}
\end{bmatrix}
\end{gather*}

\begin{figure}[hpt]
\centering
\includegraphics[width=0.8\textwidth]{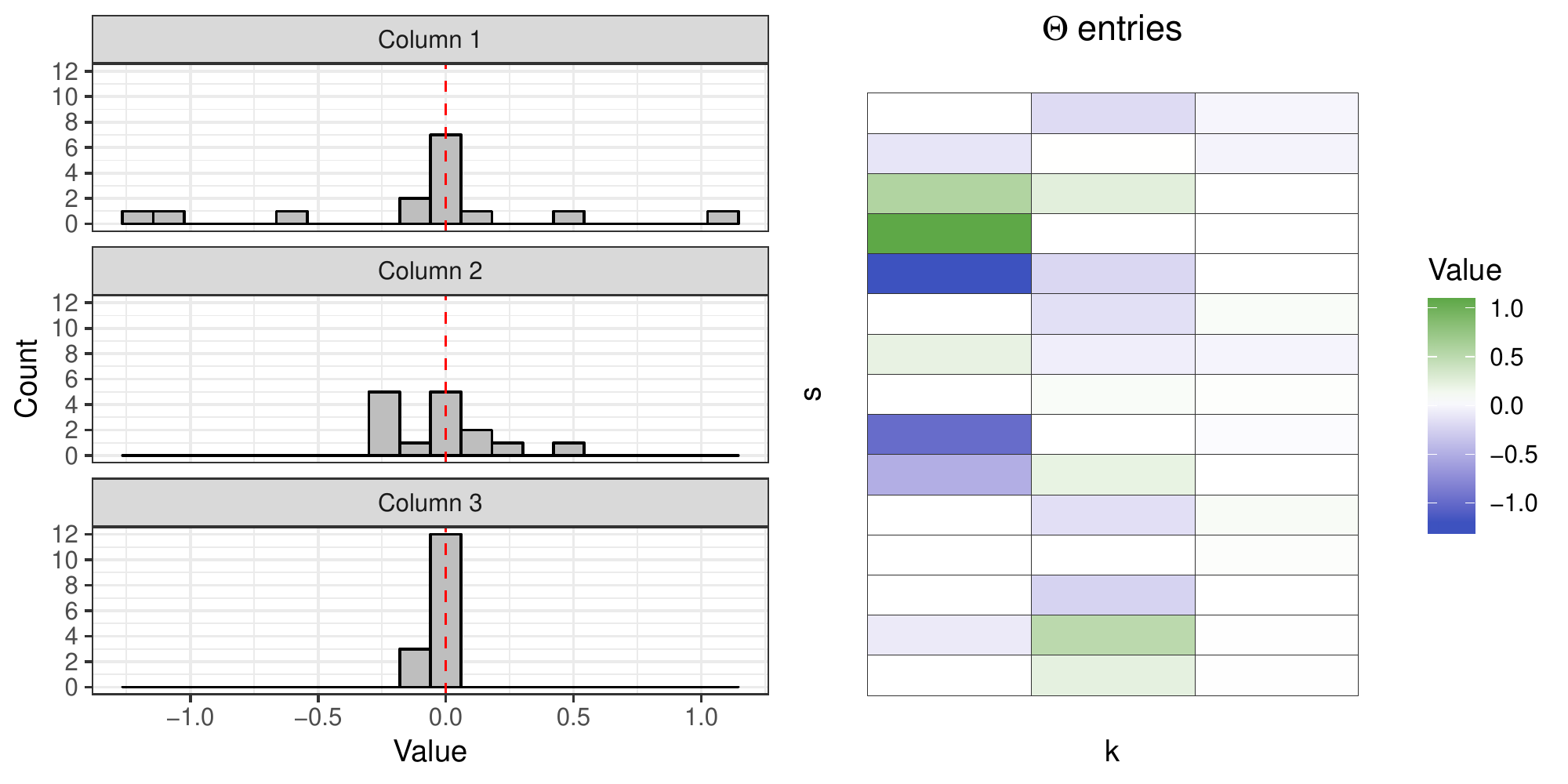}
\caption{Visualization of elementwise sparsity and the shrinkage of columns of an example $S=15$ by $K=3$ loadings matrix $\Theta$. The effect is an automatic truncation of the number of factors in the model and a learned latent dimension. In this figure columns of $\Theta$ had column-specific variance $\tau_k^{-1}$ being $1, \frac{1}{25},\frac{1}{400}$ for $k=1,2,3.$} 
\label{fig:BandDTheta}
\end{figure}

There is a very rich literature proposing elaborate shrinkage and sparsity priors for factor loadings (e.g., \citep{yoshida2010bayesian, meng2010uncovering, knowles2011nonparametric, pati2014posterior}). We opt for a horseshoe prior \citep{carvalho2010horseshoe} modified for simple sampling \citep{makalic2016simple} on entries $\theta_{sk}$ of $\Theta$:
\begin{align}
\begin{split}
    \theta_{sk} & \sim \text{N}(0,\beta^2 \gamma_{sk}^2 \tau_k^{-1}), \\     
    \beta^2 | t & \sim \text{IG}(1/2,1/t) \\
    \gamma_{sk}^2|b_{sk} & \sim \text{IG}(1/2,1/b_{sk}) \\
    \{b_{sk}\}, t & \sim \text{IG}(1/2,1), \\
    s &= 1,\ldots S, \quad k=1,\ldots K.
\end{split}
\end{align}
The horseshoe component of this prior is in the hierarchical hyper-prior on global variance term $\beta^2$ and local variance term $\gamma_{sk}^2$. The column-specific variance $\tau_k^{-1}$ applies stochastically increasing shrinkage as column index increases. See Figure \ref{fig:BandDTheta} for a visualization of $\Theta$ and column shrinkage.

Recall $\text{Cov}(z_{is},y_{i}(d)) = \sum_k \theta_{sk} \lambda_k(d).$ The shrinkage prior on entries of $\Theta$ allows for the possibility that a given chemical feature may not have a contribution from some or any of the latent factors (i.e., that $\theta_{sk}$ for such feature/factor combinations is near 0). In this case, the covariance between the feature and points along the dose response profile will not depend on those factor(s). In the extreme case that the $s$th feature is unrelated to toxicity, $\theta_{sk}$ should be shrunk toward zero for all $k$, leading $z_{is}$ and $y_{i}(d)$ to be uncorrelated.

\textbf{The dimension of the latent space underlying the dose response curves and of that underlying toxicity-relevant features should be the same.}

The same column-specific precision $\tau_k^{-1}$ is used for both $\Theta$ and $\Lambda,$ so relative column shrinkage is applied consistently across the two matrices. The result is the same effective truncation on the number of latent factors in the joint space. This specification, along with the common $\bm{\eta}_i,$ is what allows the shared directions of variability between $\bm{y}_i$ and $\bm{x}_i$ to be learned. Distance can be defined over the $\bm{\eta}$-vector with elements $\eta_k$ weighted by precision $\tau_k^{-1},$ to give a sense of closeness in ``$\bm{\eta}$ space'' that reflects the true amount of information contained in each latent direction.

\textbf{There is additional variability in chemical structure beyond that which impacts chemical activity.} Unless all chemical features were carefully hand-selected to be toxicity relevant for the specific assay considered in the model (a tall task, and unlikely to be completely true no matter how careful the selection), accounting for variability in $\bm{z}_i$ shared with $\bm{y}_i$ will not capture all structured variability in chemical features. 

After accounting for the variability in $\bm{z}_i$ shared with $\bm{y}_i$ (via latent factor $\bm{\eta}_i$), $\bm{z}_i$ may still have structured variation due to individual latent factor $\bm{\nu}_i.$ Let $\bm{z}_i^*=\bm{z}_i-\Theta \bm{\eta}_i.$ Then $\bm{z}_i^*=\Xi \bm{\nu}_i + \bm{e}_i,$ which once again looks like a traditional factor model. The direct application of priors in \citet{bhattacharya2011sparse} is used for elements of $\Xi$. Specifically elements $\xi_{s,j}$ of $\Xi$ are given prior
\begin{align}
\begin{split}
\xi_{s,j}|\kappa_{sj},\omega_j \sim \text{N}(0,\kappa_{sj}^{-1}\omega_j^{-1}), \quad \omega_j = \prod_{h=1}^j \zeta_h, \quad s=1, \ldots S, \quad j=1, \ldots J,\\
\kappa_{sj} \sim \text{Ga}(g_{\kappa}/2, g_{\kappa}/2), \quad 
\zeta_1 \sim \text{Ga}(m_1,1),\quad
\zeta_h \sim \text{Ga}(m_2,1), h \geq 2. 
\label{eq:shrink_xi}
\end{split}
\end{align}
Stochastic column-specific shrinkage via the $\omega_j^{-1}$ term removes the need to select an ideal number of factors $J$ and allows for simply selecting $J$ ``large enough.'' This formulation also allows for efficient Gibbs sampling of the posterior. Following the note by \citep{durante2017note} on hyperparameter selection, set $m_1=2.1$ and $m_2=3.1$ in equation (\ref{eq:shrink_xi}). The value of $g_{\kappa},$ the hyperparameter for the entry-level precision terms of $\Xi$, should be chosen to reflect the scale of the data.

If there is in fact no additional variability in $\bm{z}_i$ beyond that shared with $\bm{y}_i,$ the following specification allows for all columns of $\Xi$ to be shrunk to 0-vectors. This case reduces to a fully joint factor model in which all variability in $\bm{z}_i$ is shared with $\bm{y}_i.$

\textbf{Chemical activity is not necessarily measured on a fully observed, regularly spaced grid.} There are a handful of common dose measurements at which the majority of chemicals are measured (see Figure \ref{miss_act}), but there are many chemicals whose observations are less regular. Furthermore, some chemicals have multiple observed dose response curves. 

The issue of irregular spacing between the unique values associated with the indices is handled automatically via the use of GPs for modeling columns of $\Lambda$. The covariance between points is defined by the kernel for any pair of input values (see equation (\ref{eq:GP_kernel})) and not dependent on a regular measurement grid.

\textbf{Toxicologists may wish to report different components and/or summaries of predicted dose-response curves.} For example, they may be interested in the dose value at which the response first exceeds some threshold, the maximum response value reached, the area under the curve, etc. Each of these summaries provides different information about the dose response relationship. An advantage of a Bayesian formulation is that we can obtain posterior samples for any functional of the dose response curve trivially, with these samples then used to obtain point and interval estimates. 

\subsection{Posterior computation}

The posterior for the $\text{BS}^3\text{FA}$ model is not available in closed form. However, closed form full conditional distributions of the parameters associated with the model allow the use of a straightforward Gibbs sampler for these draws. Samples obtained directly from this Markov chain Monte Carlo (MCMC) algorithm allow for the calculation of posterior means, simultaneous bands \citep{meyer2015bayesian}, and credible intervals for identifiable model components, including the predicted mean, covariance, and noise variance of $X$ and $Y$. A post-processing step to resolve rotational ambiguity and account for label/sign switching allows for identifiability of the individual model components, including the factor scores $\eta$ and loadings $\Lambda$ and $\Theta$ (code modified from \url{https://github.com/poworoznek/sparse_bayesian_infinite_factor_models}). Full details on the Gibbs sampler steps and initialization are included in the Supplemental Materials.

\subsection{Code base and reproducibility}

Code for simulating data and sampling from the $\text{BS}^3\text{FA}$ model, along with a user manual, are made available at \url{https://github.com/kelrenmor/bs3fa}. A hands-on demonstration of the package is available at \url{https://www.youtube.com/watch?v=qLyxBQ-sVcY}. Code specific to this paper (i.e., to reproduce the simulations, figures, and results) is provided in the online supplementary material. 

\subsection{Simulation study}

Simulation studies were performed in order to assess the ability of $\text{BS}^3\text{FA}$ to learn the true toxicity-relevant distance between chemicals, its predictive performance, and the model fit. Two broad categories of simulations were performed: first, those in which the true data generating process aligns with model assumptions (i.e., when data are simulated from a partially shared latent factor model) and when it does not (i.e., when data are simulated from something other than a factor model). In the former category, we also assess how well model sub-components can be learned. 

For all simulations, 25\% of simulated ``chemicals'' are held out. That is, rather than withholding 25\% of dose-response observations across chemicals, we withhold all dose-response data for each hold-out chemical. For distance performance, $\text{BS}^3\text{FA}$ is compared to Euclidean distance using all features, using the features selected via the decision analytic approach pioneered by \citet{hahn2015decoupling} in the B-FOSR model from the \textbf{fosr} \textbf{\textsf{R}} package, using the principal component scores explaining 95\% of the variability in the data found in PCA, and using the selected variables from a frequentist FOSR using the \texttt{fosr.vs()} function (FOSR-VS) from the \textbf{refund} \textbf{\textsf{R}} package with the PC scores used as inputs. PC scores were used instead of the full set of features for FOSR-VS because the \texttt{fosr.vs()} function returned an error  saying the dimension of $\bm{y}_i$ was not high enough relative to $\bm{x}_i$ (i.e., that $D$ was too small relative to $S$) when the full feature set was used. The performance of each method is compared by computing the correlation between predicted and true distance, a choice made because the scale of distance varies across methods and, by virtue of the structure of the simulated data, we don't expect any extreme outliers in this space. For predictive performance, $\text{BS}^3\text{FA}$ is compared to the BAABTP model \citep{wheeler2019bayesian}, B-FOSR, FOSR-VS, and least absolute shrinkage and selection operator (LASSO) using each covariate, dose, and all pairwise interactions. 

For simulations in which the true data generating process aligns with model assumptions, the true dimension of the latent toxicity-relevant space $K$ was varied, taking values 1, 3, and 5. For each $K,$ the true dimension of the latent toxicity-irrelevant space $J$ was varied from 0 to 20 in intervals of 5. At each combination of $K$ and $J,$ 100 data sets were simulated with $N=300$, $D=10$, $S=40$. For roughly half of the chemicals in each data set, $\bm{\eta}_i$ was set to a zero vector for that chemical (i.e., for each simulated data set, there was a 50\% chance that any given chemical was inactive). Further simulation details are provided in the Supplemental Materials. Overall, the model does quite well at capturing the structure of the noise variance and the true components $\Lambda$ and $\Theta$. 

\begin{figure}[!htpb]
\centering
\includegraphics[width=0.85\textwidth]{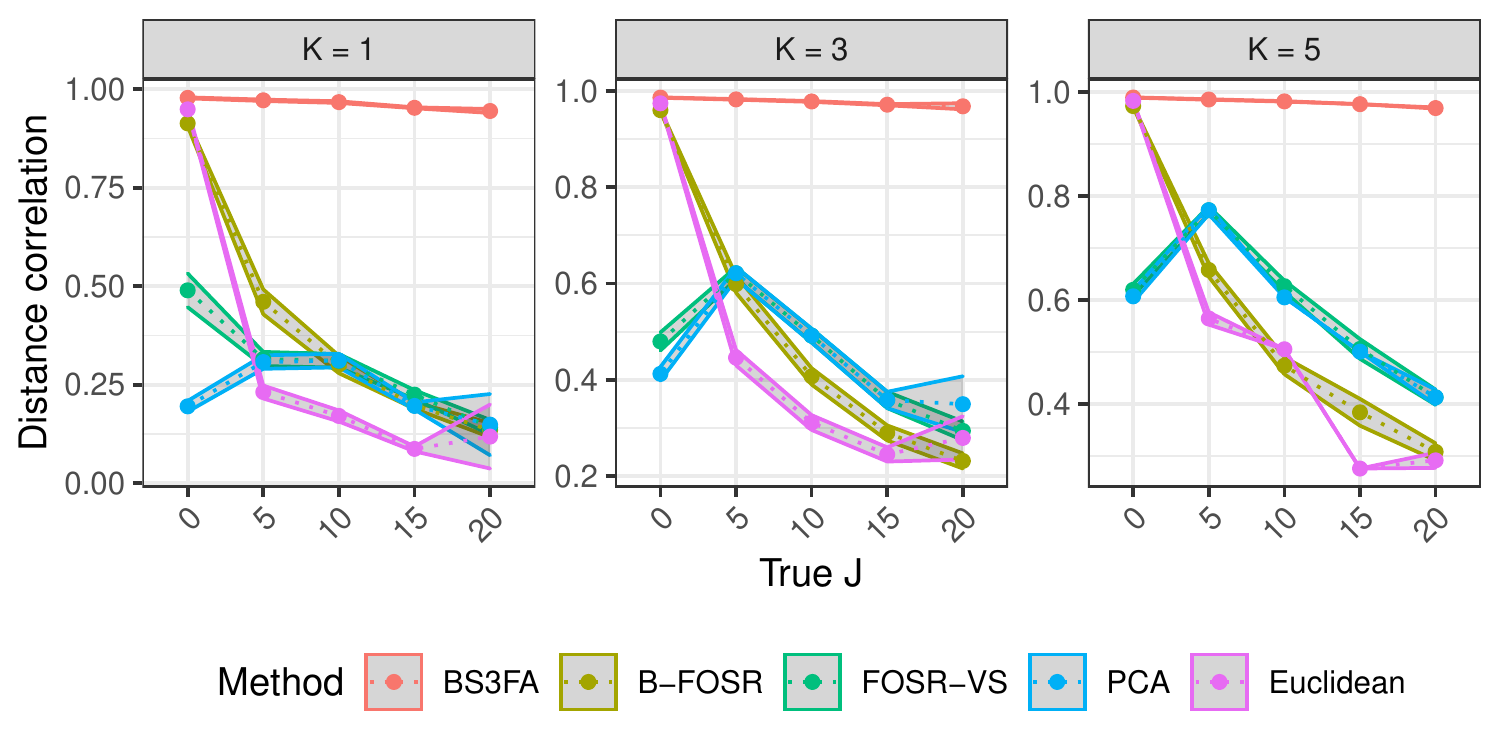}
\caption{Correlation between entries in the true pairwise distance matrix (i.e., the Euclidean distance between true latent factors $\eta$) and the predicted pairwise distance matrix for holdout chemicals. Each subplot shows the result of 100 simulations per $J$ across methods for a given true shared subspace dimension $K$.} 
\label{simResDist}
\end{figure}

Figure \ref{simResDist} shows the correlation between entries in the true pairwise distance matrix (i.e., the Euclidean distance between true latent factors $\eta$) and the predicted pairwise distance matrix for holdout chemicals. We see that even in the case of small $J,$ performing PCA on the $S-$dimensional $X$ matrix obscures the true distance in the latent space. As $J$ increases, the correlation between the chemical distance in the true $\eta$ and that in either PCA space or Euclidean space drops quickly. This phenomenon also occurs for distances computed via variable selection using B-FOSR or FOSR-VS. The $\text{BS}^3\text{FA}$ has stable high correlation across all values of $J$ and $K$. 

\begin{figure}[!htpb]
\centering
\includegraphics[width=0.85\textwidth]{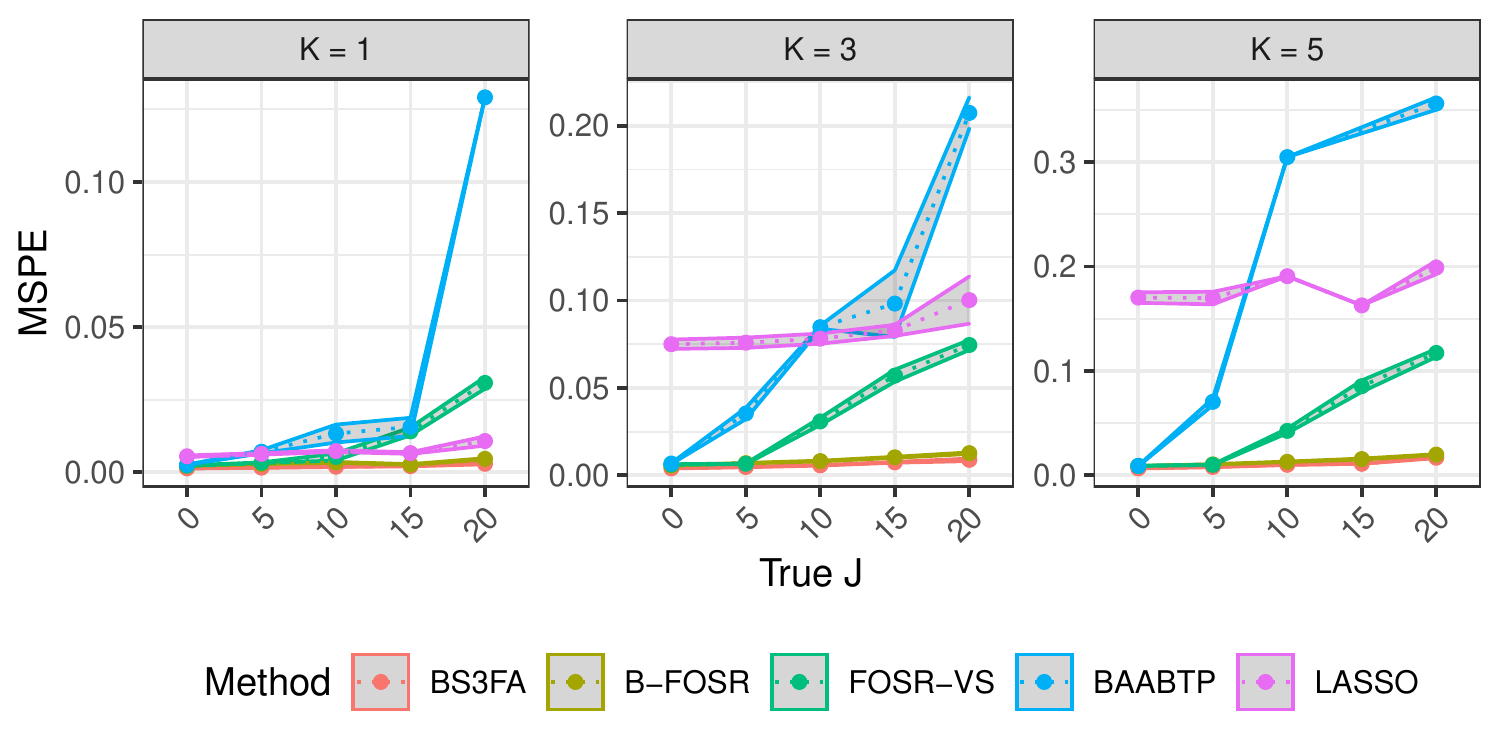}
\caption{Mean squared predictive error (MSPE) for the hold-out chemicals' dose-response mean functions. Each subplot shows the result of 100 simulations per $J$ across methods for a given true shared subspace dimension $K$.} 
\label{simResMSPE}
\end{figure}

Figure \ref{simResMSPE} shows the mean squared predictive error (MSPE) for the simulated hold-out chemicals' dose-response mean functions. Although the performance of all models deteriorates as the amount of ``superfluous'' information in $X$ increases (i.e., as $J$ increases), the $\text{BS}^3\text{FA}$ and B-FOSR models are the most robust, showing superior performance across all values of $K$ and $J$. The $\text{BS}^3\text{FA}$ consistently outperforms the B-FOSR model, with MSPE of the latter on average 1.5 times higher than that of the former (the 95\% quantile of this multiplicative factor ranges from 1.1 to 2.4 across all simulations and values of $K$ and $J$). The BAABTP model appears most sensitive to the value of $J,$ with MSPE near that of $\text{BS}^3\text{FA}$ model when $J$ is small, but among the worst MSPE when $J$ is high. The LASSO model is able to perform fairly well when $K$ is small, but as $K$ increases it is unable to learn the more complicated relationship between $\bm{x}_i$ and $\bm{y}_i$. The FOSR-VS model becomes increasingly sensitive to $J$ as $K$ increases. 

Coverage of the simultaneous bands for the $\text{BS}^3\text{FA}$ model (shown in detail in the Supplemental Materials) remains close to nominal across all values of $K$ and $J$. The coverage of the BAABTP model decreases as $J$ increases, another reflection of its overall poor predictive performance. The coverage of the B-FOSR model is much higher than nominal, nearing 1 across all values of $K$ and $J$. 

In the ToxCast analysis following this section, we discuss results from one possible method of deeming a chemical active. Namely, call a chemical active if its global Bayesian p-value \citep{meyer2015bayesian} is less than 0.05. The null hypothesis in this case is that the dose response mean function equals 0 everywhere, i.e. that a chemical is inactive. If alternatively activation (suppression) is of specific interest, one could use the additional requirement that the direction of observed effect be positive (negative) in order to reject the null. The true positive rate (TPR), false positive rate (FPR), and false discovery rate (FDR) of this method on the simulated data are shown in Table \ref{tab:TPR_FPR_FDR}. Across all values of $K$ and $J$ the TPR is generally high, and the FPR and FDR are near zero. Note that the TPR increases with $K$ and decreases with $J,$ while FPR and FDR remain stable across values of $K$ and $J$. That is, the model seems more likely to identify chemicals as active as the dimension of the latent toxicity-relevant space increases, but loses sensitivity when there is more toxicity-irrelevant information. Results for the B-FOSR method, provided in the Supplemental Materials, show slightly improved TPRs but at the cost of much higher FPRs and FDRs (whereas for BS$^3$FA the FPR and FDR are consistently near zero, these values range from 0.1 to 0.5 across $K$ and $J$ for B-FOSR).

\begin{table}[!phtb]
    \begin{tabular}{l c c c c c c }\toprule
        & & $J=0$ & $J=5$ & $J=10$ & $J=15$ & $J=20$ \\\midrule
        \rule{0pt}{4ex}& $K=1$ & 0.72 (0.08) & 0.66 (0.08) & 0.64 (0.07) & 0.58 (0.06) & 0.52 (0.09) \\
		TPR & $K=3$ & 0.98 (0.03) & 0.97 (0.03) & 0.97 (0.03) & 0.95 (0.04) & 0.93 (0.05) \\
 		& $K=5$ & 1.00 (0.01) & 1.00 (0.01) & 1.00 (0.01) & 0.99 (0.01) & 0.98 (0.02) \\
        \rule{0pt}{4ex}& $K=1$ & 0.00 (0.00) & 0.00 (0.00) & 0.00 (0.01) & 0.00 (0.00) & 0.00 (0.00) \\
		FPR & $K=3$ & 0.00 (0.01) & 0.00 (0.01) & 0.00 (0.01) & 0.00 (0.01) & 0.00 (0.01) \\
 		& $K=5$ & 0.00 (0.01) & 0.00 (0.01) & 0.00 (0.01) & 0.00 (0.01) & 0.00 (0.01) \\
        \rule{0pt}{4ex}& $K=1$ & 0.00 (0.01) & 0.00 (0.01) & 0.00 (0.01) & 0.00 (0.00) & 0.00 (0.01) \\
		FDR & $K=3$ & 0.00 (0.01) & 0.00 (0.01) & 0.00 (0.01) & 0.00 (0.01) & 0.00 (0.01) \\
 		& $K=5$ & 0.00 (0.01) & 0.00 (0.01) & 0.00 (0.01) & 0.00 (0.01) & 0.00 (0.01) \\
        \\ \bottomrule
    \end{tabular}
    \caption{True positive rate (TPR), false positive rate (FPR), and false discovery rate (FDR) for the proposed method of assessing whether a chemical is active. A perfect classifier has a TPR of 1 and an FPR/FDR of 0.}\label{tab:TPR_FPR_FDR}
\end{table} 

When there is misalignment between the structure assumed by the $\text{BS}^3\text{FA}$ model and the true data generating process, $\text{BS}^3\text{FA}$ is still able to predict similarly to or better than the competitors. As with the well-aligned simulation, $\text{BS}^3\text{FA}$ is robust to increasing ``superfluous'' information in $X.$ A similar story can be seen in the coverage and distance results for the misaligned simulation. Even when the assumed latent factor model is not the model from which data are simulated, the coverage of $\text{BS}^3\text{FA}$ is close to nominal. B-FOSR, on the other hand, suffers from much higher-than nominal coverage across all simulations while BAABTP suffers from much lower-than nominal coverage as ``superfluous'' information in $X$ increases. $\text{BS}^3\text{FA}$ is still able to recover a distance metric that is highly correlated (mean correlation 0.97, range 0.94 to 0.99 across simulations) with the distance in the true relevant $X$ dimensions, although B-FOSR more perfectly learns such a metric. Visual results are shown in the Supplemental Materials. 

\section{Relating chemical structure to toxicological response}
\label{Sec:ToxApp}

Data pre-processing steps and results of the analysis of the ToxCast ATG PXR assay are discussed in the following subsections. The structure of $\text{BS}^3\text{FA}$ allows for learning about structured variability in both the feature and response space, prioritizing chemicals for future evaluation, and predicting chemical activity for as-yet-unobserved chemicals.

\begin{figure}[!htpb]
\centering
\includegraphics[width=1.0\textwidth]{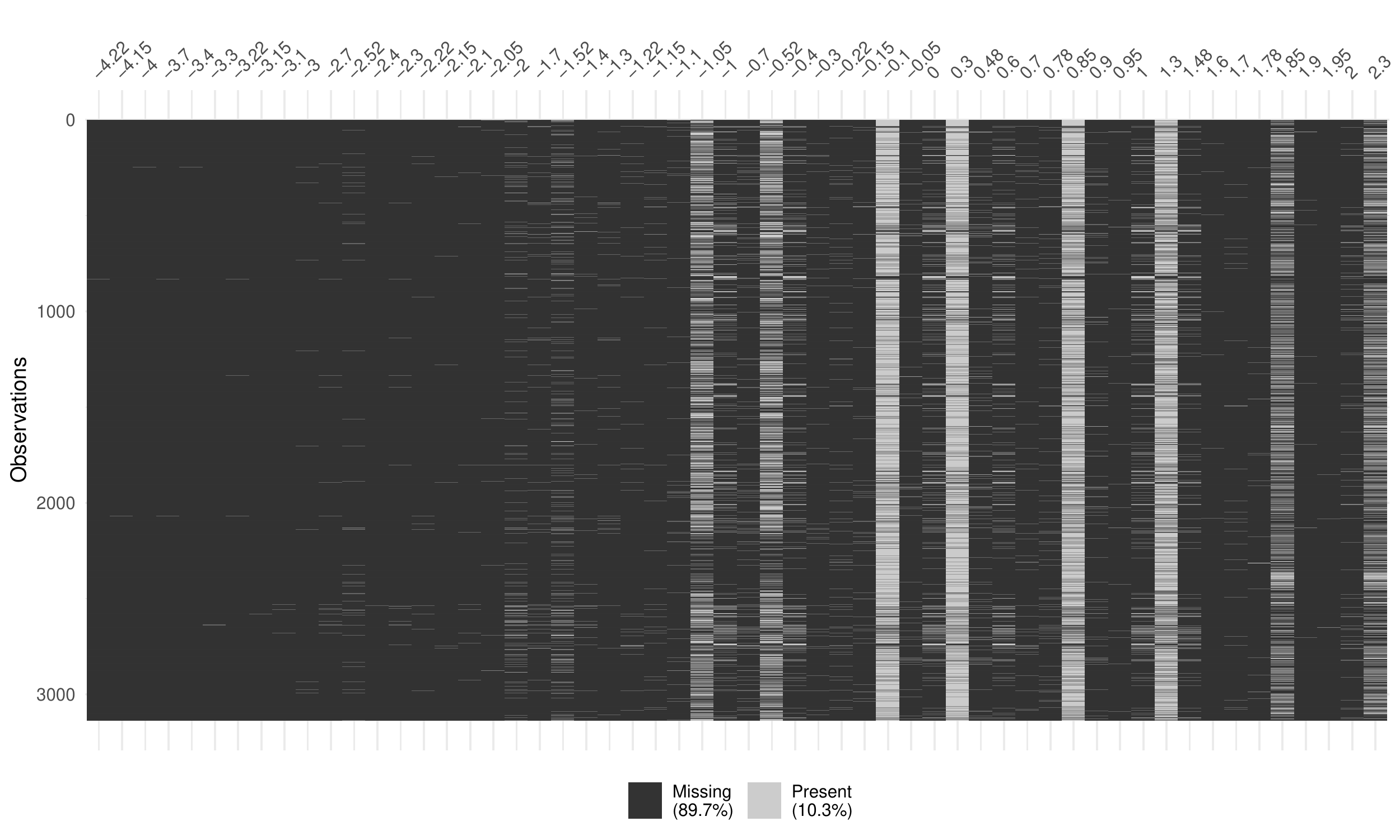}
\caption{Missingness by dose for each chemical in the analysis. Each row is corresponds to a chemical, and each column to a dose value (sorted in ascending order). Note that very few chemicals have observations below $-2$ log uM.}\label{miss_act}
\end{figure}

\subsection{ToxCast setup}
\label{Sec:ToxApp_data_processing}

Observations below the cytotoxicity limit for $3540$ Phase 1, Phase 2, and e1k chemicals tested in the AttaGene PXR assay are included in our data analysis. The structure information for each chemical is summarized by 777 Mold2 chemical features \citep{hong2008mold2}. As discussed previously, $\text{BS}^3\text{FA}$ has the advantage of being able to effectively ignore toxicity-irrelevant features via shrinkage on elements of $\Theta,$ making the careful curation of a feature set unnecessary.


Chemicals having no provided SMILES information ($n=405$) were omitted from further analysis. The result is $3135$ chemicals having Mold2 descriptions of their chemical structure. Note that some chemicals have identical Mold2 output. These sets of chemicals are in effect considered a single chemical (i.e., treated as multiply observed dose response curves by the model). For analysis the number of `unique' chemical sets (i.e. the number of unique SMILES represented across the $3135$ chemicals) is $N=3070$. The 8 most common dose concentrations are -1.05, -0.52, -0.1, 0.3, 0.85, 1.3, 1.85, 2.3 log uM (see the vertical white bands in Figure \ref{miss_act}), but $\text{BS}^3\text{FA}$ allows both common and unique doses. As the bulk of chemicals have no information about extremely low ($<-2$ log uM) dose activity (see Figure \ref{miss_act}), the data considered are the $D=38$ unique doses $\geq -2$ log uM, out of the $56$ total unique doses. Approximately 4\% of chemicals have multiply observed dose response curves, e.g. Allethrin and Clorophene from Figure \ref{fig:ExDRChems}. While such repeated measurements may come from different experimental runs, unfortunately the data set does not provide information on which point is associated with which run so we assume independence in this respect rather than, e.g., adding a run-specific random effect term.

\begin{figure}[!htpb]
\centering
\includegraphics[width=0.8\textwidth]{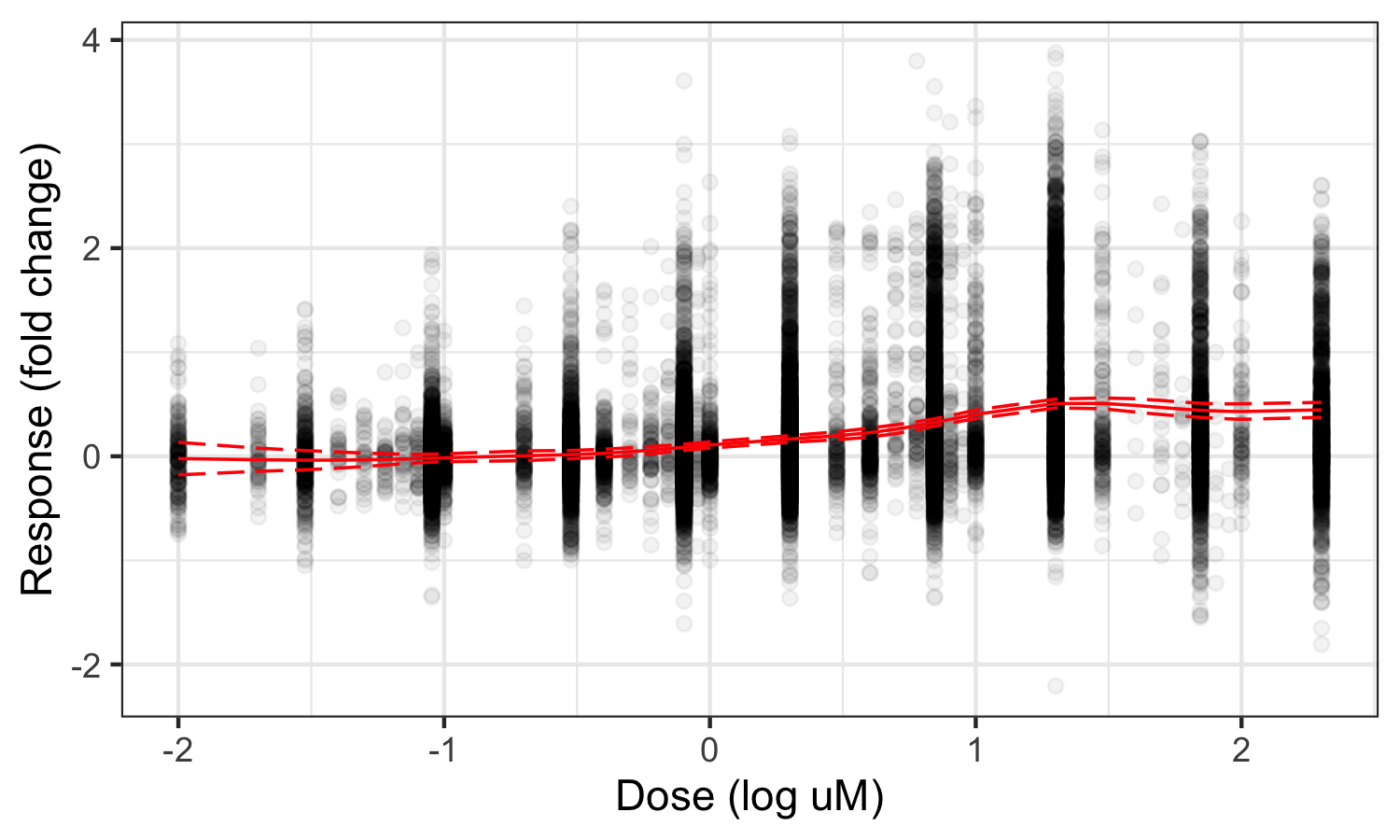}
\caption{All response values at each dose in the data set. The solid line shows the model-predicted global mean $\bm{\mu}^y$ and the dashed lines show the model-predicted 95\% simultaneous credible band about $\bm{\mu}^y$.}\label{avgDR}
\end{figure} 

Mold2 is used to generate a set of 777 numeric molecular descriptors associated with each chemical. As noted above, some chemicals exactly shared these descriptors due to Mold2's inability to capture certain differentiating structures (see the Supplemental Materials for an example); these ``identical'' chemicals were treated as multiply observed chemicals. After removing features having no variability (99 total, including, e.g., features equalling 0 for all chemicals such as number of 11-membered rings and number of Argon), features with duplicated entries (16), or features having $>99$\% of chemicals sharing a feature value (99, e.g. only one chemical has any aromatic group urea derivatives), $S = 563$ features remain. As a further pre-processing step, the continuous variables are scaled to have mean 0 and variance 1. Further information on creating and using Mold2 descriptors is included in the Supplemental Materials. 



In order to mimic the scenario of using the $\text{BS}^3\text{FA}$ model to prioritize chemicals for further screening, we hold out all of the dose-response observations for 25\% of chemicals in the data set. That is, we provide the model with these chemicals' structure but not their dose-response curves. Of interest will be how similar these unobserved chemicals are to known active chemicals, and the AC50 (the dose at 50\% of maximum activity) for unobserved chemicals predicted by the model to be activity-increasing. Recall for the purpose of this analysis, we consider a chemical activity-increasing if the global Bayesian p-value \citep{meyer2015bayesian} of its dose response profile is less than 0.05 and the direction of observed effect is positive. 

A global mean term $\bm{\mu}^y$ is included in the model (see Figure \ref{avgDR}) to account for a nonzero average profile rather than mean centering, as the low number of observations at less common doses leads to a `noisy' center. Additionally, the dose-response matrix is scaled by a multiplicative factor so that the scaled $X$ and $Y$ matrices have the same Frobenius norm (a measure of total variation); specifically set $Y= ||X||_F \ \frac{Y}{||Y||_F}$. This rescaling keeps larger matrices from dominating when learning the shared column-specific shrinkage terms $\{\tau_k\}$ or the shared score vectors $\{\bm{\eta}_i\}$ (e.g., when $S$ is much larger than $D$, as in this setting).

We ran the sampler for 40,000 iterations. After an initial burn-in of 20,000 iterations, every 10th sample was saved. Computation time was approximately 8 hours on a 2016 MacBook pro with a 2.9 GHz Intel Core i7 processor. Trace plots of model predictions show good mixing; these, along with those of model components and an assessment of the sufficiency of the chosen $K$ and $J$ values, are available in the Supplemental Materials. 

\subsection{Model components}

\begin{figure}[!htpb]
\centering
\includegraphics[width=0.9\textwidth]{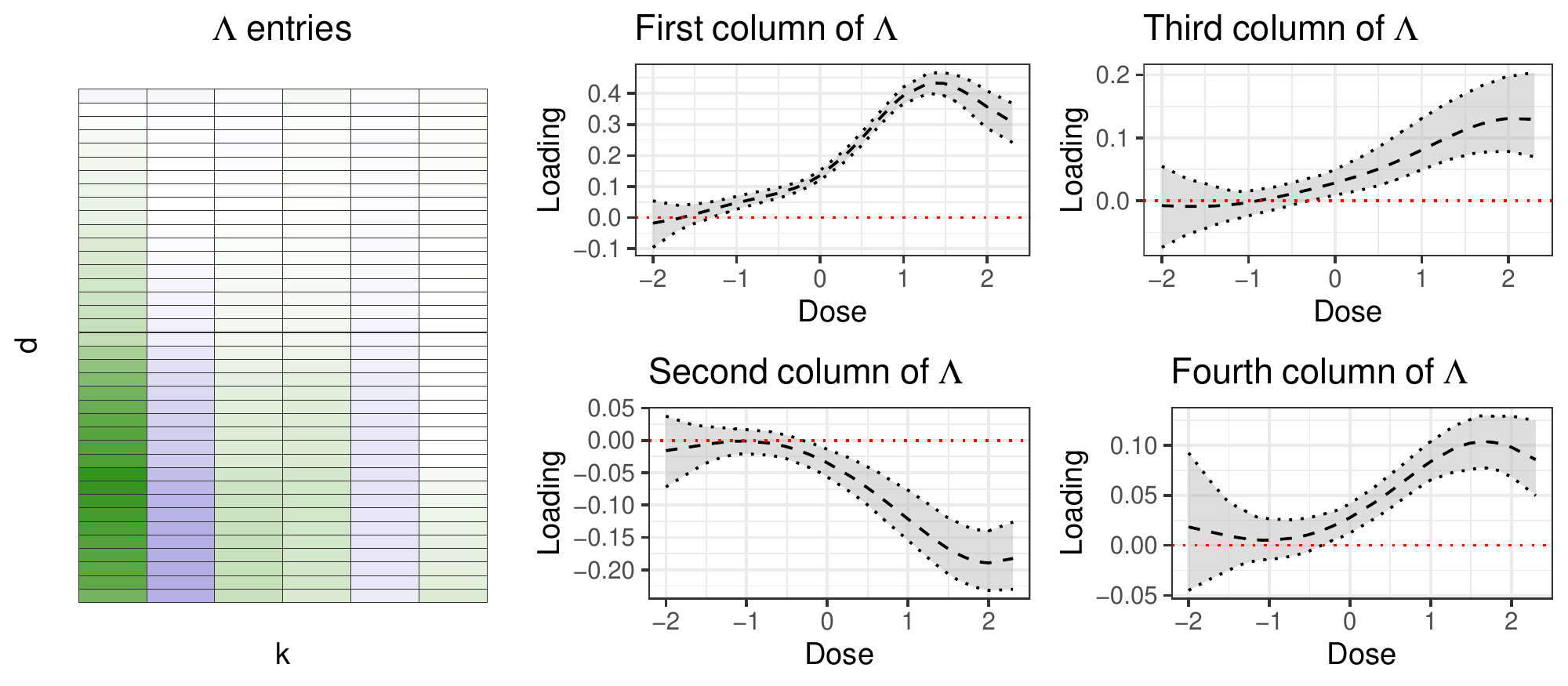}
\caption{Left: First 6 columns of the predicted mean of $\Lambda$. Each row represents a unique dose value. Middle and right: Mean and 95\% simultaneous bands for the first four columns of $\Lambda$. The dose values are given on the $x$-axis.}
\label{fig:res_inference_Lambda}
\end{figure}

The learned matrix $\Lambda,$ shown on the left side of Figure \ref{fig:res_inference_Lambda}, provides a snapshot of the directions of structured variation present in the dose-response data. The first column of $\Lambda,$ shown in the top middle of Figure \ref{fig:res_inference_Lambda}, is the dominant factor loading (i.e., the factor loading having the largest estimated norm). Unsurprisingly, this vector takes the shape of a prototypical dose response curve. Later columns of $\Lambda$ act to provide smooth deviations from this prototypical shape. For example, the third column characterizes a flat profile until around -1 log uM followed by a near-linear increase until leveling off at a dose value just under 2 log uM, while the fourth column shows an initial start point above zero followed by a gentle U-shaped dip. 

\begin{figure}[!htpb]
\centering
\includegraphics[width=0.6\textwidth]{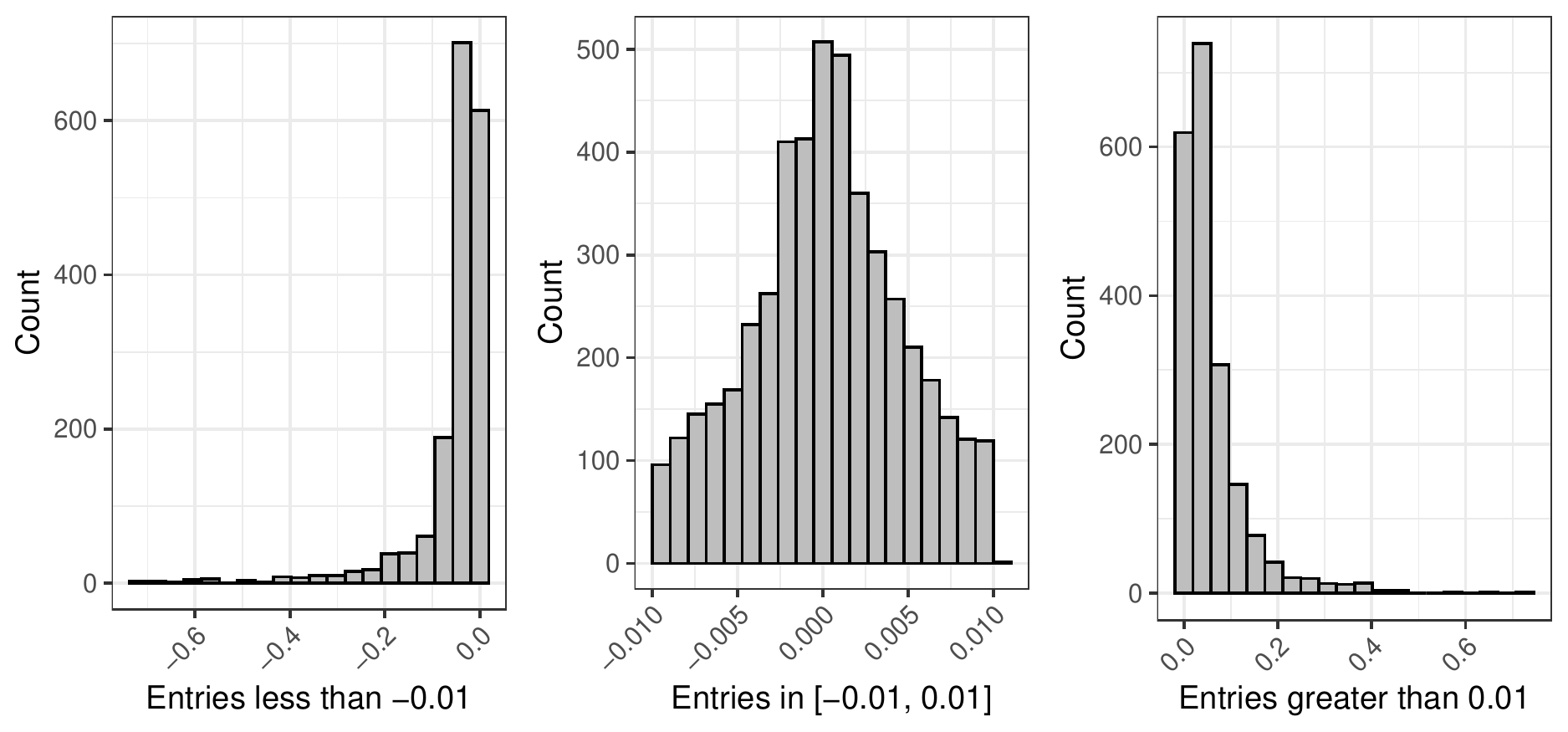}
\caption{Histogram showing entries of the predicted mean of $\Theta$. The bulk of the estimated entries are close to 0 due to the shrinkage effect of the horseshoe prior.}
\label{fig:res_inference_Theta}
\end{figure}

The learned matrix $\Theta,$ the values of which are shown in Figure \ref{fig:res_inference_Theta}, provides a snapshot of the directions of structured toxicity-relevant variation present in the feature data. The bulk of the estimated entries are very close to 0 due to the shrinkage effect of the horseshoe prior. The $\text{BS}^3\text{FA}$ model structure allows us to interpret the nonzero entries of a given column of $\Theta$ as being those related to the particular structure present in the corresponding column of $\Lambda.$ For example, the significantly non-zero entries of the first column of $\Theta$ are those associated with the prototypical activity profile seen in the top right of Figure \ref{fig:res_inference_Lambda}. By absolute magnitude, the largest such features include the number of group esters, the number of group X-C on aromatic ring, the sum of the topological distance between the vertices O and Cl, the number of Chlorine, and the sum eigenvalue weighted by polarizability distance matrix.

The chemicals having extreme positive values of $\eta_1$ will be those for which the dose-response profile has a large component due to $\lambda_1$ and a large chunk of toxicity-relevant molecular variability described by the first column of $\Theta.$ In the training set, the chemicals having the largest expected value for $\eta_1$ are Calcium bromide, Nickel(II) chloride, Dibromomethane, Zinc chloride, and Dimethylamine. All are known toxins. Several of these chemicals have in common the presence of Cl, so it is unsurprising that features involving Cl appeared amongst the high-value loadings components for the first column of $\Theta.$

\subsection{Distance learning}

\begin{figure}[!htb]
\centering
\includegraphics[width=0.8\textwidth]{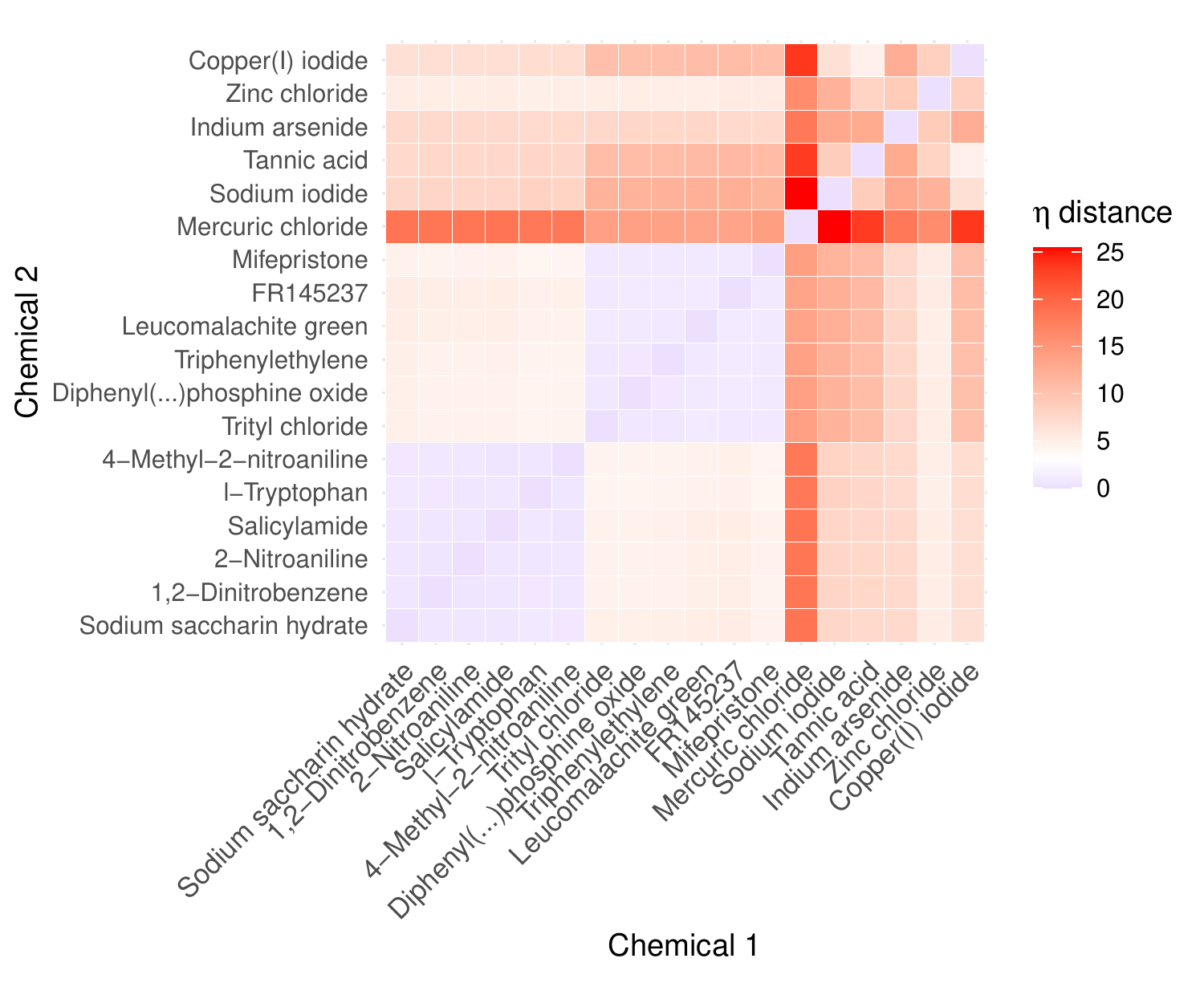}
\caption{Expected distance in $\bm{\eta}$ space for a select set of training chemicals.}
\label{fig:chem_dist}
\end{figure}

Figure \ref{fig:chem_dist} shows the predicted pairwise distance matrix between a set of example chemicals chosen as clusters of chemicals in the training set closest to specific recognizable hold-out chemicals. Included are a cluster of similar low-activity chemicals (the training chemicals nearest to hold-out chemical Saccharin) in the bottom left block, and a cluster of similar high-activity chemicals (the training chemicals nearest to hold-out chemical Clomiphene citrate (1:1)) in the central block. A handful of miscellaneous chemicals that are fairly isolated in $\bm{\eta}$ space relative to the other included chemicals are shown in the top right block.

When selecting future chemicals for prioritization, one can either seek to ``fill in'' the space around chemicals of known toxicity relevance, or to ``venture out'' into spaces not near any currently tested chemicals. While addressing this experimental design problem is outside the scope of this work, we assume for the sake of exposition that both possible avenues are of interest. We further assume that the set of hold-out chemicals represents the space of options for further \textit{in vitro} testing.

Assuming the central cluster of chemicals in Figure \ref{fig:chem_dist} is of interest for more targeted exploration, we could select the hold-out chemicals closest to that set to test further. In terms of average distance between each cluster member, the four closest chemicals in the hold-out set are Clomiphene citrate (1:1) (i.e., our `seed' chemical for this training group, for which the model predictive performance can be seen in Figure \ref{fig:res_inference_yPred_doubly}), 4-Hydroxytamoxifen, Tamoxifen, and Tamoxifen citrate. Research has suggested similarity in action between Clomiphene citrate and Tamoxifen \citep{dhaliwal2011tamoxifen, seyedoshohadaei2012comparison}, lending credence to this distance measure's accounting of activity-relevant similarity.

\begin{figure}[!htpb]
\centering
\includegraphics[width=0.8\textwidth]{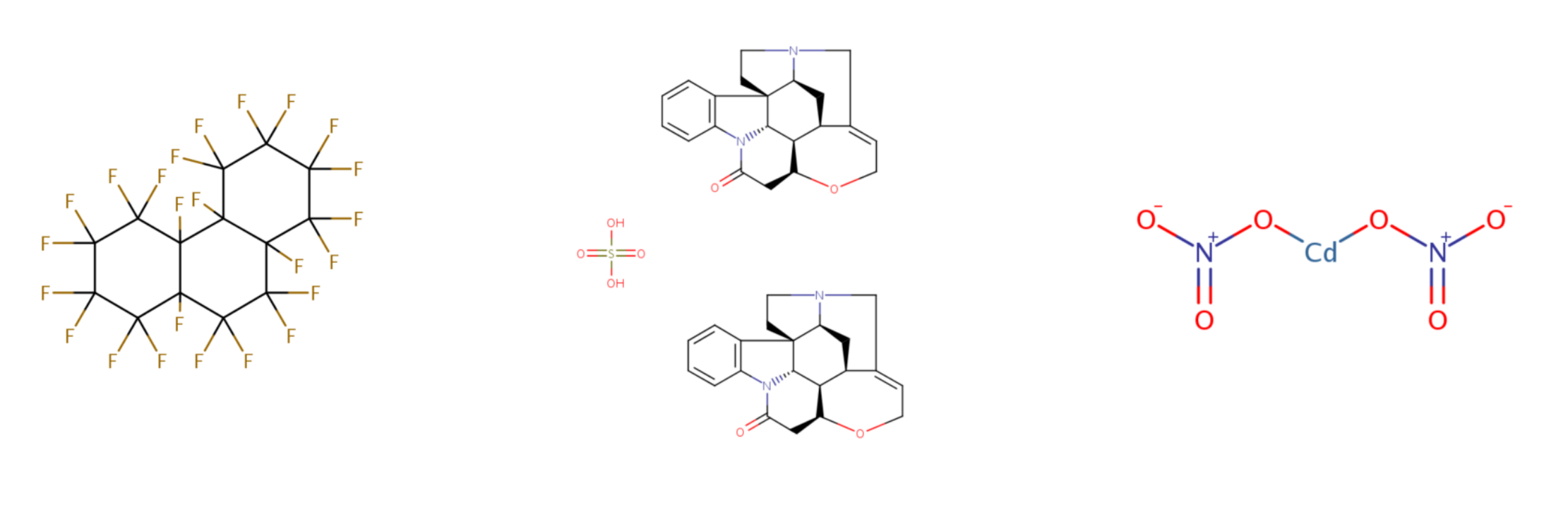}
\caption{From left to right: Tetracosafluorotetradecahydrophenanthrene, Strychnine hemisulphate salt, and Cadmium dinitrate.}
\label{fig:far_testchems}
\end{figure}

If, on the other hand, our goal was to test new chemicals that are least similar to observed chemicals, then we could choose the hold-out chemicals having the largest minimum distance to a training chemical. Assuming we chose such chemicals iteratively, we would select Tetracosafluorotetradecahydrophenanthrene (a solvent used in the preparation of certain polymers), Strychnine hemisulphate salt (a pesticide used for rodent and bird control), and Cadmium dinitrate (a colorant and photographic flash powder component). That these chemicals are the most distant in $\bm{\eta}$ space from both the training set and each other suggests that the model considers them to have distinctive activity relevant variability. See Figure \ref{fig:far_testchems} for their structure diagrams; the predicted activity for Cadmium dinitrate is also shown in Figure \ref{fig:res_inference_yPred_singly}. 

\subsection{Prediction}

Overall, the MSE between the data and the predicted dose-response profiles is 0.24 for the training chemicals and 0.30 for the hold-out chemicals. For comparison, the straw man model using the mean of the training data at a given dose to predict the hold-out data at that dose leads to an MSE of 0.37, and predicting all hold-out dose response profiles to be inactive (i.e. 0 everywhere) yields an MSE of 0.47. LASSO using dose, feature, and the interaction between dose and feature gives an MSE of 0.31 for the training chemicals and 0.40 for the hold-out chemicals. The B-FOSR model, as run using the \texttt{fosr()} function in the \textbf{fosr} package, had unstable predictions as the amount of missingness in the training data was too high (recall Figure \ref{miss_act}) for the imputation scheme used in the code -- the MSE was in the 100s. Limiting the training data to only include the most commonly observed dose levels did not resolve this issue. The \texttt{fosr.vs()} function simply returned an error due to the amount of missingness in the training data. The BAABTP model was not run for the toxicity data for computation time reasons.

The 95\% prediction intervals for the $\text{BS}^3\text{FA}$ model run cover 94.2\% of the training data and 92.6\% of the hold-out data, respectively. Unsurprisingly, hold-out chemicals having lower coverage also tend to have higher MSE. The coverage for specific hold-out chemicals is inversely related to the minimum distance between that chemical and its closest neighboring training chemical, while the MSE is directly related to the minimum distance between that chemical and its closest neighboring training chemical. That is, as a hold-out chemical moves farther away from other training chemicals, on average its coverage and MSE become worse. This finding suggests that increasing the amount of training data with specific care to the lesser-known regions of the chemical feature space would likely improve the model's MSE and coverage. 

\begin{figure}[!htpb]
\centering
\includegraphics[width=0.8\textwidth]{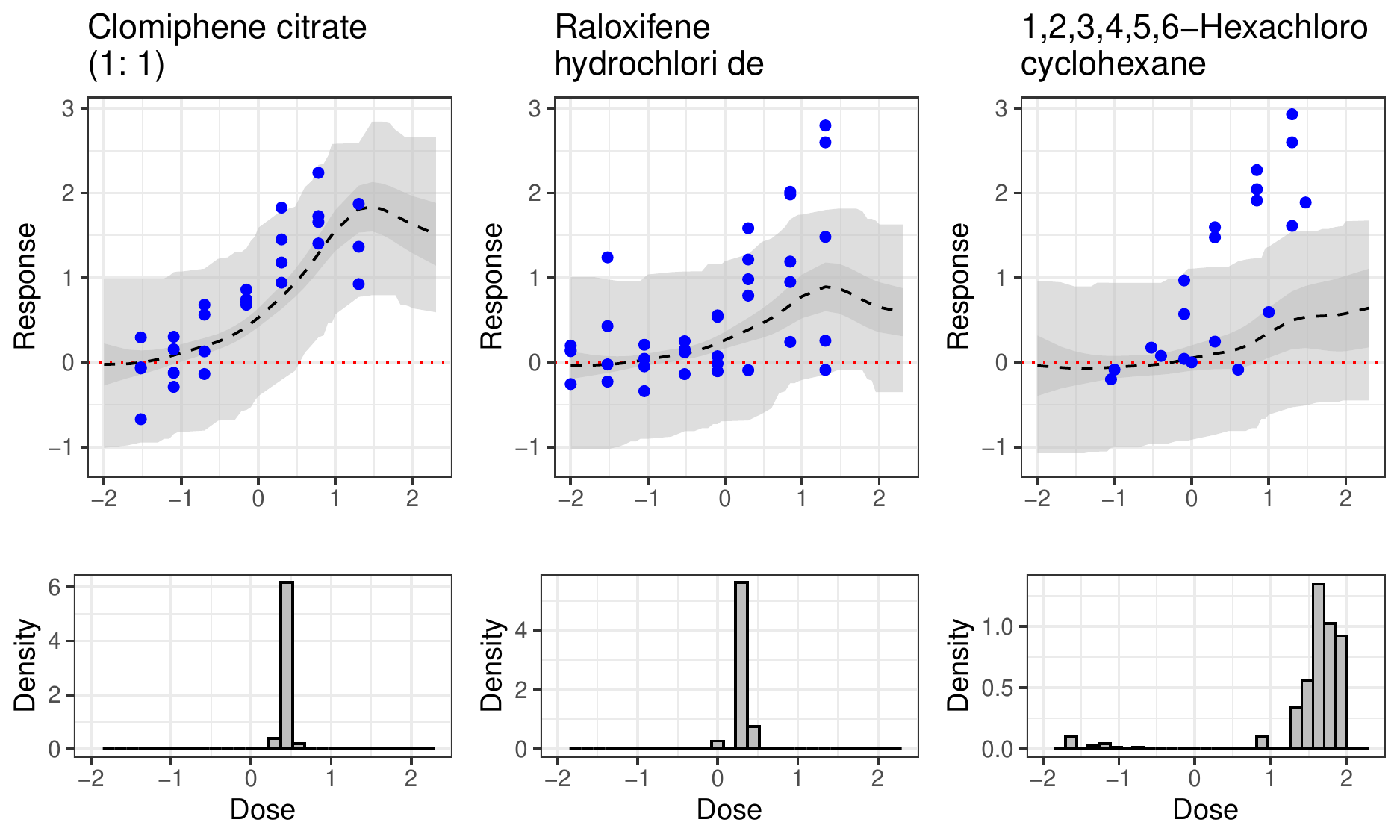}
\caption{Results for heavily tested hold-out chemicals predicted by the model to be activating. MSEs from left to right are 0.21, 0.49, and 1.44. Top: Predicted average dose-response curve (dashed black line), 95\% simultaneous band for expected dose-response curve (darker grey ribbon), and 95\% simultaneous band for observed data (lighter grey ribbon). Data (held out in training) are solid blue points. Bottom: Posterior samples of the AC50 value, i.e. the dose at which the dose response curve is at half of its maximal value.}
\label{fig:res_inference_yPred_doubly}
\end{figure}

\begin{figure}[!htpb]
\centering
\includegraphics[width=0.8\textwidth]{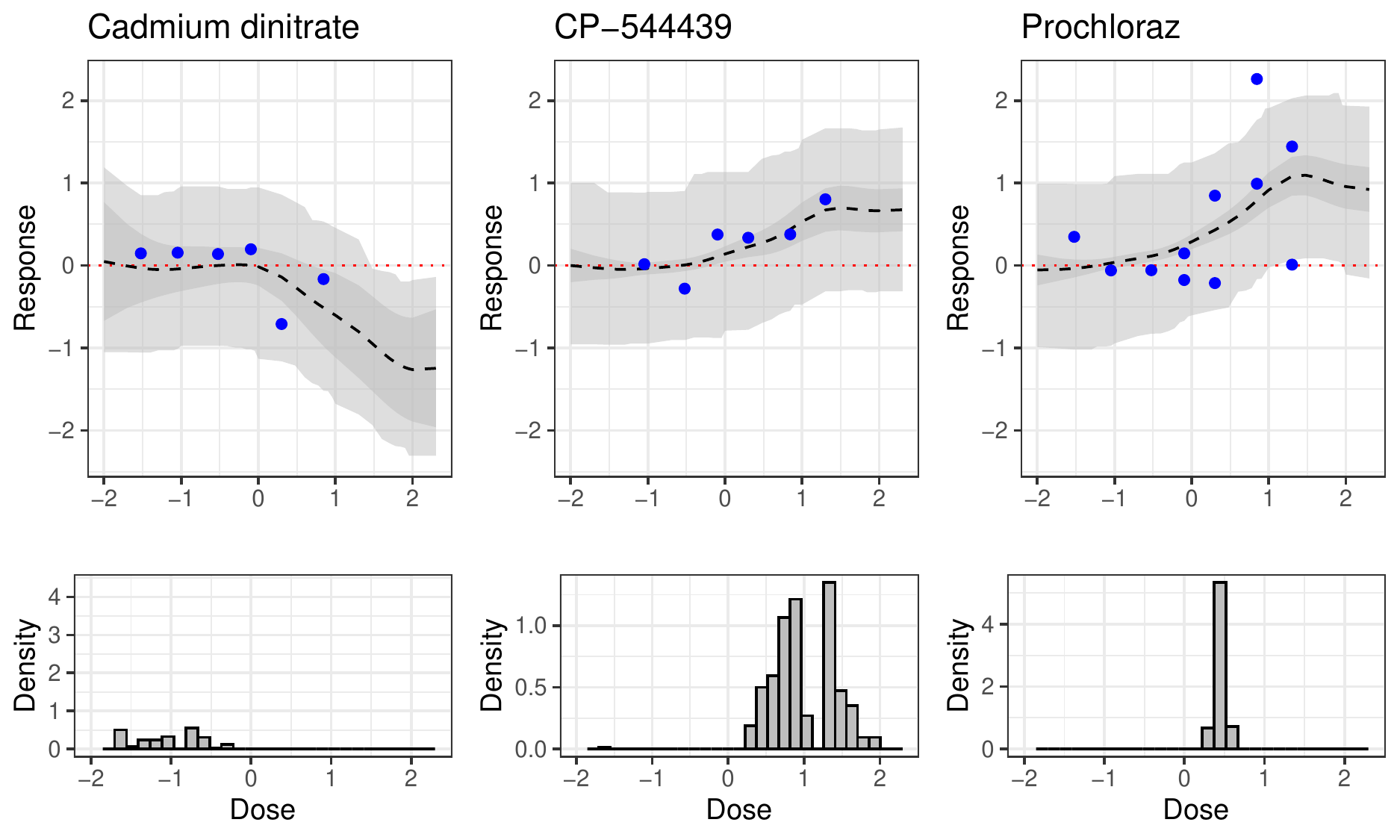}
\caption{Results for hold-out chemicals predicted by the model to be active (activity decreasing in the case of Cadmium dinitrate). MSEs from left to right are 0.10, 0.03, and 0.40. Top: Predicted average dose-response curve (dashed black line), 95\% simultaneous band for expected dose-response curve (darker grey ribbon), and 95\% simultaneous band for observed data (lighter grey ribbon). Data (held out in training) are solid blue points. Bottom: Posterior samples of the AC50 value, i.e. the dose at which the dose response curve is at half of its maximal value.}
\label{fig:res_inference_yPred_singly}
\end{figure}

We say a chemical is predicted to be activating, i.e. to increase activity, if the global Bayesian p-value \citep{meyer2015bayesian} for the predicted dose response profile is less than 0.05 and the 95\% simultaneous bands exceed 0 for at least one dose value. Figures \ref{fig:res_inference_yPred_doubly} and \ref{fig:res_inference_yPred_singly} show model predicted mean dose-response (MDR) curves along with samples of the AC50 value, i.e. the dose at which the dose response curve is at half of its maximal value, for activating hold-out chemicals. The predicted mean and 95\% posterior bands for the MDR curves are smooth due to the underlying structure of $\Lambda.$ The proportion of hold-out chemicals deemed activating by our model is nearly twice as high among the population of chemicals that are heavily tested (i.e., that have 10 or more observations). Since chemicals that are known to have toxic effects tend to be more heavily tested, this finding is suggestive of the model's capability to detect activity.

Although the predictive ability of the model is imperfect (e.g., 1,2,3,4,5,6−Hexachlorocyclohexane in Figure \ref{fig:res_inference_yPred_doubly} is under-predicted), overall performance appears reasonable. Particularly poorly predicted chemicals, examples of which are shown in the Supplemental Materials, tend to have shapes that differ from the common profiles and/or be farther in $\bm{\eta}$ space from training data than well-predicted curves. Also note that although the observations above the reported cytotoxicity limit were removed prior to running the model, the hormesis shape (i.e., the downturn at the end of the predicted dose response profile) remains prominent in highly active chemicals because it is a feature of the first column of $\Lambda,$ which has the highest column norm and drives large scale variation across profiles. 

\begin{figure}[!htpb]
\centering
\includegraphics[width=0.8\textwidth]{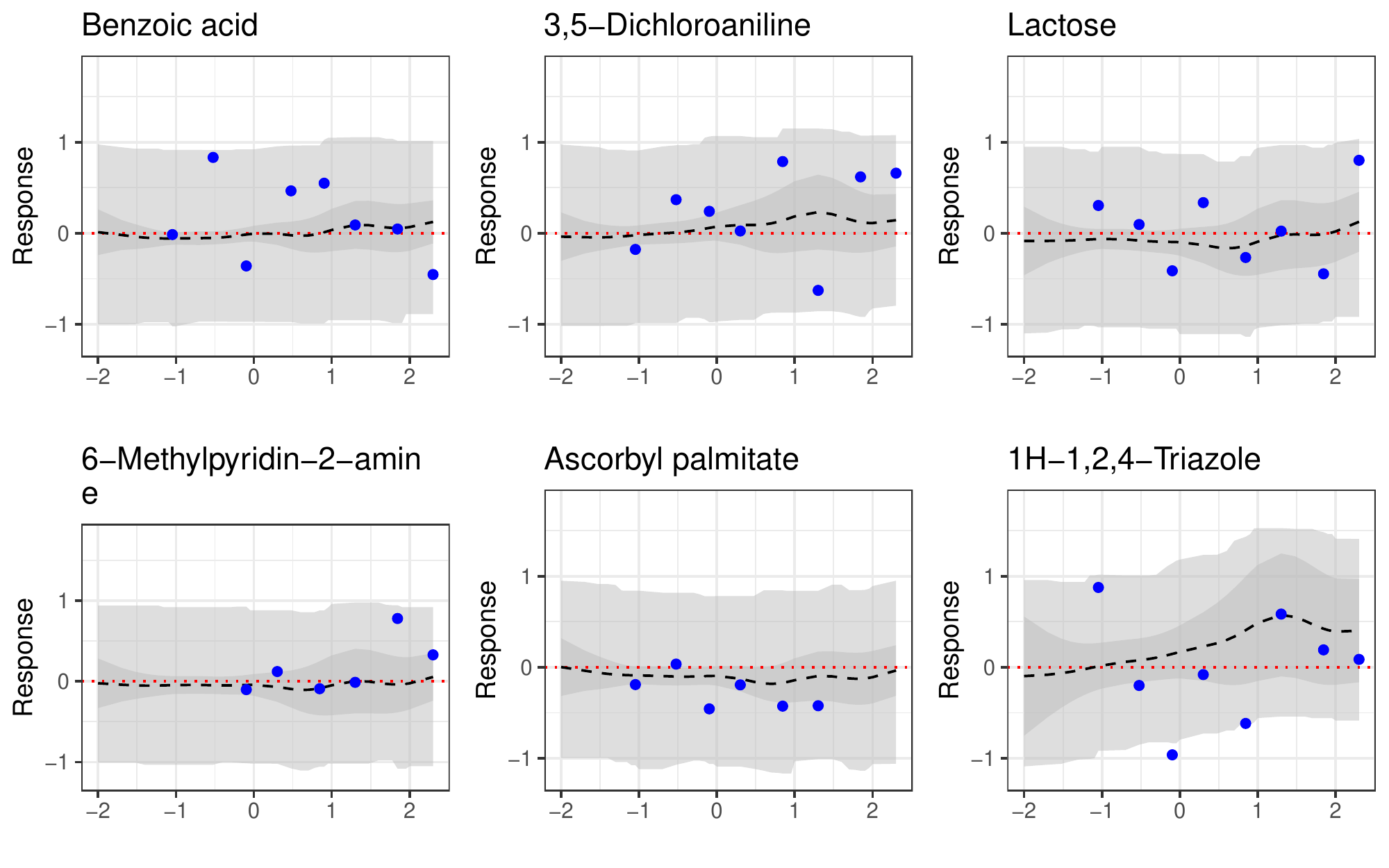}
\caption{Results for select hold-out chemicals predicted by the model to be inactive. MSEs from left to right, top to bottom, are 0.22, 0.23, 0.14, 0.13, 0.06, and0.42. Shown are predicted average dose-response curve (dashed black line), 95\% simultaneous band for expected dose-response curve (darker grey ribbon), and 95\% simultaneous band for observed data (lighter grey ribbon). Data (held out in training) are solid blue points.}
\label{fig:res_inference_yPred_inactive}
\end{figure}

In the context of this experiment, we fail to reject the null hypothesis that a chemical is inactive if its global Bayesian p-value \citep{meyer2015bayesian} is greater than 0.05 (i.e., if the bounds of the 95\% posterior simultaneous bands for the predicted MDR curve include zero at all points). For convenience, we refer to these chemicals as being predicted to be inactive by the model. Figure \ref{fig:res_inference_yPred_inactive} shows model predicted inactive dose-response curves for hold-out chemicals. As before, the predictions are smooth and appear reasonable relative to the true data. On average, hold-out chemicals deemed inactive under the criteria outlined above have a lower true maximum observed response than those deemed activity-increasing.

The choice of how to prioritize chemicals for future evaluation is flexible. Assuming there are in fact no dose-response data for the hold-out chemicals, a simple scheme by which chemicals could be selected for \textit{in vitro} study based on their $\text{BS}^3\text{FA}$ predictions would be to screen all chemicals for which the lower limit of the $1-\alpha$ posterior simultaneous bands for the MDR curve exceeds some threshold (e.g., 0). The value of $\alpha$ could be selected with attention to the resources available-- a larger $\alpha$ would lead to more chemicals being screened, whereas a smaller $\alpha$ would mean only those chemicals the model is most confident about would be screened. Once the set of chemicals are selected for further testing, the order of screening could be determined by chemicals' expected AC50 value, by the maximum value of their predicted MDR curves, by the highest value taken by the lower $\alpha/2$ simultaneous band for expected response, by their proximity in latent space to known toxic chemicals, or by another metric of interest.

For example, if one chooses $\alpha=0.05,$ then 68\% of hold-out chemicals have predicted dose response curve lower bound that at some point exceeds 0. The 5 highest priority chemicals using the max lower bound method would be Basic Blue 7, Ergocalciferol, Toremifene citrate, 4-(2-Phenylpropan-2-yl)-N-[4-(2-phenylpropan-2-yl)phenyl]aniline, and 2,4-Bis(1-methyl-1-phenylethyl)phenol. As discussed previously, Toremifene citrate is a close relative of Clomiphene citrate (1:1) (activity shown in Figure \ref{fig:res_inference_yPred_doubly}), and a selective estrogen receptor modulator used to treat ovulatory dysfunction in women trying to become pregnant. Basic Blue 7 is labeled as corrosive, an irritant, acutely toxic, and an environmental hazard. The closest neighbors to Basic Blue 7 in the training set are other colorants, Gentian Violet and Malachite green, that have known toxic effects \citep{docampo1990metabolism, srivastava2004toxicological}.

\section{Conclusion}


We have focused on the utility of distance learning for designing future chemical test sets, but these pairwise distance matrices could be used in place of Euclidean distance in any distance-based statistical analysis.  This would include distance-based clustering of chemicals, as well as kernel and Gaussian process-based models. As a specific example, the authors believe these activity-relevant distances have the potential to improve main effects estimates in mixture models for human health outcomes. It is likely that incorporating knowledge about similarity in activity-relevant space (e.g., by using the toxicity-relevant pairwise distance matrix to inform a group penalized regression model) would provide stabilization for main effects, and in turn allow for better estimation of the interaction effects. 

The designation of active vs. inactive in the $\text{BS}^3\text{FA}$ model is based on a posterior summary of the predicted dose-response profiles; there is no direct incorporation of the concept of a chemical being inactive in the model itself. It may be desirable, particularly when considering assays having very few chemicals presenting with any activity, to probabilistically model inactivity. For example, the dose response profile could be modeled as a mixture between the zero-vector and the $\text{BS}^3\text{FA}$ factor model, with a learned weight on the zero vector corresponding to the probability of inactivity.

The $\text{BS}^3\text{FA}$ model deals with the structured decomposition of a single assay and a single feature data set. In reality, any sort of model hoping to extend to human health outcomes will need to utilize information from multiple sources. In the ToxCast data set, there is not just one dose-response curve per chemical. There are many assay endpoints of potential relevance to human toxicity. Furthermore, there are potentially many useful chemical feature descriptors (we used Mold2, but others include MACCS keys, Daylight Fingerprints, or Morgan Fingerprints, to name a few). Future work will link the ideas in $\text{BS}^3\text{FA}$ to those in \citet{wilson2014hierarchical} to allow for a more direct active/inactive assignation and to hierarchically describe variability across multiple assays. Extending even further, the holy grail of toxicity modeling would be an explicit linking of multiple assay endpoints to human health data, such that human health outcomes could be predicted from chemical structure alone.


\section{Acknowledgements}

This work was partially supported by the National Institute of Environmental Health Sciences of the United States National Institutes of Health (grants 1R01ES028804-01 and 5R01ES027498-02) and the Department of Energy Computational Science Graduate Fellowship (grant DE-FG02-97ER25308). The funders had no role in study design, data collection and analysis, decision to publish, or preparation of the manuscript. The authors would like to thank Evan Poworoznek and Bora Jin for helpful comments. 

\bibliographystyle{imsart-nameyear}
\bibliography{bibliography}

\end{document}


\begin{frontmatter}
\title{Supplement to: Bayesian joint modeling of chemical structure and dose response curves}
\runtitle{Supplement to: Modeling of chemical structure and dose response curves}

\begin{aug}
\author{Kelly R. Moran, David Dunson, Matthew W. Wheeler, and Amy H. Herring}

\runauthor{K. R. Moran et al.}

\end{aug}


\end{frontmatter}

\beginsupplement

\section{Gibbs sampler}\label{gibbs_sampler}

To initialize the sampler normally distributed mean 0 variables, constrained to be positive when necessary, can be sampled for all parameters (the authors have also found that choosing a small variance helps keep the sampler from moving in unstable directions). Alternatively, the SVD of [$Y,X$] can be used to initialize [$\Lambda$,$\Theta$], $\eta$ can be set to $\Lambda^T Y$, and $\Xi$ $X-\Theta \eta$, and the noise variance terms can be initialized based on the residuals when using these initial values.

Assume for notational convenience that the data have been mean-centered prior to analysis (i.e., that $\bm{\mu}^x$ and $\bm{\mu}^y$ are fixed to 0-vectors). Let $O^Y$ denote the $D \times N$ matrix with the $(d,i)$th entry giving the number of observations at dose $d$ for chemical $i$. Throughout let $\{V\}^{-a}$ denote the set of elements of vector $V$ excepting the $a$-th entry. Assuming initial values as specified above, the sampler proceeds as follows:

\textbf{\textit{Step 1.}  Sample $Y$-specific factor loading matrix $\Lambda$ and associated hyper parameters.}

Sample columns of $\Lambda$ one at a time, letting $\bm{\lambda}_k$ denote the $k$th column of $\Lambda$. For $k\in 1,\ldots,K$: 
\begin{itemize}
\item[--] Set $Y_k^*$ to be an $D \times N$ matrix with $i$th column $y^*_i= (\bm{y}_i-\sum_{h \neq k} \lambda_h \eta_{h,i}) / \eta_{k,i}.$ Define $S_k^*$ to be a length-$D$ vector with the $d$th entry being $\frac{ \sum_{i=1}^N y^*_{d,i} \eta_{k,i}^2 O^Y_{d,i}/\sigma^2_Y }{ \sum_{i=1}^N \eta_{k,i}^2 O^Y_{d,i}/\sigma^2_Y }$. In other words, $S_k^*$ is an inverse variance weighted mean of $Y_k^*$.
\item[--] Let $C_k$ denote the $D \times D$ GP covariance matrix at all unique dose values, formally defined by the kernel $c_k(d,d') = \alpha_k^2 e^{-\frac{(d-d')^2}{2\ell^2}}$. Let $C^*_k$ be analagous to $C_k$ but with the additive inverse variance component along the diagonal, i.e. $c^*_k(d,d') = c_k(d,d') + \mathbbm{1}_{d=d'} \frac{1}{\sum_{i=1}^N \eta_{k,i}^2 O^Y_{d,i}/\sigma^2_Y}.$
\item[--] Sample $\bm{\lambda}_k|\{\lambda_h\}_{h \neq k}, Y, \eta, \alpha_k, \ell, \sigma^2_Y \sim \text{N}(C_k (C^*_k)^{-1} S_k^*,C_k - C_k (C^*_k)^{-1} C_k) .$
\end{itemize}

Sample hyper parameters associated with $\Lambda,$ including the shared $\{\delta_k\}$ that are also used by $\Theta$:
\begin{itemize}
\item[--] Let $C^{-\alpha}_k$ denote the $D \times D$ covariance matrix at all unique dose values without the function variance term, formally defined by the kernel $c^{-\alpha}_k(d,d') = e^{-\frac{(d-d')^2}{2\ell^2}}$. 
\item[--] Assume $l \in L,$ where $L$ is a discrete set of possible length-scale values. Sample $\Pr(\ell=l \ | \ -) \ = \ c^* \ \prod_{k=1}^K \{\text{det}(C^{l}_k) \}^{-0.5} \exp{} \{ -0.5 \ \tau_k^{-1} \bm{\lambda}_k^T (C^{l}_k)^{-1} \bm{\lambda}_k \},$ where $C^{l}_k$ is defined by covariance kernel $c^{l}_k(d,d') = e^{-\frac{(d-d')^2}{2l^2}}$ and $c^* \ = \sum_{l' \neq l} \Pr(\ell = l' \ | \ -)$ is the normalizing constant.
\item[--] Sample $\phi \sim \text{Ga}(\frac{g_{\phi}+DK}{2}, \frac{\sum_{k=1}^K [\tau_k^{-1} \bm{\lambda}_k^T (C^{-\alpha}_k)^{-1} \bm{\lambda}_k ] }{2}).$
\item[--] Define $\tau_{k}^{(-h)} = \prod_{t=1,t\neq h}^{k} \delta_{t}$ for $h=1,\ldots,K$.
\clearpage
\item[--] Sample 
\begin{align*}
\begin{split}
\delta_{1} \ | \ \text{all} & \sim \text{Ga}( a_{1} + K(D+S)/2 , 1+\frac{1}{2} \sum_{k=1}^K [   \tau_{k}^{(-1)} \phi \bm{\lambda}_k^T (C^{-\alpha}_k)^{-1} \bm{\lambda}_k + \tau_{k}^{(-1)} \beta^{-2} \sum_{s=1}^S \gamma_{s,k}^{-2} \Theta_{s,k}^2  ] ), \\
\delta_{h} \ | \ \text{all} & \sim \text{Ga}( a_{2} + (K-h)(D+S)/2 , 1+\frac{1}{2} \sum_{k=1}^K [   \tau_{k}^{(-h)} \phi \bm{\lambda}_k^T (C^{-\alpha}_k)^{-1} \bm{\lambda}_k + \tau_{k}^{(-h)} \beta^{-2} \sum_{s=1}^S \gamma_{s,k}^{-2} \Theta_{s,k}^2  ] ), \\
& h=2,\ldots,K.
\end{split}
\end{align*}
\end{itemize}

Finally, set $\tau_k = \prod_{h=1}^k \delta_h$ and $\alpha_k^2 = \big(\phi \tau_k\big)^{-1}$ for $k=1,\ldots,K.$

Note that the posterior over $\ell$ stems from a uniform prior over a discrete grid of possible length scale values. The authors have found a grid size of 100 ranging over ``reasonable'' length-scale values (i.e., those not implying an effective range smaller than the difference between the closest two dose value, or larger than the range of the data) to be sufficient in the simulation and application described in the paper. However, for larger $D$ or grid sizes it may be computationally preferable to use a Metropolis Hastings step here instead.

\textbf{\textit{Step 2.}  Sample latent variable $Z$ corresponding to any non-continuous entries of $X$.}

Let $E^X$ be the $S \times N$ matrix of expected values of $X,$ i.e. $E^X = \Theta \eta + \Xi \nu.$ Then, for $s=1,\ldots,S$ and $i=\1,\ldots,N$ such that $x_{s,i}$ is not continuous sample $z_{s,i}$ as follows:
\begin{itemize}
\item[--] For binary $x_{s,i},$ sample
\begin{equation*}
z_{s,i} \ | \ E^X_{s,i} \sim \begin{cases} 
\mbox{N}_{+}(E^X_{s,i},1), & \mbox{if } x_{s,i} = 1 \\ 
\mbox{N}_{-}(E^X_{s,i},1), & \mbox{if } x_{s,i} = 0
\end{cases}
\end{equation*}
\item[--] For count $x_{s,i},$ sample
\begin{equation*}
z_{s,i} \ | \ E^X_{s,i} \sim \begin{cases} 
\mbox{N}_{[t-1,t)}(E^X_{s,i},1), & \mbox{if } x_{s,i} = t \\ 
\mbox{N}_{(-\infty,0]}(E^X_{s,i},1), & \mbox{if } x_{s,i} = 0.
\end{cases}
\end{equation*}
\end{itemize}
Note categorical variables should have already been pre-transformed into multiple binary indicators prior to running the Gibbs sampler.

\textbf{\textit{Step 3.}  Sample $X$-specific toxicity-irrelevant components, including factor loadings matrix $\Xi$, scores $\nu$, and associated hyper parameters.} Define $D=X-\Theta \eta$ to be the $S \times N$ `residual' toxicity-irrelevant feature matrix.

First sample the $J \times N$ matrix $\nu$, i.e. the matrix of $X-$specific toxicity-irrelevant factors, by column:
\begin{itemize}
\item[--] Define $S \times J$ matrix $\Xi^* = \biggl[ \begin{smallmatrix} \sigma^{-2}_1 \Xi_{\text{ row 1}} \\ \ldots \\ \sigma^{-2}_S \Xi_{\text{ row S}}\end{smallmatrix}\biggr]$, then set $J \times J$ matrix $R^* = ( \Xi^T \Xi^*  +\text{diag}(1,\ldots,1) )^{-1}$.
\item[--] For $i=1,\ldots,N,$ sample $\bm{\nu}_i \ |\ \text{all} \sim \text{N}(R^* (\Xi^*)^T D_i, R^*).$
\end{itemize}

Next sample the $S \times J$ matrix $\Xi$, i.e. the matrix of $X-$specific toxicity-irrelevant factor loadings, by row. For $s=1,\ldots,S:$
\begin{itemize}
\item[--] Define $J \times J$ matrix $R_s^*= (\sigma^{-2}_{X,s} \nu \nu^T + \text{diag}(\kappa_{s,1}\omega_1,\ldots,\kappa_{s,J}\omega_J) )^{-1}$.
\item[--] Sample $\Xi_{\text{ row }s} \ | \ \text{all} \sim \text{N}(R_s^* \nu (D_{\text{ row }s})^T / \sigma^2_{X,s},R_s^*).$
\end{itemize}

Next sample the local and column-specific shrinkage parameters $\{\kappa_{s,j}\}$ and $\{\zeta_j\}$:
\begin{itemize}
\item[--] For $s=1,\ldots,S$ and $j=1,\ldots,J$, sample $\kappa_{s,j} \ | \ \xi_{s,j},\omega_j,g_{\kappa} \ \sim \text{Ga}(\frac{g_{\kappa}+1}{2},\frac{g_{\kappa}+\xi^2_{s,j} \omega_j}{2}).$
\item[--] Define $\omega_{j}^{(-h)} = \prod_{t=1,t\neq h}^{j} \zeta_{t}$ for $h=1,\ldots,J$.
\item[--] Sample 
\begin{align*}
\begin{split}
\zeta_{1} \ | \ \text{all} & \sim \text{Ga}( a_{1} + JS/2 , 1+\frac{1}{2} \sum_{j=1}^J \omega_{j}^{(-1)} \sum_{s=1}^S \kappa_{s,j}^{-2} \Xi_{s,j}^2  ), \\
\zeta_{h} \ | \ \text{all} & \sim \text{Ga}( a_{2} + (J-h)S/2 , 1+\frac{1}{2} \sum_{j=1}^J \omega_{j}^{(-h)} \sum_{s=1}^S \kappa_{s,j}^{-2} \Xi_{s,j}^2  ] ), \\
& h=2,\ldots,J.
\end{split}
\end{align*}
\end{itemize}

Finally set $\omega_j = \prod_{h=1}^j\zeta_j.$

\textbf{\textit{Step 4.}  Sample $X$-specific toxicity-relevant components, including factor loadings matrix $\Theta$ and its associated shrinkage hyper parameters.} Define $D=X-\Xi \nu$ to be the $S \times N$ toxicity-relevant feature matrix, i.e. the data matrix with the `residual' toxicity-irrelevant features removed.

First sample the $S \times K$ matrix $\Theta$, i.e. the matrix of $X-$specific toxicity-relevant factor loadings, by row. For $s=1,\ldots,S:$
\begin{itemize}
\item[--] Define $K \times K$ matrix $R_s^*= (\sigma^{-2}_{X,s} \eta \eta^T + \text{diag}(\beta^{-2}\gamma_{s,1}^{-2}\tau_1,\ldots,\beta^{-2}\gamma_{s,K}^{-2}\tau_K) )^{-1}$.
\item[--] Sample $\Theta_{\text{ row }s} \ | \ \text{all} \sim \text{N}(R_s^* \eta (D_{\text{ row }s})^T / \sigma^2_{X,s},R_s^*).$
\end{itemize}

Next sample the global and local shrinkage parameters $beta^2$ and $\{\gamma_{s,k}\}$ (note that the shared column-specific shrinkage parameter $\delta_k$ was already sampled in the update step for components of $\Lambda$):
\begin{itemize}
\item[--] Sample $\beta^2 \ | \ \{\theta_{s,k}\},\{\tau_k\},\{\gamma^2_{s,k}\} \ \sim \text{Ga}(\frac{SK+1}{2},\frac{1}{t}+\frac{\sum_{k=1}^K\tau_k \sum_{s=1}^S \theta^2_{s,k}/\gamma^2_{s,k}}{2}).$
\item[--] Sample $\gamma^2_{s,k} \ | \ \theta_{s,k},\tau_k,\beta^2 \ \sim \text{Ga}(1,\frac{1}{b_{s,k}}+\frac{\tau_k \theta^2_{s,k}}{2\beta^2})$ for $s=1,\ldots,S$ and $k=1,\ldots,K.$
\end{itemize}

Finally sample $\{b_{s,k}\}$ and $t,$ the hyperparameters from the horseshoe prior imposed on elements of $\Theta:$
\begin{itemize}
\item[--] Sample $ t \ | \ \beta^2 \ \sim \text{Ga}(1,1+\frac{1}{\beta^2}).$
\item[--] Sample $ b_{s,k} \ | \ \gamma^2_{s,k} \ \sim \text{Ga}(1,1+\frac{1}{\gamma^2_{s,k}})$ for $s=1,\ldots,S$ and $k=1,\ldots,K.$
\end{itemize}

\textbf{\textit{Step 5.}  Sample shared toxicity-relevant factor matrix $\eta$.} Define concatenated $(D+S) \times K$ loadings matrix $\Omega = \biggl[ \begin{smallmatrix} \Lambda \\ \Theta \end{smallmatrix}\biggr]$ and $(D+S) \times N$ data matrix $W = \biggl[ \begin{smallmatrix} Y^{\text{sum}} \\ X - \Xi \nu \end{smallmatrix}\biggr]$ where $Y^{\text{sum}}$ is a $D \times N$ matrix with entry $[h,i]$ giving the sum across replicates of the response values for chemical $i$ at dose index $h$ (or left as missing for chemical $i$ with no observations at dose index $h$). Explicitly, $Y^{\text{sum}}[h,i] = \sum_{r=1}^{R[i]} \bm{y}_{i[r], h}.$ Let $O_{h,i}$ denote the number of observations of chemical $i$ at dose index $h.$ For example, if chemical $i$ has two replicates observed at dose indices $\{1, 3, 5\}$ and $\{1, 2, 5, 6\}$, respectively, then (assuming $D=6$) $\bm{O}_i = [O_{1,i}, \ldots, O_{6,i}]'$ equals $[2,1,1,0,2,1]'.$ 

Sample the $K \times N$ matrix $\eta$ by column. For $i =1,\ldots,N$:
\begin{itemize}
\item[--] Let $U_i^* = \{d^*_1,\ldots,d^*_U\}$ be the set of $U\leq D$ unique dose values at which chemical $i$ has at least one observation (i.e., $d^*_u \in U_i^*$ iff $O^Y_{d^*_u,i}>0$). Use $M_{[U_i^*]}$ to denote the matrix with only the relevant doses selected (across either rows or columns, depending on which is the relevant dimension). For example, $\Omega_{[U_i^*]}$ is the $(U+S) \times K$ matrix with the first $U$ rows being the subset of the original first $D$ rows for doses in $U_i^*.$
\item[--] Define $\sigma_{Y,i}^2$ to be a length-$D$ vector with entry $\sigma_{Y,i[d]}^2 = \begin{cases} 
\sigma_{Y}^2/O^Y_{d,i}, & \mbox{if } O^Y_{d,i} > 0 \\ 
\sigma_{Y}^2, & \mbox{if } O^Y_{d,i} = 0
\end{cases}$.
\item[--] Set $\Omega_i^*$ to be the $(D+S) \times K$ matrix with row $p$ being the $p$th row of $\Omega$ times $1/\sigma_{Y,i[d]}^2$ for the first $D$ rows, and the $p$th row of $\Omega$ times $1/\sigma_{X,p-D}^2$ for the remaining $S$ rows. 
\item[--] Define the $K \times K$ matrix $R^*=( \Omega_{[U_i^*]}^T \Omega^*_{i[U_i^*]} + \text{diag}(1,\ldots,1) )^{-1}.$
\item[--] Sample $\bm{\eta}_i \ |\ \text{all} \sim \text{N}(R^* (\Omega^*_{i[U_i^*]})^T W_{[U_i^*],i}, R^*).$
\end{itemize}

\textbf{\textit{Step 6.}  Sample entries of noise variance parameters $\Sigma_X$ and $\Sigma_Y$.} Let $E^Y=\Lambda \eta$ and $E^X=\Theta \eta + \Xi \nu$ denote the mean of $Y$ and $X,$ respectively.

First sample $\sigma^2_Y$, the common noise variance term for the dose response curves:
\begin{itemize}
\item[--] Let $D_i^*$ denote the set of all (not necessarily unique) doses at which chemical $i$ has observations. E.g., $D_i^*$ could be $\{0,0,0.1,0.1,0.2,0.3\}$. 
\item[--] Define $\text{RSS}_Y = \sum_{i=1}^N \sum_{d\in D_i^*} (Y_{d,i} - E^Y_{d,i})^2 $ and $N_Y = \sum_{i=1}^N |D_i^*|.$
\item[--] Sample $\sigma^2_Y \ |\ \text{all} \sim \text{Ga}(\frac{a_{\sigma_Y}+N_Y}{2},\frac{b_{\sigma_Y}+\text{RSS}_Y}{2}).$
\end{itemize}

Next sample $\{\sigma^2_{X,s}\}$, the feature-specific noise variance terms, for $s =1,\ldots, S$:
\begin{itemize}
\item[--] Define $\text{RSS}_{X,s} = \sum_{i=1}^N (X_{s,i} - E^X_{s,i})^2 $.
\item[--] Sample $\sigma^2_{X,s} \ |\ \text{all} \sim \text{Ga}(\frac{a_{\sigma_X}+N}{2},\frac{b_{\sigma_X}+\text{RSS}_{X,s}}{2}).$
\end{itemize}

Note that an informative prior on $\sigma^2_Y$ can utilize the variance of low-dose observations in related assays, i.e. observations for which no activity is expected. Such a prior encourages the model to learn structure in the curves rather than simply learning a large noise variance.

\section{Additional simulation information and results}
\label{SI_sec:sim_additionalinfo}

\subsection{Sampling details}

In the $\text{BS}^3\text{FA}$ model, the B-FOSR, and the BAABTP model, four chains are run with the initial 20000 samples discarded as burn-in and every 10th draw of the subsequent 20000 samples saved. For the $\text{BS}^3\text{FA}$ model the number of toxicity relevent and irrelevant factors is set to be the true number of factors plus 5, the intention being to mimic the act of providing a conservative `upper bound' in the real data scenario. For the BAABTP model the number of mixture components is set to 5. For LASSO, the dose levels, all chemical features, and all pairwise interactions are included as covariates. 

\subsection{Distance performance}

For distance performance, $\text{BS}^3\text{FA}$ is compared to Euclidean distance in the full feature space and in PCA space (i.e., with no supervision). The distance between chemicals in $\eta$ space is used as the truth when assessing model chemical similarity performance. Correlation between model predicted distance and true distance is used as a scale-invariant measure of model performance. Predicted distance is a metric defined by the latent factors in each method (i.e., by model-predicted $\eta$ for $\text{BS}^3\text{FA}$ and by the principal component scores in PCA). Coverage is assessed for our model, B-FOSR, BAABTP, and a straw man model in which the standard deviation about low-dose response values is used to approximate a 95\% confidence interval around the observed training mean. Note that JIVE was not included as a competitor algorithm because the amount of missingness introduced by omitting the observations of $Y$ for the test data is is too high for the algorithm (the use of the \texttt{jive()} function in the \textbf{\textsf{R}} package \textbf{r.jive} resulted in an error message: ``Unable to complete matrix, too much missing data'').

\subsection{Inputs provided for each model}

The inputs to B-FOSR are all chemical features. The inputs to LASSO are the dose, chemical features, and all pairwise interactions therein. The features used in the BAABTP model are $\text{PC}_{95}$. The LASSO model was trained using the \texttt{cv.glmnet} function from the \textbf{glmnet} package and predictions were made using the \textt{predict} function from the resulting object with \texttt{s=`lambda.1se'}. 

\subsection{Simulated data structure with truth-model alignment}

For all simulations, data on 300 `chemicals' was created, comprised of dose-response curves and structure information. The number of unique doses was set to $D=10$, and the number of chemical features was set to $S=20$. The noise terms were set to be homoscedastic with $\sigma_{Y}=0.2$ and $\sigma_{X}=0.1$. The true dimension of the latent toxicity-relevant space was varied $K\in\{1,3,5\}$, as was that of the latent toxicity-irrelevant space $J\in\{0,5,10,15,20\}$. In each simulated data set, 25\% of the chemical dose-response curves are hidden from the models as test data, while the remaining 75\% are used as training data (note that feature data are available for all chemicals, not just training data). At each simulation setting, 100 simulated data sets are created. 

The toxicological `data' were simulated so as to mimic realistic dose response curves. A smooth factor loadings matrix $\Lambda$ is created in each simulation by first smoothing the real Attagene PXR data using a functional factor model, then randomly sampling 500 smoothed dose response curves from this set and setting $\Lambda$ to be the first $K$ loadings from the SVD of this smooth subset of curves. A sparse factor loadings matrix $\Theta$ is created for each simulation by creating a $K \times S$ standard normal matrix $M$, then using the \texttt{spEigen()} function from the \textbf{sparseEigen} \textbf{\textsf{R}} package to compute sparse orthogonal eigenvectors of the covariance matrix of $M$ (i.e., setting $\Theta$ equal to \texttt{spEigen(t(M) \%*\% M, q=K, rho=0.2)\$vectors}). 

\subsection{Simulated data structure with misalignment between the model and the true data generating process}\label{sec:SI_sim_misaligned}

We also perform a simulation study in which the structure assumed by $\text{BS}^3\text{FA}$ is \textit{not} the true data generating process, to mimic the (likely) scenario of a misalignment between our assumptions and the truth. The hope is that $\text{BS}^3\text{FA}$ is still able to perform as well as competitors under such a misalignment.

Define $S_\text{relevant}$ to be the number of toxicity-relevant features and $S_\text{irrelevant}$ to be the number of toxicity-irrelevant features. Then $S=S_\text{relevant} + S_\text{irrelevant}$ is the total number of features in $X.$ To assess how the model performs when the true data generating process does not match the data generating process assumed by $\text{BS}^3\text{FA},$ data were simulated assuming a polynomial relationship with dose. The misalignment simulation sets $y_{i,d} = \sum_{m=1}^{S_\text{relevant}} x_{i,m} \ d^m$. That is, the response is a polynomial function of dose with the parameters controlling the shape of the polynomial being the true ``toxicity-relevant'' entries in $\bm{x}_i.$ Thus the shape of the dose response curves does depend on some features in $\bm{x}_i,$ but not in the way assumed by $\text{BS}^3\text{FA}$ (although we note that when $S_\text{relevant}=1$ the polynomial is linear and can in fact be represented by the $\text{BS}^3\text{FA}$ model). We vary the number of toxicity-relevant features $S_\text{relevant}\in\{1,2,3\}$ and the number of toxicity-irrelevant features $S_\text{irrelevant}\in\{0,5,10,20,30\}.$ 

For all simulations and choices of $S_\text{relevant}$ and $S_\text{irrelevant}$, data on 300 `chemicals' was created. The number of unique doses was set to $D=10$. The noise about the true dose response mean is assumed to be homoscedastic with $\sigma_{Y}=0.2$. In each simulated data set, 25\% of the chemical dose-response curves are hidden from the models as test data, while the remaining 75\% are used as training data (note that feature data are available for all chemicals, not just training data). At each simulation setting, 100 simulated data sets are created.


\subsection{Additional results for model-truth-aligned simulation}
\label{sec:SI_sim_res_covgdist}

Within a plot throughout this section, each point shows the mean of the metric of interest from the 100 simulated data sets having the specified settings. The lower and upper bands around this point give the 2.5 and 97.5 percentiles of the performance values across the 100 simulations, respectively.

\begin{figure}[!htpb]
\centering
\includegraphics[width=0.6\textwidth]{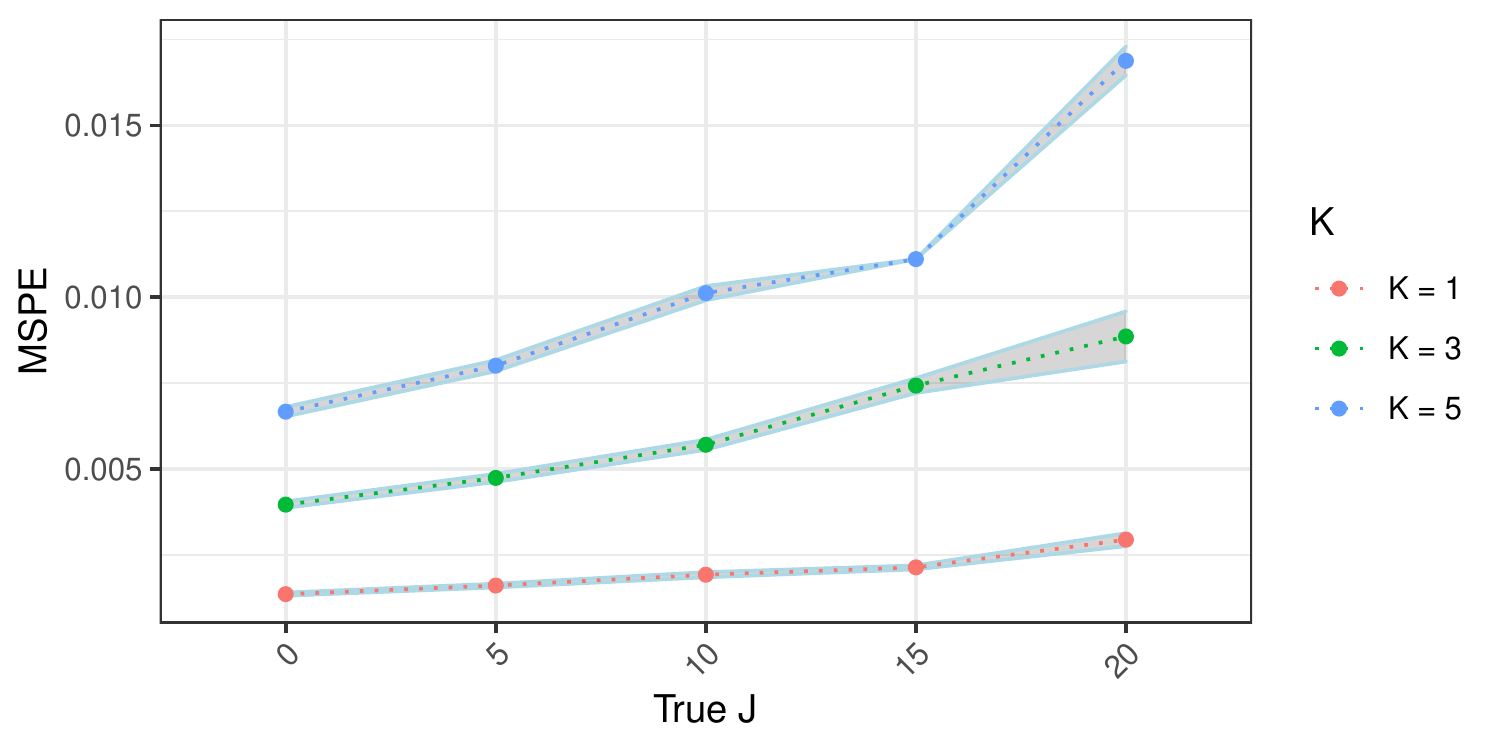}
\caption{MSPE between $\text{BS}^3\text{FA}$-predicted dose-response profile and true dose-response profile for hold-out ``chemicals''.}
\label{simResMSEbs3fa}
\end{figure}

\begin{figure}[!htpb]
\centering
\includegraphics[width=0.6\textwidth]{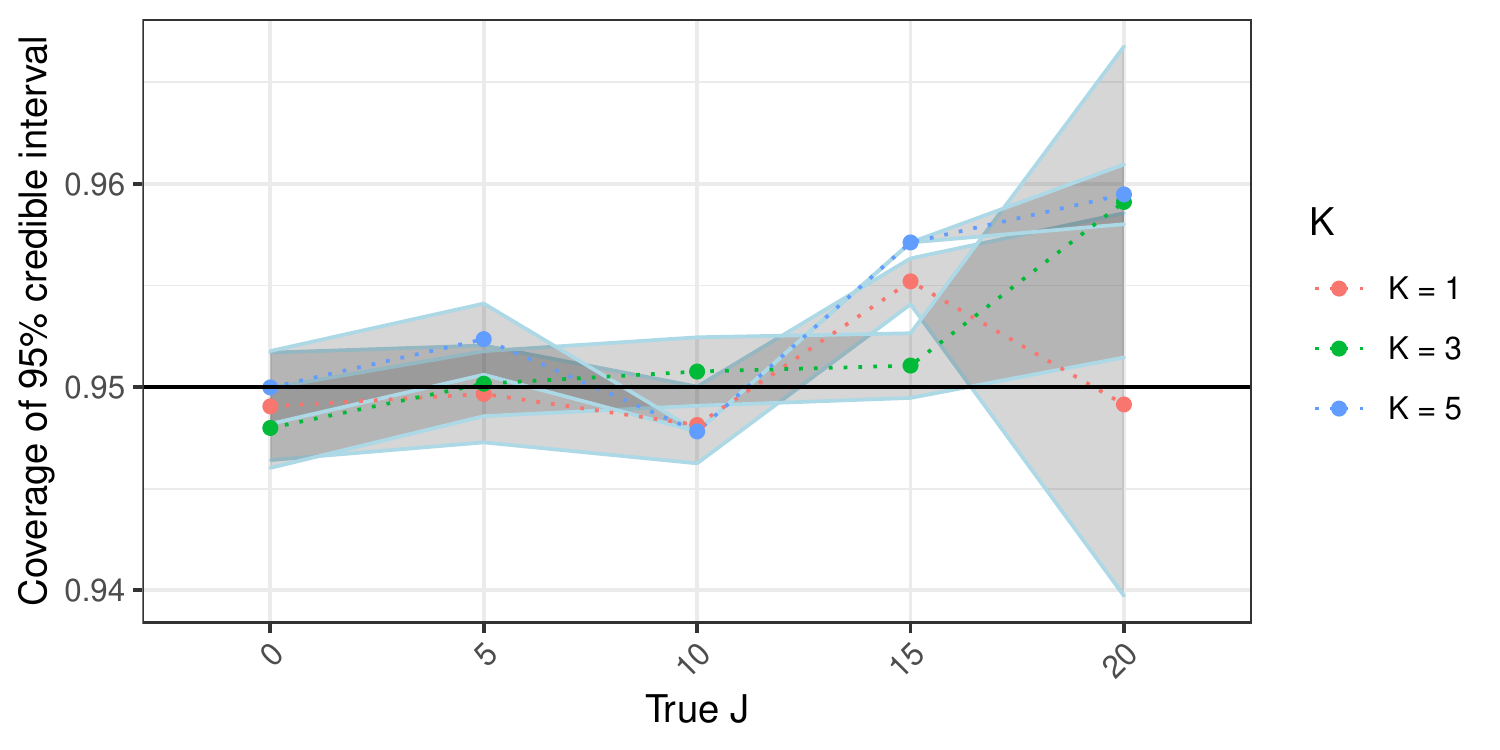}
\caption{Proportion of hold-out chemicals' (noisy) data points covered by the $\text{BS}^3\text{FA}$ 95\% credible/confidence intervals.}
\label{simResCovgbs3fa}
\end{figure}

\begin{figure}[!htpb]
\centering
\includegraphics[width=0.6\textwidth]{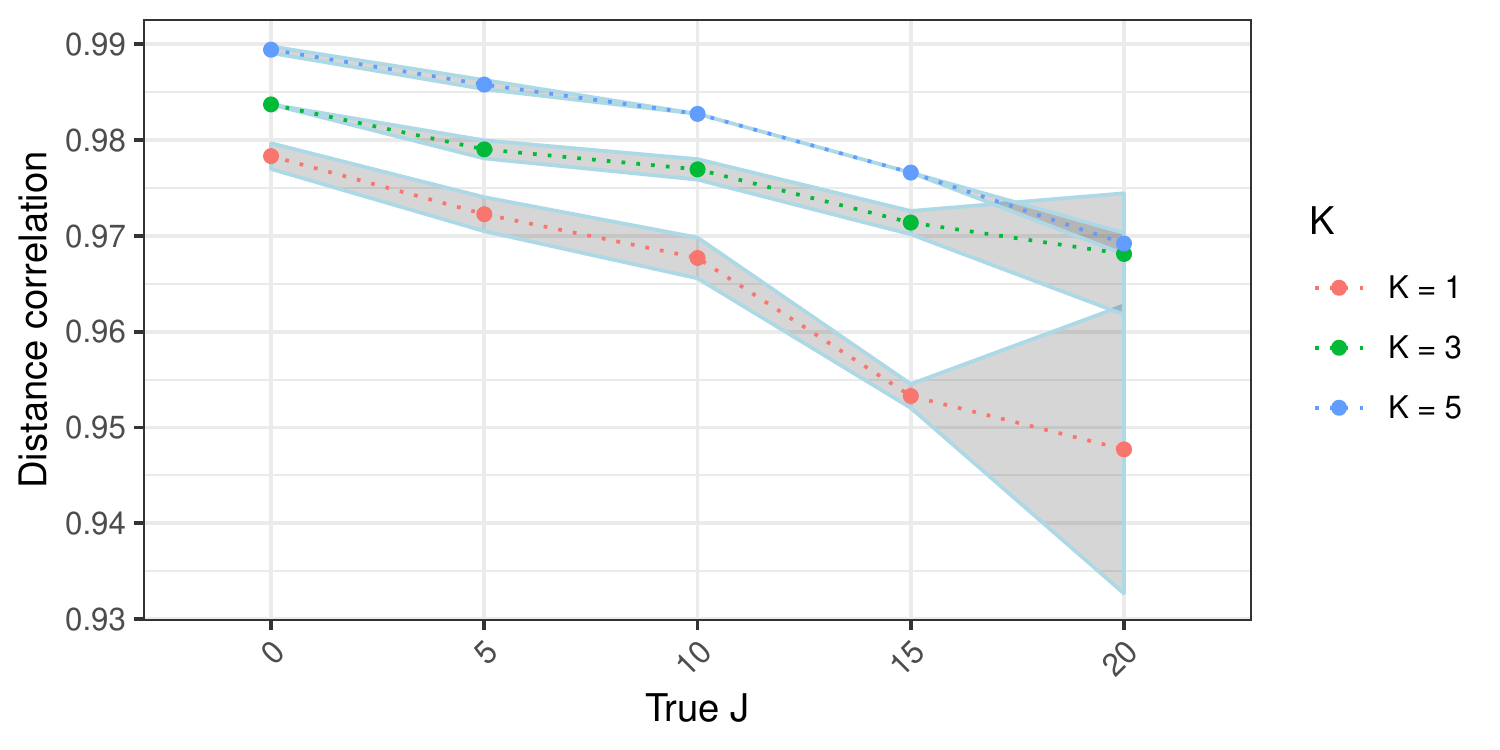}
\caption{Correlation between $\text{BS}^3\text{FA}$-predicted distances in $\bm{\eta}$ space and true distance in $\bm{\eta}$ space.}
\label{simResdistbs3fa}
\end{figure}

Figures \ref{simResMSEbs3fa} through \ref{simResdistbs3fa} show visualizations of MSE, correlation, and distance coverage results for the $\text{BS}^3\text{FA}$ model alone. It is clear that the model does best when the dimension of both $\bm{\eta}$ and $\bm{\nu}$ are smallest. That is, performance degrades as $K$ and/or $J$ increase. However, this degradation is relatively minimal, and the model performance overall is still quite good even with large $K$ and/or $J.$

\begin{figure}[!htpb]
\centering
\includegraphics[width=0.7\textwidth]{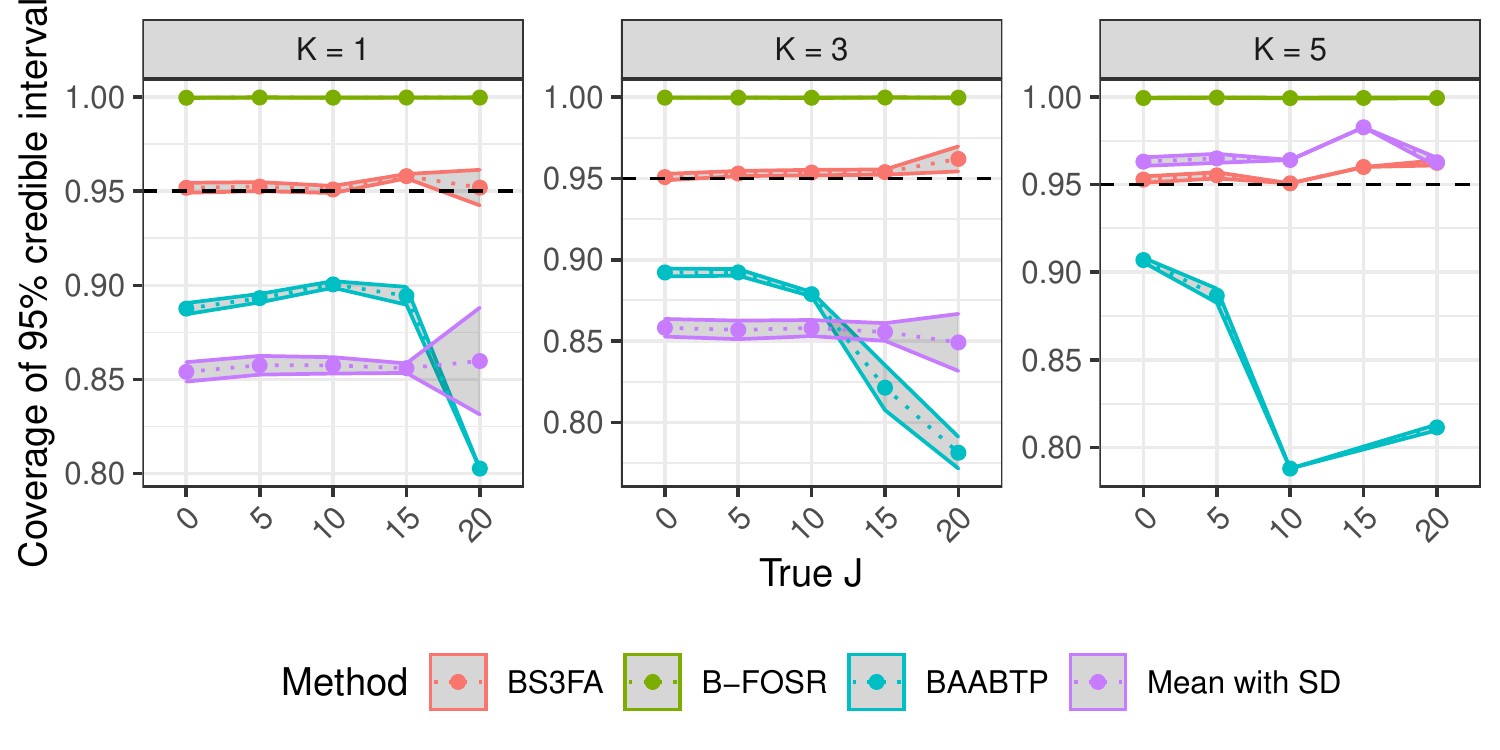}
\caption{Proportion of hold-out chemicals' (noisy) data points covered by the 95\% credible/confidence intervals for each of the methods. Each subplot shows the result across methods for a given true shared subspace dimension $K$.} 
\label{simResCovg}
\end{figure}

Figure \ref{simResCovg} shows the coverage for the hold-out chemicals' dose-response values. The $\text{BS}^3\text{FA}$ model generally has close-to-nominal coverage and is robust to increasing ``superfluous'' information in $X,$ whereas the BAABTP model is harmed by its presence. This phenomenon is likely due to the distance based kernel and the flexibility of the BAABTP model; the kernel cannot separate relevant from irrelevant features, which leads to overfitting of the training data and poor performance on holdout data.

Of course, predictive ability and ability to characterize distance in $\bm{\eta}$ space are not the only model capabilities of interest. Model components may be interrogated for interpretation as well. Table \ref{table:sim_inference_results} provides an assessment of the fit for various model components of $\text{BS}^3\text{FA}$. The MSE for $\sigma_x^2$ is poor when $J$ is small because there is a lack of identifiability between the non-toxicity-relevant structure component $\Xi \bm{\nu}$ and the noise term $\sigma_x^2.$ Since the latent dimension $J$ is set to an over-estimate when running $\text{BS}^3\text{FA},$ there is ``room'' in $\Xi \bm{\nu}$ to absorb $\sigma_x^2.$ As $J$ increases, the model becomes greedier and needs to use up the $\Xi \bm{\nu}$ component to explain non-noise variability in $X$ not shared with $Y.$ On the other hand, the terms $\Lambda \Lambda'$, $\Theta \Theta'$, and $\sigma_y^2$ are all identifiable and are consistently well estimated by our model.

\begin{table}[!phtb]
    \begin{tabular}{l c c c c c c }\toprule
        & & $J=0$ & $J=5$ & $J=10$ & $J=15$ & $J=20$ \\\midrule
        \rule{0pt}{4ex}& $K=1$ & 4e-05 (3e-05) & 4e-05 (2e-05) & 4e-05 (2e-05) & 4e-05 (2e-05) & 4e-05 (2e-05) \\
		$\Lambda \Lambda'$ & $K=3$ & 3e-05 (0) & 5e-05 (1e-05) & 7e-05 (0) & 8e-05 (4e-05) & 0.00011 (5e-05) \\
 		& $K=5$ & 4e-05 (2e-05) & 5e-05 (2e-05) & 5e-05 (0) & 3e-05 (0) & 8e-05 (3e-05) \\
        \rule{0pt}{4ex}& $K=1$ & 9e-05 (3e-05) & 0.00011 (6e-05) & 1e-04 (6e-05) & 7e-05 (3e-05) & 0.00015 (0.00011) \\
		$\Theta \Theta'$ & $K=3$ & 3e-05 (0) & 3e-05 (0) & 4e-05 (0) & 4e-05 (1e-05) & 4e-05 (2e-05) \\
 		& $K=5$ & 2e-05 (0) & 2e-05 (0) & 1e-05 (0) & 2e-05 (0) & 3e-05 (1e-05) \\
        \rule{0pt}{4ex}& $K=1$ & 0.00011 (0.00015) & 0.00011 (0.00017) & 2e-05 (2e-05) & 1e-05 (1e-05) & 1e-05 (2e-05) \\
		$\sigma_y^2$ & $K=3$ & 1e-05 (0) & 2e-05 (0) & 5e-05 (0) & 3e-05 (3e-05) & 3e-05 (4e-05) \\
 		& $K=5$ & 3e-05 (3e-05) & 3e-05 (3e-05) & 1e-05 (0) & 3e-05 (1e-05) & 8e-05 (7e-05) \\
        \rule{0pt}{4ex}& $K=1$ & 0.4350 (0.0373) & 0.0369 (0.0052) & 0.0121 (0.0014) & 0.0064 (0.0002) & 0.0045 (0.0003) \\
		$\sigma_x^2$ & $K=3$ & 0.2065 (0.0267) & 0.0257 (0.0029) & 0.0103 (0.0009) & 0.0064 (0.0003) & 0.0051 (0.0001) \\
 		& $K=5$ & 0.1123 (0.0182) & 0.0216 (0.0029) & 0.0091 (0.0008) & 0.0056 (0.0003) & 0.0060 (0.0001) \\
        \\ \bottomrule
    \end{tabular}
    \caption{MSE between the true and estimated values for each model component. These components characterize the structured toxicity-relevant directions of variation and the noise variance in the simulated data.}\label{table:sim_inference_results}
\end{table} 

Recall that a chemical is deemed active if its global Bayesian p-value is less than 0.05. Table \ref{tab:TPR_FPR_FDR} shows the true positive rate (TPR), false positive rate (FPR), and false discovery rate (FDR) of the B-FOSR method on the simulated data. Across all values of $K$ and $J$ the TPR is quite high, and the FPR and FDR are middling. Note that the TPR, FPR, and FDR all increase with $K$ and seem fairly invariant to changes in $J.$ That is, the model loses specificity when the dimension of the latent toxicity-relevant space increases.

\begin{table}[!phtb]
    \begin{tabular}{l c c c c c c }\toprule
        & & $J=0$ & $J=5$ & $J=10$ & $J=15$ & $J=20$ \\\midrule
        \rule{0pt}{4ex}& $K=1$ & 0.82 (0.07) & 0.82 (0.07) & 0.81 (0.08) & 0.82 (0.04) & 0.76 (0.07) \\
		TPR & $K=3$ & 1.00 (0.01) & 1.00 (0.01) & 0.99 (0.01) & 0.99 (0.01) & 0.99 (0.02) \\
 		& $K=5$ & 1.00 (0.00) & 1.00 (0.00) & 1.00 (0.00) & 1.00 (0.00) & 1.00 (0.00) \\
        \rule{0pt}{4ex}& $K=1$ & 0.13 (0.09) & 0.21 (0.12) & 0.21 (0.09) & 0.11 (0.09) & 0.18 (0.09) \\
		FPR & $K=3$ & 0.31 (0.10) & 0.38 (0.10) & 0.36 (0.09) & 0.38 (0.10) & 0.39 (0.11) \\
 		& $K=5$ & 0.45 (0.10) & 0.48 (0.13) & 0.49 (0.09) & 0.48 (0.09) & 0.49 (0.08) \\
        \rule{0pt}{4ex}& $K=1$ & 0.13 (0.07) & 0.19 (0.09) & 0.20 (0.08) & 0.11 (0.08) & 0.18 (0.09) \\
		FDR & $K=3$ & 0.24 (0.06) & 0.28 (0.07) & 0.26 (0.06) & 0.27 (0.07) & 0.28 (0.08) \\
 		& $K=5$ & 0.30 (0.07) & 0.31 (0.07) & 0.33 (0.06) & 0.32 (0.06) & 0.33 (0.05) \\
        \\ \bottomrule
    \end{tabular}
    \caption{True positive rate (TPR), false positive rate (FPR), and false discovery rate (FDR) for the B-FOSR model under the proposed method of assessing whether a chemical is active. A perfect classifier has a TPR of 1 and an FPR/FDR of 0.}\label{tab:TPR_FPR_FDR}
\end{table}

\subsection{MSPE, coverage, and distance results for model-truth-misaligned simulation}\label{sec:SI_sim_res_misaligned}

\begin{figure}[!htpb]
\centering
\includegraphics[width=0.7\textwidth]{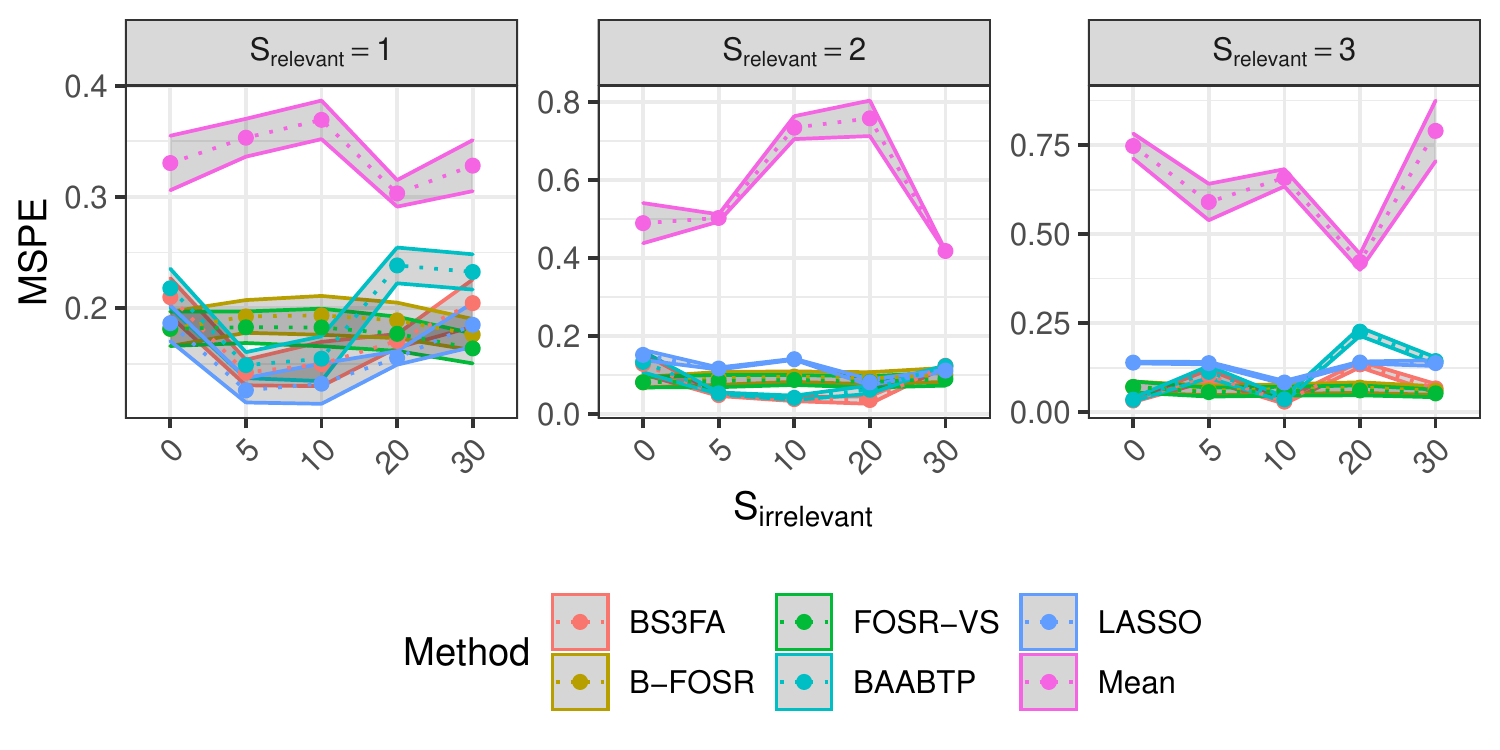}
\caption{Mean squared predictive error (MSPE) for the hold-out chemicals' dose-response mean functions under the polynomial model.} 
\label{fig:simResMSPE_misspec}
\end{figure}

Figure \ref{fig:simResMSPE_misspec} shows the mean squared predictive error (MSPE) for the hold-out chemicals' dose-response mean functions when there is misalignment between the structure assumed by the $\text{BS}^3\text{FA}$ model and the true data generating process. In spite of this misalignment, $\text{BS}^3\text{FA}$ is able to predict similarly to or better than the competitors. As with the well-aligned simulation, $\text{BS}^3\text{FA}$ appears robust to increasing ``superfluous'' information in $X.$ 

\begin{figure}[!htpb]
\centering
\includegraphics[width=0.7\textwidth]{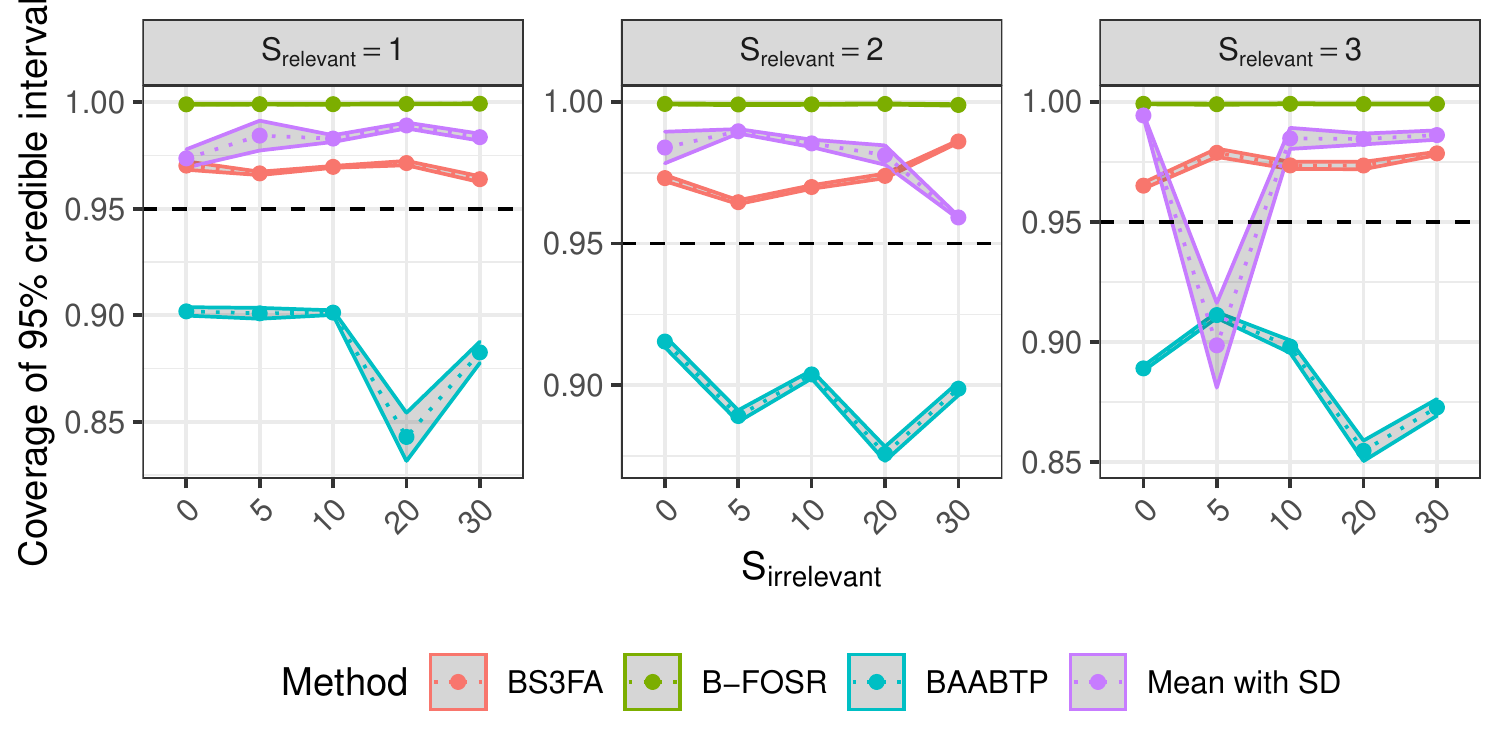}
\caption{Proportion of hold-out chemicals' (noisy) data points covered by the 95\% credible/confidence intervals for each of the methods under misalignment between the true and $\text{BS}^3\text{FA}$-assumed structure.} 
\label{simResCovg_missp}
\end{figure}

\begin{figure}[!htpb]
\centering
\includegraphics[width=0.7\textwidth]{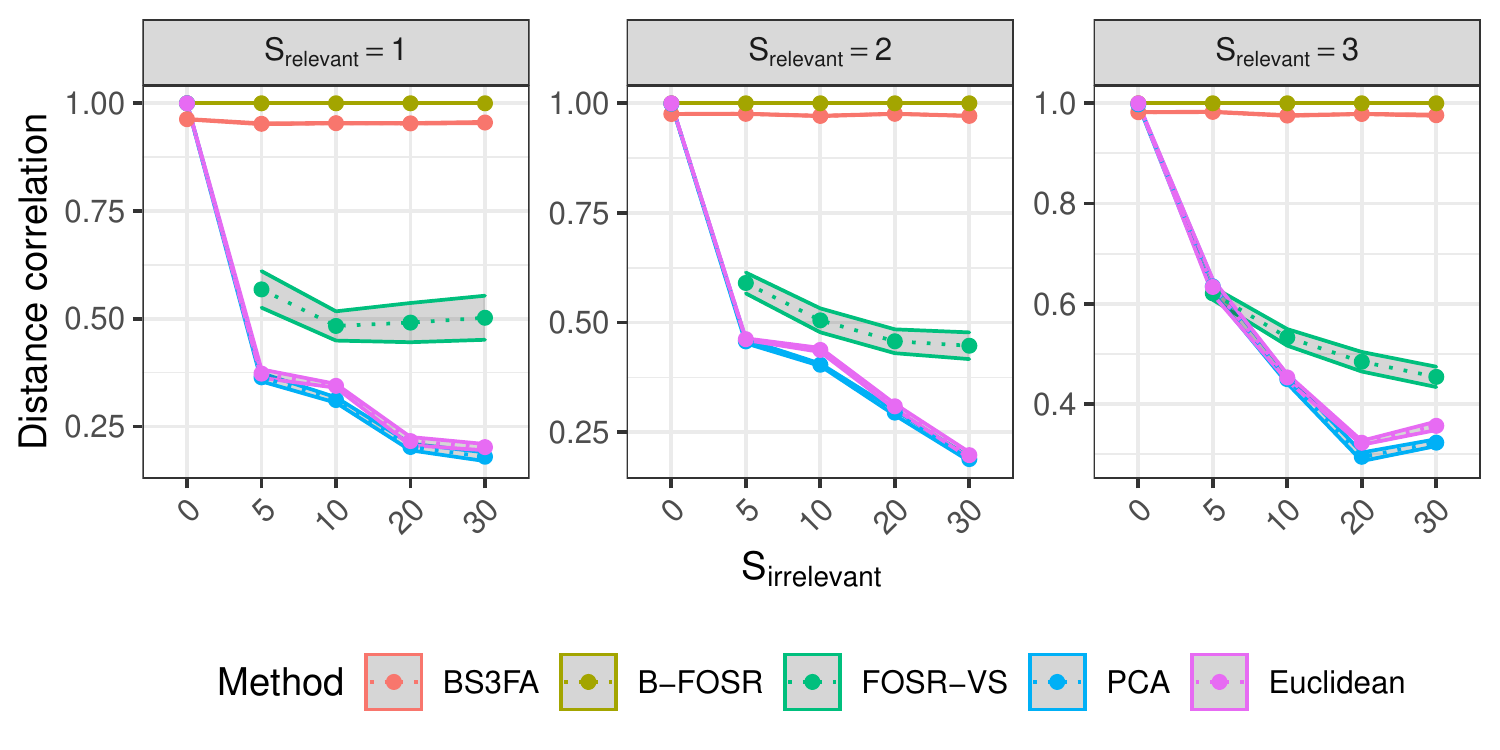}
\caption{Correlation between entries in the true pairwise distance matrix (i.e., the Euclidean distance between true relevant dimensions of $X$) and the predicted pairwise distance matrix for holdout chemicals under misalignment between the true and $\text{BS}^3\text{FA}$-assumed structure.} 
\label{simResDist_missp}
\end{figure}

A similar story can be seen in the coverage and distance results (Figures \ref{simResCovg_missp} and \ref{simResDist_missp}) for the misaligned simulation as for the model-truth-aligned simulation. Namely, $\text{BS}^3\text{FA}$ has  closest-to-nominal (albeit still somewhat high) coverage, while other methods are less well calibrated. $\text{BS}^3\text{FA}$ has very stable, high correlation between the true Euclidean distance matrix for the relevant parts of $X,$ and that of the model-predicted $\eta.$ Direct PCA, FOSR-VS, and Euclidan distance suffer immediately and harshly when $S_{\text{irrelevant}}$ exceeds 0.

\section{Additional ToxCast run information}

\subsection{Data download}\label{data_download}

A pre-cleaned \textbf{\textsf{R}} data file holding a subset of the information from the ToxCast database was created by Dr. Matthew Wheeler and can be downloaded from \url{https://1drv.ms/ u/s!AoYFThhStiORt0YjFGEQDKjBa4BZ}. The download, \texttt{gain.Rdata}, is a file containing the dose-response information for all Phase 1, Phase 2, and e1k chemicals tested across all ToxCast assays. For this analysis, only results for the the Attagene PXR assay (i.e., the assay having value 135 for the variable \texttt{aeid}) were saved and are provided with this manuscript as the file \texttt{atg\_pxr\_data.Rdata}. The full ToxCast data are available for download from \url{https://www.epa.gov/ chemical-research/exploring-toxcast-data-downloadable-data}.

\subsection{Results and diagnostics}
\label{SIsec:tox_run_more}

\begin{figure}[!htpb]
\centering
\includegraphics[width=0.85\textwidth]{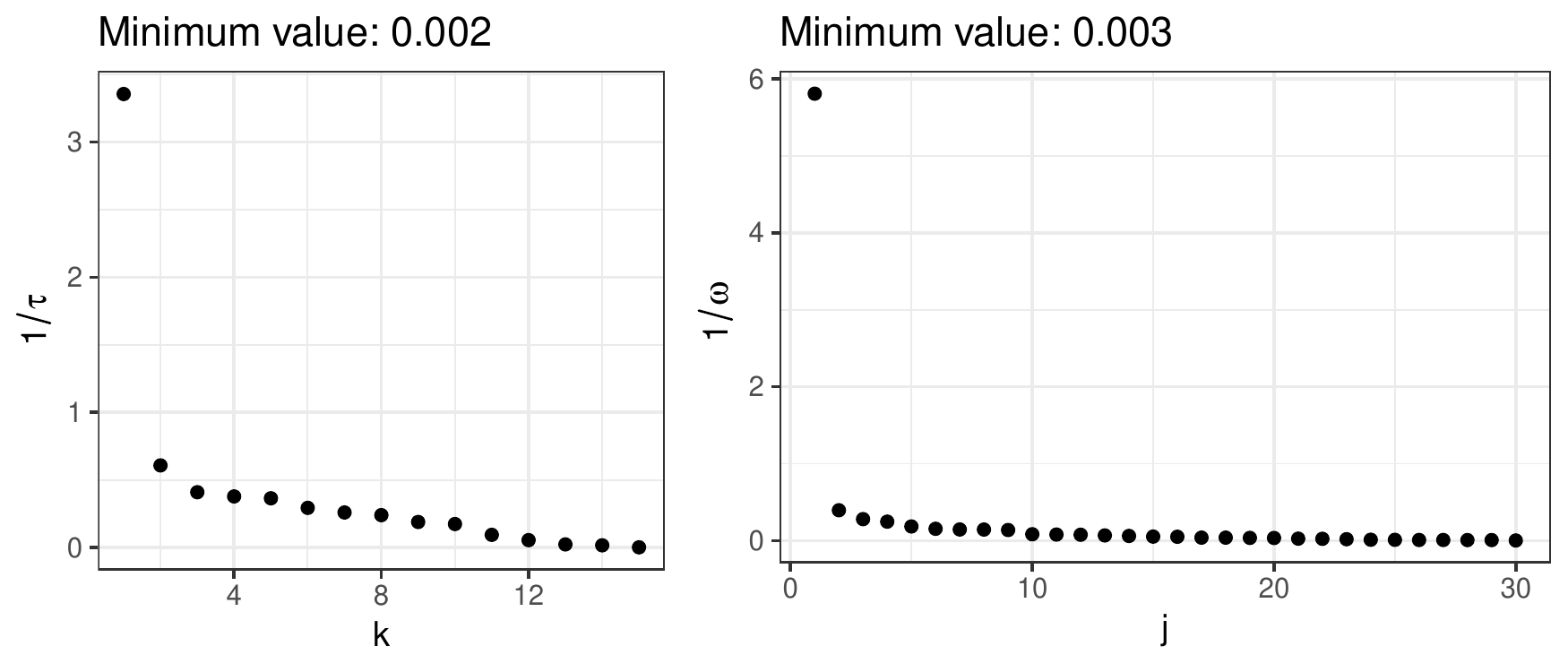}
\caption{Ordered variance elements driving column-specific shrinkage of the factor loadings matrices $\Lambda$ and $\Theta$ (left), and $\Xi$ (right).}
\label{fig:norm_diagnostic_tau}
\end{figure}

We consider $K$ to be chosen ``large enough'' if the smallest value of $1/\tau_j$ is below 0.01. Similarly, we consider $J$ to be chosen ``large enough'' if the smallest value of $1/\omega_j$ is below 0.01. Figure \ref{fig:norm_diagnostic_tau} shows the ordered values $1/\tau_j$ and $1/\omega_j$ for the Tox run.

\begin{figure}[!htpb]
\centering
\includegraphics[width=0.88\textwidth]{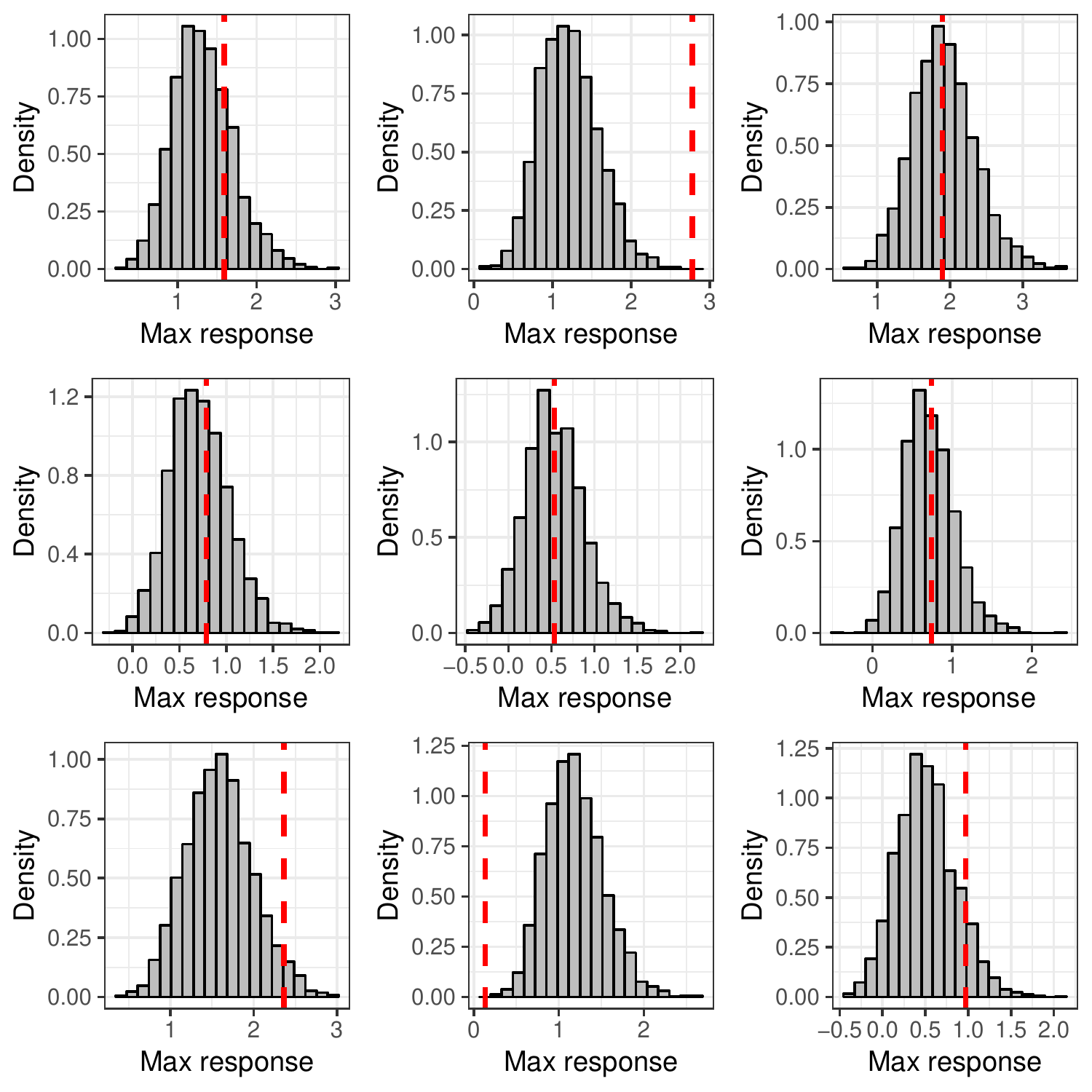}
\caption{Posterior predictive plots for the maximum observed response value for select test chemicals. The true maximum observed response value is indicated by a dashed vertical line. The posterior predictive does not appear misaligned with the data except in the case of poorly fitting curves generally or apparent data outliers. For example, the bottom left plot corresponds to general model under prediction of the dose response curve for that chemical.}
\label{fig:postpred_diagnostic}
\end{figure}

Figure \ref{fig:postpred_diagnostic} shows posterior predictive plots for the maximum observed response value for select test chemicals with the actual maximum observed response value denoted by a dashed vertical line. The chemicals shown in the first two rows are the same hold-out chemicals as those shown in Figure 19 and the first row of Figure 20 in the main paper; the final row shows three chemicals having particularly high MSPE (i.e. poor predictions). Unsurprisingly, the poorer the overall fit of the predicted dose-response profile to the data, the farther toward the boundary of the posterior predictive distribution the observed data are. For example, the bottom left subplot, in which model posterior predictions are all much lower than the observation, reflects the fact that this chemical was chronically under-predicted by the model. When the predicted dose response profile seems well aligned with the data the posterior predictive plots show minimal signs of misalignment between the posterior and the data.
 
\begin{figure}[!htpb]
\centering
\includegraphics[width=0.8\textwidth]{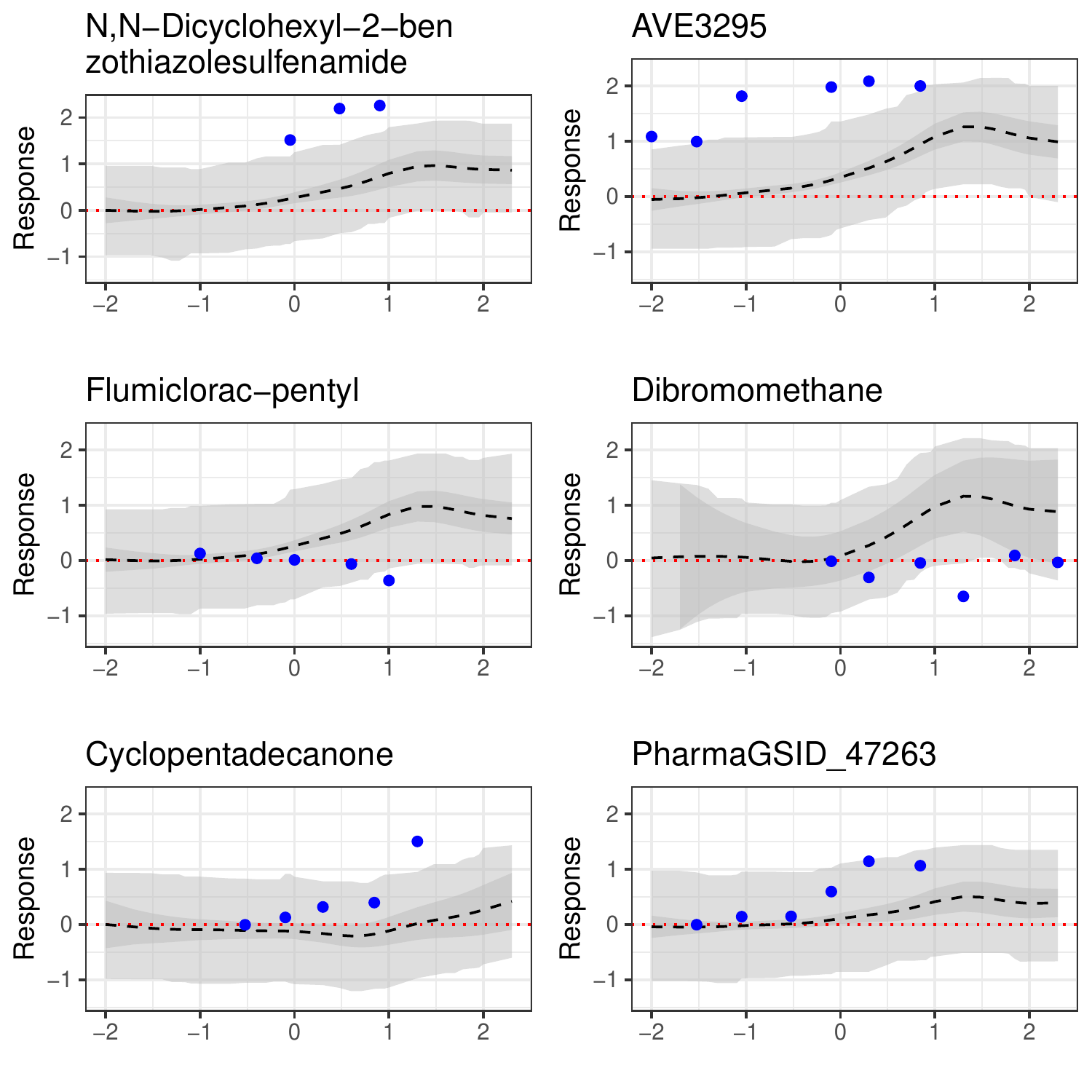}
\caption{Results for select hold-out chemicals for which the fit between predicted dose response curve and observed data is particularly poor. Shown are predicted average dose-response curve (dashed black line), 95\% simultaneous interval for expected dose-response curve (darker grey ribbon), and 95\% credible interval for observed data (lighter grey ribbon). Data (held out in training) are solid blue points. Top: both chemicals have abnormally high response values. While $\text{BS}^3\text{FA}$ predicts both to be activating, it does not predict the height of the activity. Middle: these chemicals both appear non-activating, but the model predicted that these chemicals were activating. Bottom: both chemicals appear activating, but the model predicted that these chemicals were non-activating or of low activity.}
\label{fig:res_inference_yPred_bad}
\end{figure}

Figure \ref{fig:res_inference_yPred_bad} shows the results for hold-out chemicals having poorly predicted dose response curves. These poor predictions may be a sign that the Mold2 structure information doesn't contain enough toxicity-relevant information to fully inform the dose response curve shapes, that we do not have enough training data, and/or that underlying model assumptions are incomplete. Further work should explore which chemicals are poorly predicted and why.

\begin{figure}[!htbp]
\centering
\includegraphics[width=0.75\textwidth]{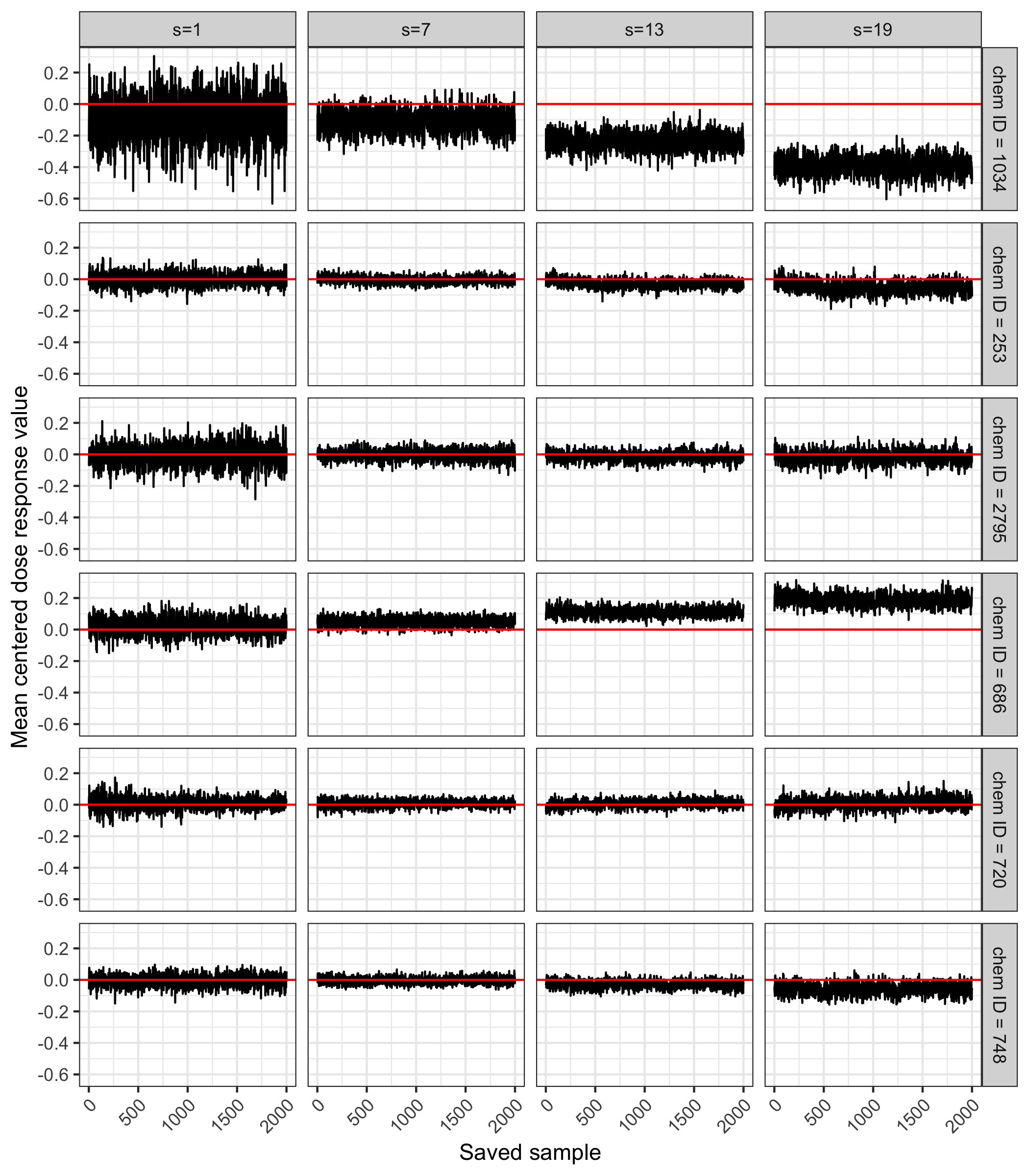}
\caption{Trace plots for randomly selected hold-out chemicals' response at multiple doses (indexed by $s$ in column headers).}
\label{fig:trace_predDR}
\end{figure}

\begin{figure}[!htbp]
\centering
\includegraphics[width=0.6\textwidth]{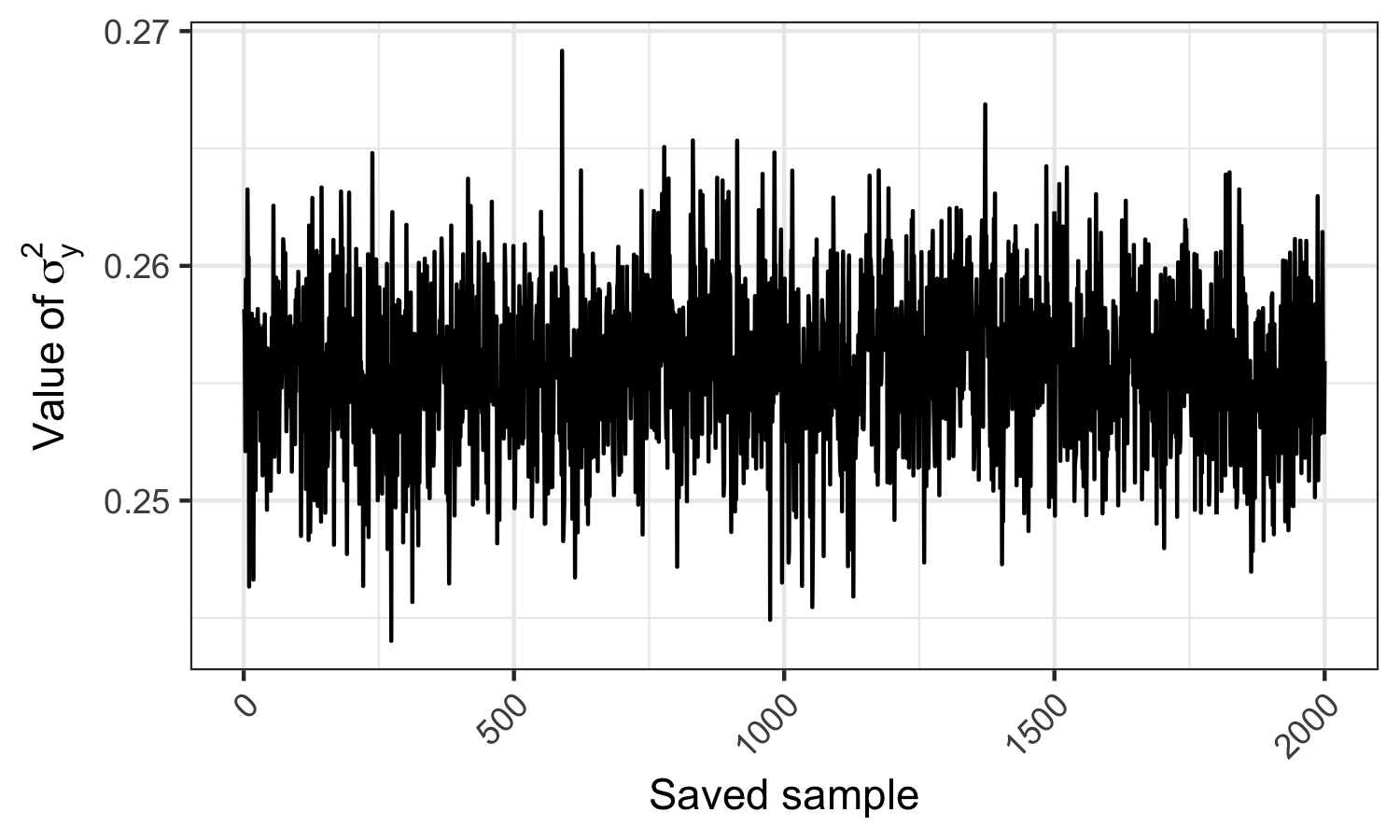}
\caption{Trace plot showing samples of the noise variance term for $Y$.}
\label{fig:trace_sigy}
\end{figure}

Trace plots for the predicted dose response profiles of hold-out chemicals show good mixing (a randomly sampled set of chemicals are shown across multiple doses in Figure \ref{fig:trace_predDR}), as do the noise variance terms for the data (samples of $\sigma_Y$ are shown in Figure \ref{fig:trace_sigy}). 

\begin{figure}[!htbp]
\centering
\includegraphics[width=0.85\textwidth]{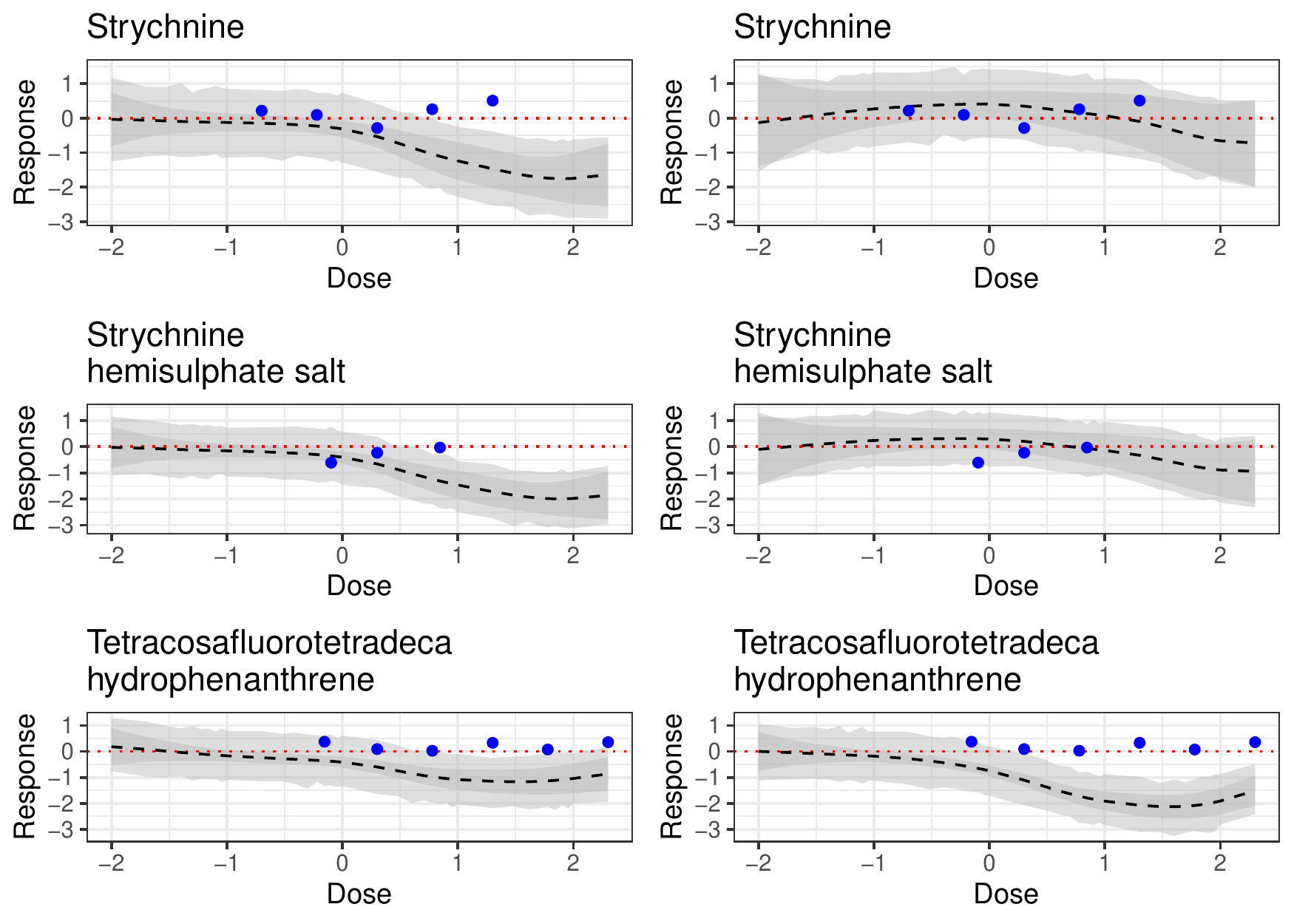}
\caption{The predicted dose response curves for hold-out chemicals having the largest divergence between predicted mean across two chains. While the exact shape of the predicted curve differs (e.g., note how in the bottom row chain 1 predicts a slightly flatter curve with less of a U shape at the end than chain 2), small regions of multi-modality do not concern us. Even in the ``worst'' behaving chemicals the general direction of effect is similar, and the chemical profiles predicted are consistent for the large majority of chemicals.}
\label{fig:pred_test_badGR}
\end{figure}

\begin{figure}[!htbp]
\centering
\includegraphics[width=0.85\textwidth]{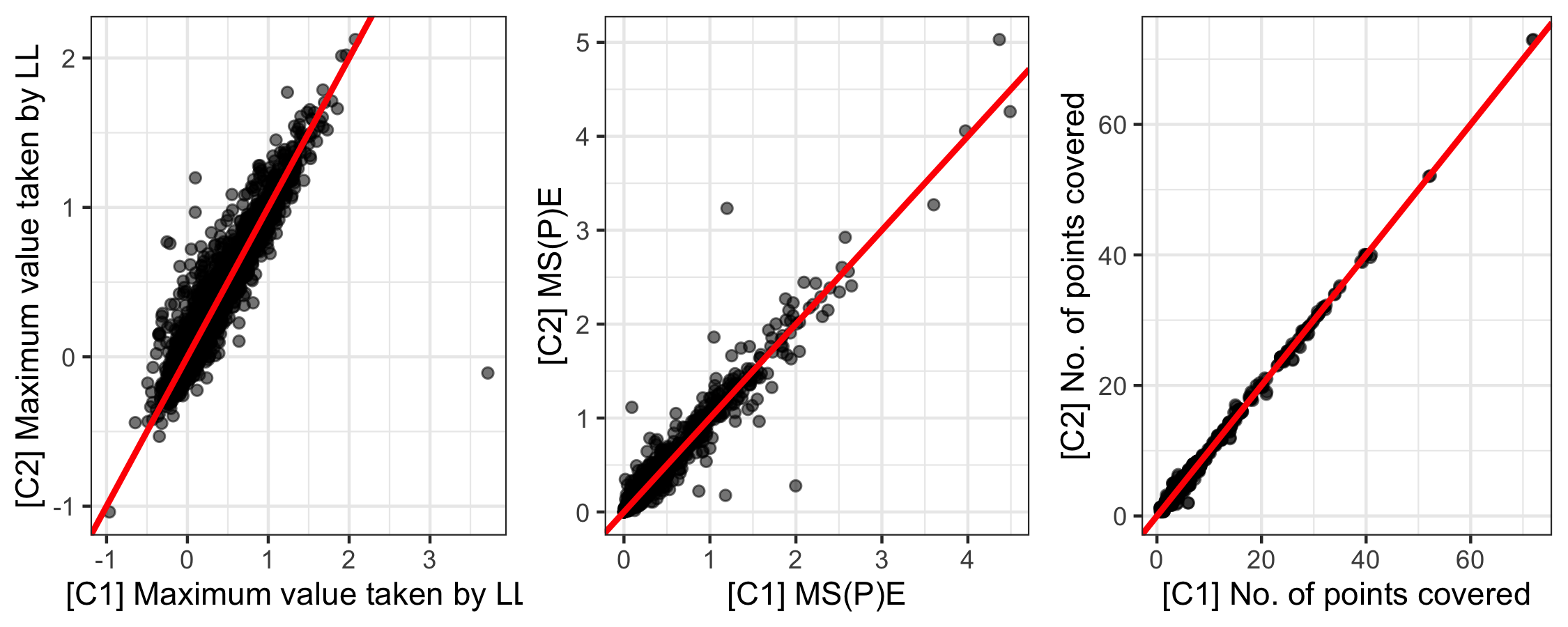}
\caption{A comparison between two chains of the following chemical-specific value. From left to right: the maximum value of the lower bound for the dose response curve; the mean square (predictive) error for training (test) chemicals; the number of points covered by the model 95\% posterior data credible interval interval. Red line denotes $x=y$.}
\label{fig:chain_res}
\end{figure}

While there are some inconsistencies between predicted values across multiple chains at higher doses (see Figure \ref{fig:pred_test_badGR}), the general curve shapes and predicted summary quantities are consistent across chains (see Figure \ref{fig:chain_res}).  

\clearpage

\begin{figure}[!htbp]
\centering
\includegraphics[width=0.55\textwidth]{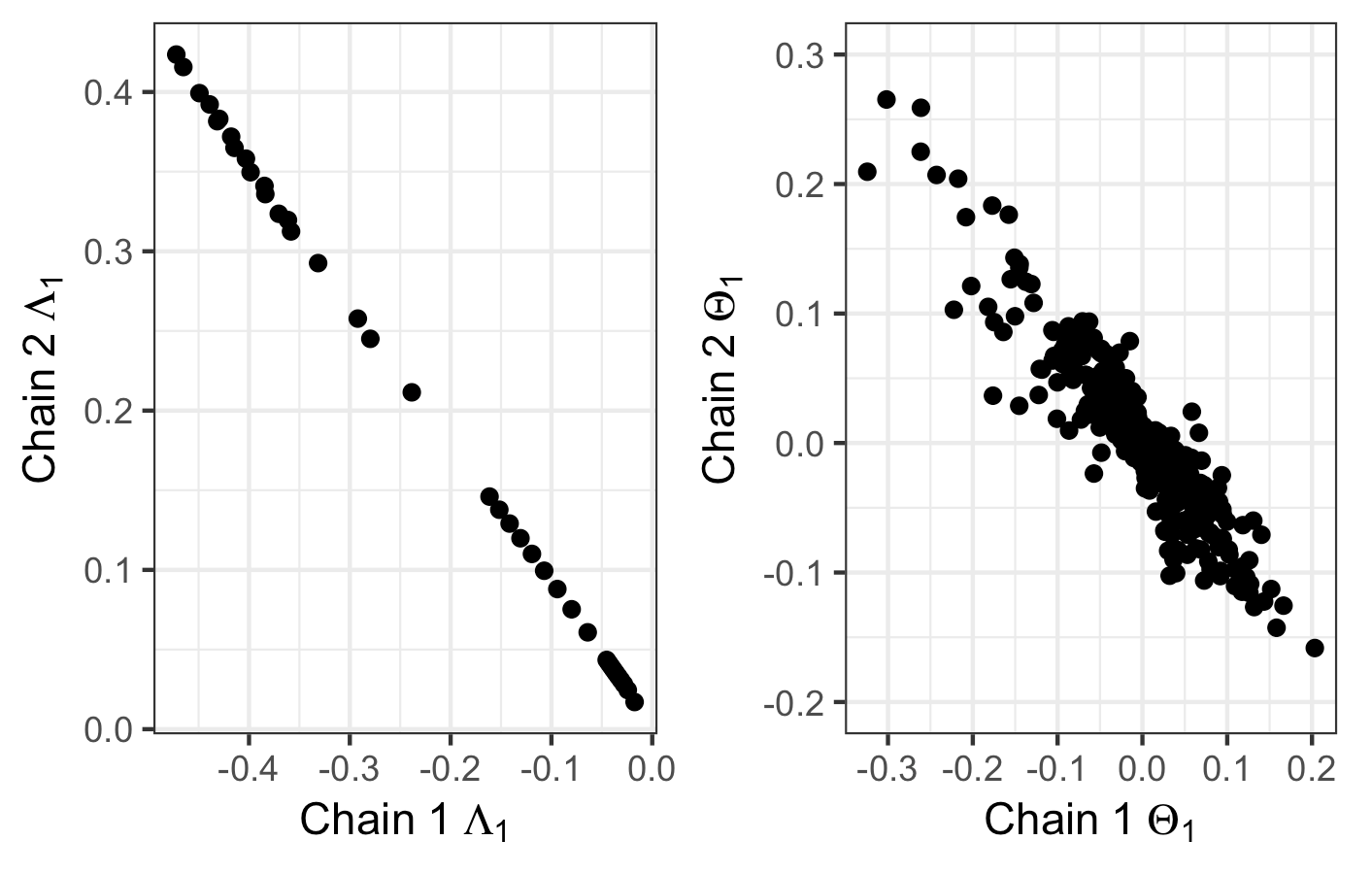}
\caption{A comparison between two chains of model predicted components. Left: the column of $\Lambda$ having the largest 2-norm. Right: the associated column of $\Theta$. Note that scales and signs differ across chains, but general relationships remain consistent.}
\label{fig:chain_res_more}
\end{figure}

\begin{figure}[!htbp]
\centering
\includegraphics[width=0.95\textwidth]{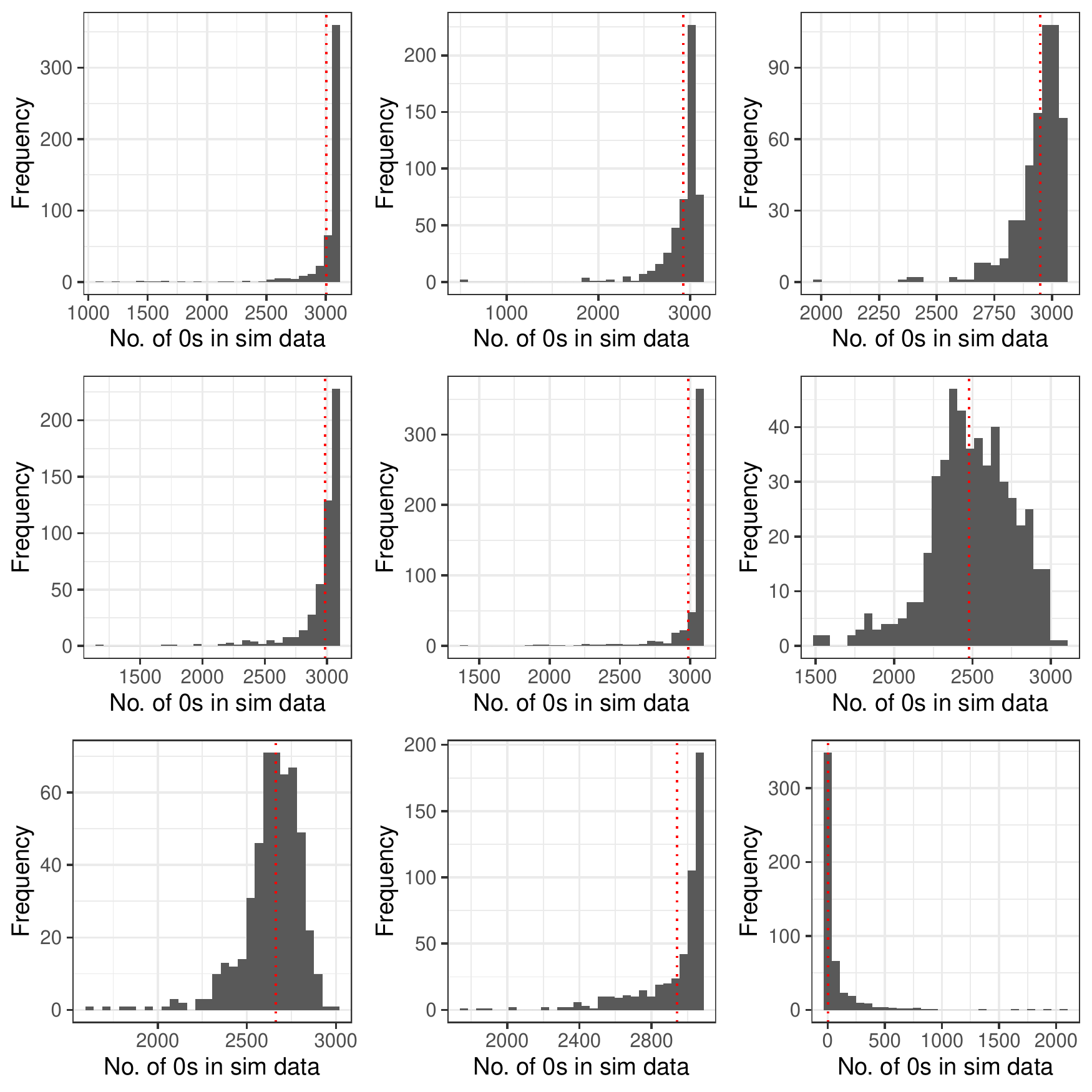}
\caption{Posterior predictive histograms for randomly selected count chemical features showing model-predicted draws of the number of 0s in simulated datasets (histogram) and the observed number of 0s in the real dataset (vertical red line).}
\label{fig:postPred_count0}
\end{figure}

\begin{figure}[!htbp]
\centering
\includegraphics[width=0.95\textwidth]{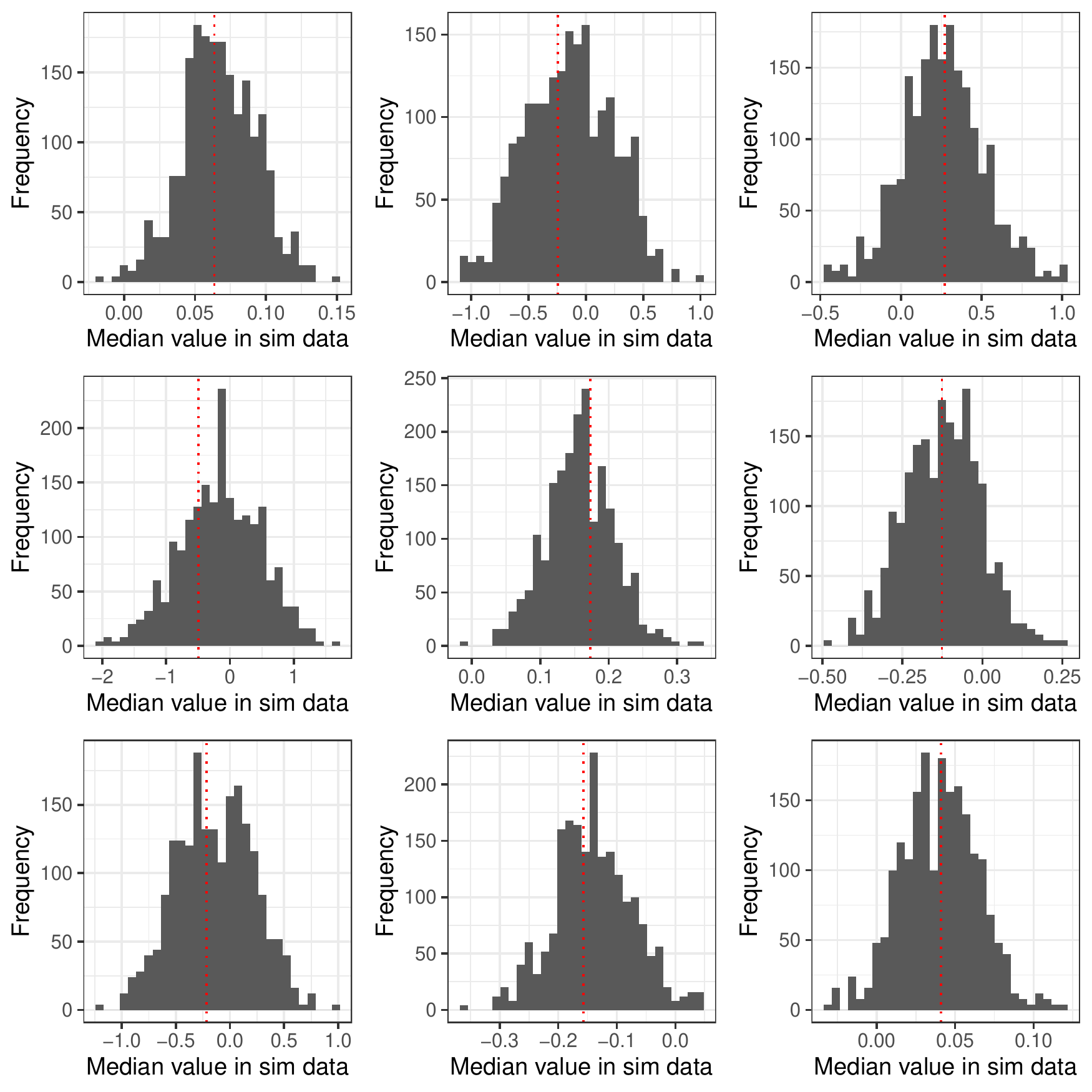}
\caption{Posterior predictive histograms for randomly selected continuous chemical features showing model-predicted draws of the median value in simulated datasets (histogram) and the observed median in the real dataset (vertical red line).}
\label{fig:postPred_median}
\end{figure}

\begin{figure}[!htbp]
\centering
\includegraphics[width=0.95\textwidth]{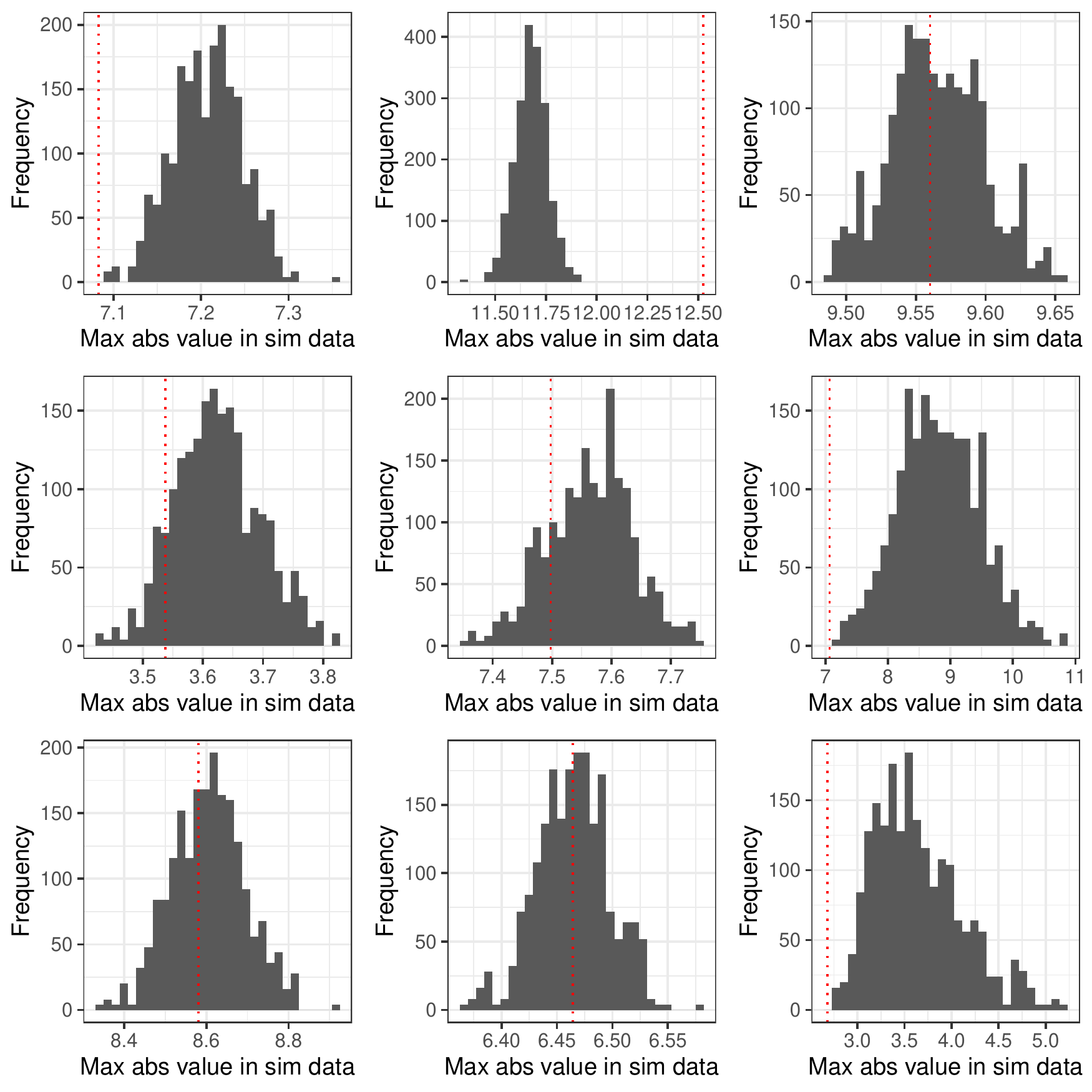}
\caption{Posterior predictive histograms for randomly selected continuous chemical features showing model-predicted draws of the maximum absolute value in simulated datasets (histogram) and the observed maximum absolute value in the real dataset (vertical red line).}
\label{fig:postPred_maxabs}
\end{figure}

Figures \ref{fig:postPred_count0} through \ref{fig:postPred_maxabs} show posterior predictive histograms for randomly selected chemical feature sets $X$. In general, the model posterior predictions are consistent with observed data. An exception to this is for a small number of continuous features having heavier-than-normal tails, for which the model posterior predicted medians are consistent with observed data but model posterior predicted maximum absolute values are less than the true maximum absolute value. Exploring allowing for heavier tails in the feature data is an avenue of future research.

\section{Data structure for chemical features}
\label{SIsec:data_struct_features}

The Mold2 software was downloaded from \url{https://www.fda.gov/scienceresearch/bioinformaticstools/mold2/default.htm}. A description of the process for generating Mold2 descriptors using the information provided by ToxCast is provided with this manuscript as the file \texttt{workflow.txt}.

\subsection{``Identical'' chemicals}
\label{SIsec:data_struct_features_identical}

Our empirical examination showed that the majority of chemicals grouped together via having the same-Mold2-output the dose-response profiles are similar up to noise (see some example output in Figure \ref{ident_chems_dr}, in which the grey/black points are chemicals having different SMILES but identical Mold2 output). We suspect that augmenting Mold2 is potentially useful, but not critical for the purpose of the illustrative application in the main paper.

\begin{figure}[!h]
\centering
\includegraphics[width=0.7\textwidth]{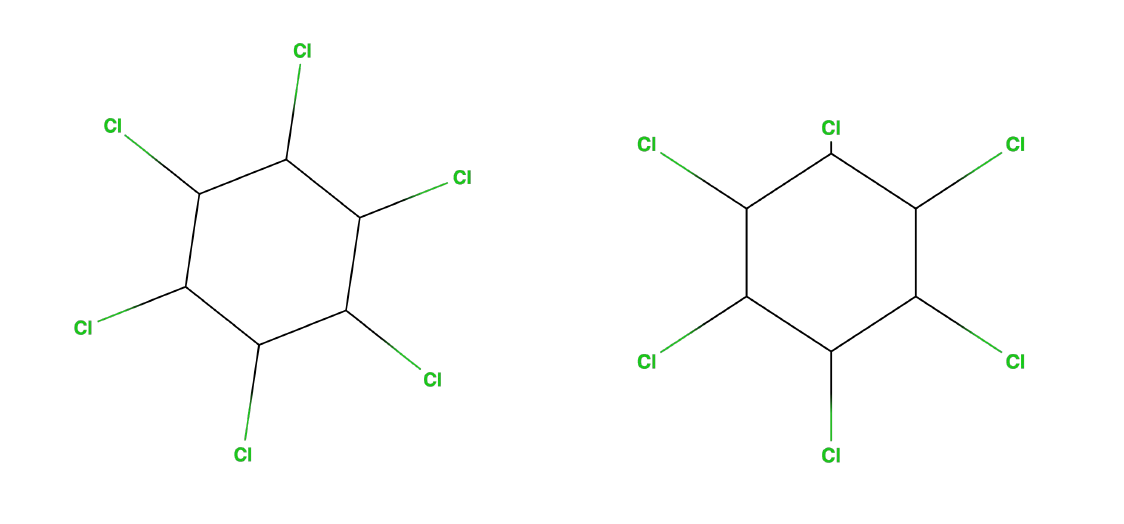}
\caption{An example of a pair of chemicals having identical Mold descriptor sets. On the left is beta-Hexachlorocyclohexane and on the right is delta-Hexachlorocyclohexane. The chemical SMILES are:\\ Cl[C@H]1[C@H](Cl)[C@@H](Cl)[C@H](Cl)[C@@H](Cl)[C@@H]1Cl and \\ Cl[C@H]1[C@H](Cl)[C@@H](Cl)[C@H](Cl)[C@H](Cl)[C@@H]1Cl, respectively.}
\label{ident_chems}
\end{figure}

\begin{figure}[!h]
\centering
\includegraphics[width=0.7\textwidth]{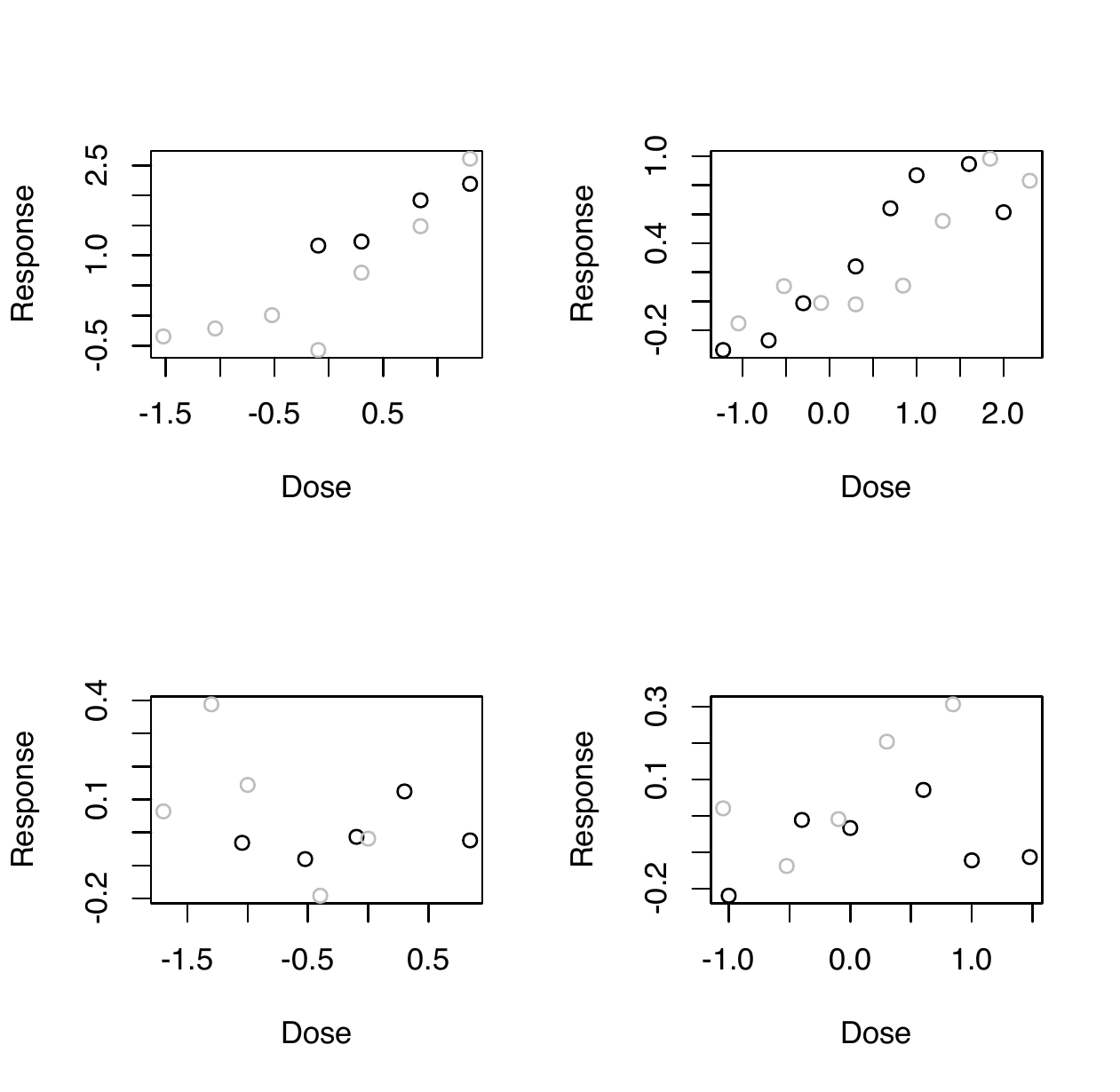}
\caption{Example dose response profiles for pairs of chemicals having identical Mold descriptor sets (point color differentiates chemical).}
\label{ident_chems_dr}
\end{figure}

\subsection{Count features}
\label{SIsec:data_struct_features_count}

The model can accommodate count data via the underlying normal assumption and rounding operator described in the main paper. However, this rounding operator is more computationally expensive than simply treating a variable as continuous, and a log transformation in many cases will allow a count feature to well-approximate a continuous normal variable.

\begin{figure}[!htbp]
\centering
\includegraphics[width=0.7\textwidth]{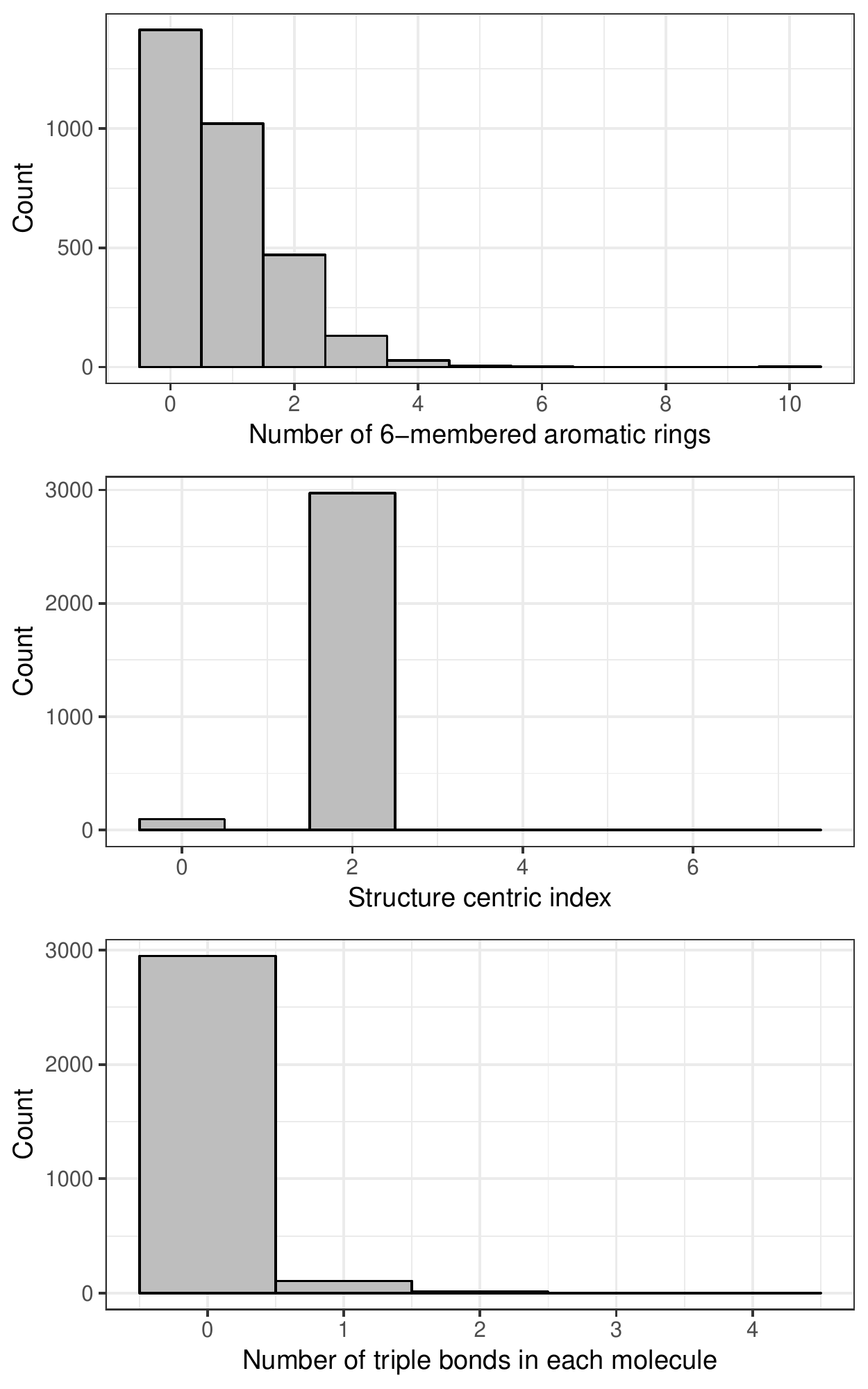}
\caption{Three examples of count data in the set of Mold2 descriptors $X$ for the included chemicals. Note that some features seem better suited to the underlying continuous assumption used in the model (i.e., the top and bottom) while others could be improved on via some other special specification based on expert input (i.e., the middle).}
\label{X_count}
\end{figure}

\begin{figure}[!htbp]
\centering
\includegraphics[width=0.8\textwidth]{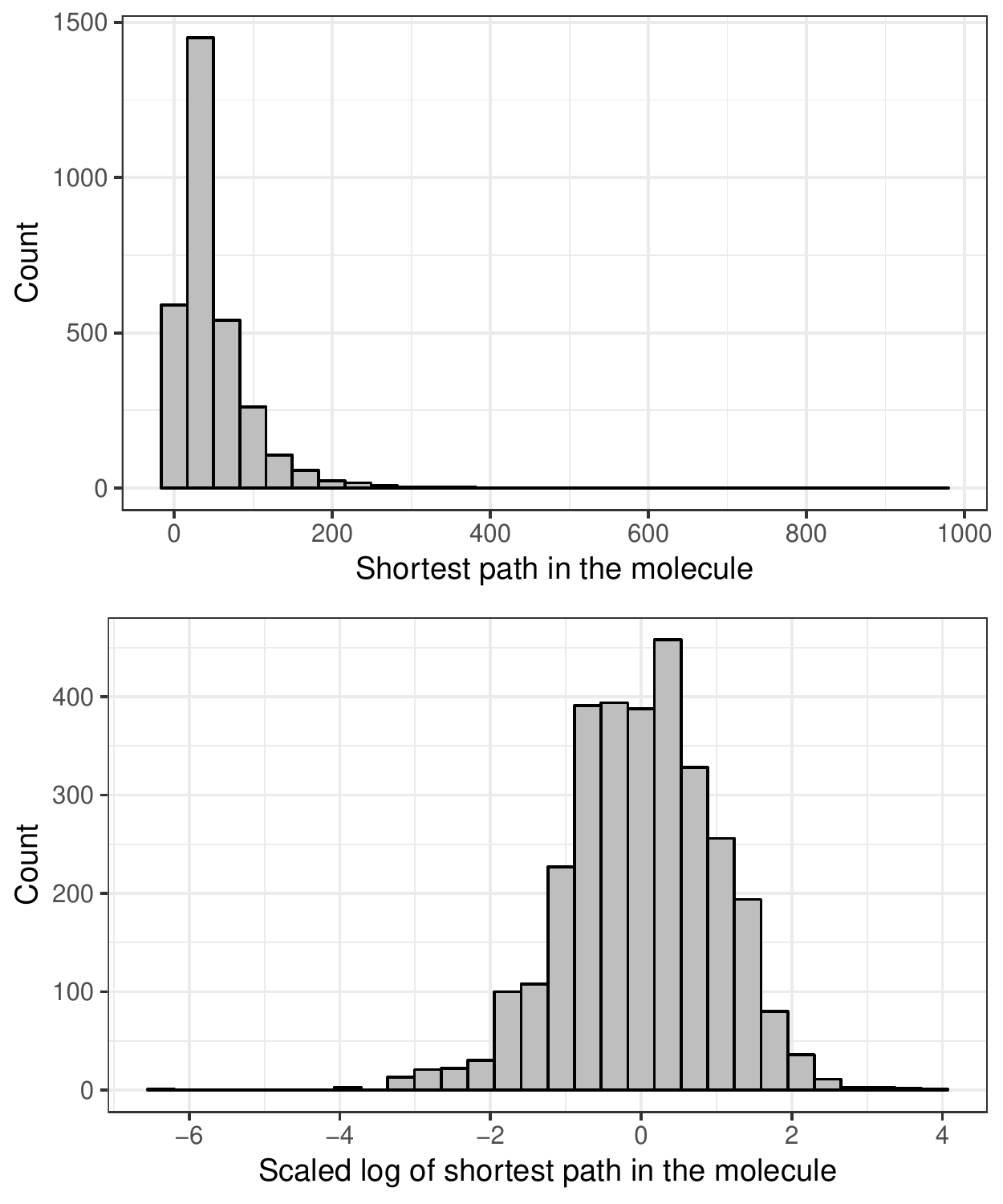}
\caption{Count variables having maximum value of greater than 10 are log transformed and then treated as continuous (i.e., scaled and centered) before their inclusion in the model. The top row shows an example of a pre-transformed feature, and the bottom row shows that same feature after taking the log, scaling, and centering.}
\label{X_count_big}
\end{figure}